\documentclass[12pt]{article}
\PassOptionsToPackage{hyphens}{url}
\usepackage[pagebackref=true,breaklinks=true,colorlinks]{hyperref}
\hypersetup{
    colorlinks,
    linkcolor={red!50!black},
    citecolor={blue!50!black},
    urlcolor={blue!80!black}
}
\usepackage{amsmath,amssymb,amsthm,mathtools,amsrefs,mathrsfs,marvosym}
\usepackage{pifont}
\usepackage{mathpazo} 
\usepackage{eulervm}
\usepackage{bbm}
\usepackage[table]{xcolor}
\usepackage{tikz}
\usetikzlibrary{arrows.meta}
\usetikzlibrary{decorations.pathmorphing,shapes}
\usepackage{comment}
\usepackage{adjustbox}
\usepackage{stmaryrd}
\usepackage{fancyhdr}
\usepackage{soul}
\usepackage{dsfont}
\usepackage[normalem]{ulem}
\usepackage{longtable}
\usepackage[bottom]{footmisc}
\usepackage{graphicx}
\usepackage{caption}
\usepackage{subcaption}
\usepackage{pictexwd, dcpic}
\usepackage{float}
\usepackage{enumerate}
\usepackage[all]{xy} 

\DeclareMathAlphabet{\mathpzc}{OT1}{pzc}{m}{it}

\usepackage{tocbasic}
\DeclareTOCStyleEntry[
  beforeskip=-0.0em plus 1pt,
  pagenumberformat=\textbf
]{tocline}{section}

\makeatletter
\newcommand*{\shifttext}[2]{%
  \settowidth{\@tempdima}{#2}%
  \makebox[\@tempdima]{\hspace*{#1}#2}%
}
\makeatother
\makeatletter
\renewcommand*\env@matrix[1][\arraystretch]{%
  \edef\arraystretch{#1}%
  \hskip -\arraycolsep
  \let\@ifnextchar\new@ifnextchar
  \array{*\c@MaxMatrixCols c}}
\makeatother

\theoremstyle{plain}
\newtheorem{theorem}[equation]{Theorem}
\newtheorem{lemma}[equation]{Lemma}
\newtheorem{proposition}[equation]{Proposition}
\newtheorem{corollary}[equation]{Corollary}
\theoremstyle{definition}
\newtheorem{definition}[equation]{Definition}
\newtheorem{construction}[equation]{Construction}
\newtheorem{question}[equation]{Question}

\newtheorem{problem}[equation]{Problem}
\newtheorem{example}[equation]{Example}
\newtheorem{exercise}[equation]{Exercise}
\newtheorem*{answer}{Answer}
\newtheorem*{solution}{Solution}
\newtheorem{remark}[equation]{Remark}

\newtheorem{notation}[equation]{Notation}
\newtheorem{noterm}[equation]{Notation and Terminology}

\newcommand\define[1]{\emph{\textbf{#1}}}

\numberwithin{equation}{section}

\let\C=\Chi

\newcommand{\be}{\begin{equation}}
\newcommand{\ee}{\end{equation}}
\def\ba{\begin{align}} 
\def\ea{\end{align}}
\newcommand{\bea}{\begin{eqnarray}}
\newcommand{\eea}{\end{eqnarray}}
\newcommand{\bx}{\begin{example}}
\newcommand{\ex}{\end{example}}
\newcommand{\bex}{\begin{exercise}}
\newcommand{\eex}{\end{exercise}}
\newcommand{\ban}{\begin{answer}}
\newcommand{\ean}{\end{answer}}
\newcommand{\bt}{\begin{theorem}}
\newcommand{\et}{\end{theorem}}
\newcommand{\bc}{\begin{corollary}}
\newcommand{\ec}{\end{corollary}}
\newcommand{\blem}{\begin{lemma}}
\newcommand{\elem}{\end{lemma}}
\newcommand{\bp}{\begin{problem}}
\newcommand{\ep}{\end{problem}}
\newcommand{\bn}{\begin{proposition}}
\newcommand{\en}{\end{proposition}}
\newcommand{\bd}{\begin{definition}}
\newcommand{\ed}{\end{definition}}
\newcommand{\bcon}{\begin{construction}}
\newcommand{\econ}{\end{construction}}
\newcommand{\bq}{\begin{question}}
\newcommand{\eq}{\end{question}}
\newcommand{\bprf}{\begin{proof}}
\newcommand{\eprf}{\end{proof}}
\newcommand{\br}{\begin{remark}}
\newcommand{\er}{\end{remark}}
\newcommand{\bs}{\begin{solution}}
\newcommand{\es}{\end{solution}}
\newcommand{\beqs}{\begin{eqnarray}}
\newcommand{\eeqs}{\end{eqnarray}}
\newcommand{\bnt}{\begin{noterm}}
\newcommand{\ent}{\end{noterm}}
\newcommand{\bnot}{\begin{notation}}
\newcommand{\enot}{\end{notation}}

\let\p=\partial \let\ov=\overline

\newcommand{\<}{\langle}
\renewcommand{\>}{\rangle}

\newcommand{\id}{\mathrm{id}}

\newcommand{\mC}{\mathcal{C}}

\newcommand{\up}{\uparrow}
\newcommand{\dn}{\downarrow}

\newcommand{\lra}{\longrightarrow}

\newcommand{\tr}{{\rm tr} }

\newcommand{\CPTP}{\mathbf{CPTP}}
\newcommand{\im}{\mathrm{im}}

\def\R{{{\mathbb R}}}
\def\C{{{\mathbb C}}}
\def\H{{{\mathbb H}}}

\def\N{{{\mathbb N}}}

\newcommand{\Ad}{\mathrm{Ad}}

\newcommand{\LS}{\mathrm{LS}}
\newcommand{\EPR}{\mathrm{EPR}}

\def\mA{{{\mathcal{A}}}}

\def\mS{{{\mathcal{S}}}}
\def\mB{{{\mathcal{B}}}}

\newcommand{\matr}{\mathbb{M}}

\renewcommand{\hom}{\mathbf{Hom}}

\def\invexcl{\rotatebox[origin=c]{180}{$!$}}
\newcommand{\bloom}{\operatorname{\invexcl}}
\newcommand{\rbloom}{\psi_{_{_{R}}}}
\newcommand{\lbloom}{\psi_{_{L}}}
\newcommand{\sbloom}{\psi_{_{S}}}
\newcommand{\pbloom}[1]{\psi^{(#1)}}
\newcommand{\spbloom}[1]{\psi^{(#1)}_{_{S}}}

\newcommand{\norm}[1]{\left\lVert#1\right\rVert}

\def\VA{\mathcal{A}}
\def\VB{\mathcal{B}}
\def\VC{\mathcal{C}}

\def\M{\mathbb{M}}

\newcommand{\Jamiol}{\mathscr{J}}

\newcommand{\diag}{\mathrm{diag}}

\newcommand{\ben}{\renewcommand{\theenumi}{\alph{enumi}} 
\renewcommand{\labelenumi}{(\theenumi)}\begin{enumerate}}
\newcommand{\een}{\end{enumerate}}
\newcommand{\xmark}{\ding{55}}%
\newcommand*{\matminus}{%
  \leavevmode
  \hphantom{0}%
  \llap{%
    \settowidth{\dimen0 }{$0$}%
    \resizebox{1.1\dimen0 }{\height}{$-$}%
  }%
}

\title{On dynamical measures of quantum information}
\author{James Fullwood and Arthur J.~Parzygnat}

\newcommand{\Addresses}{{
  \bigskip
  \footnotesize

  J.~Fullwood, \textsc{School of Mathematical Sciences, Shanghai Jiao Tong University, 800 Dongchuan Road, Shanghai 200240, People's Republic of China}\par\nopagebreak
  \textit{E-mail address}, J.~Fullwood: \texttt{fullwood@sjtu.edu.cn}

    \medskip

    A.~Parzygnat, \textsc{Graduate School of Informatics, Nagoya University, Chikusa-ku, 464-8601 Nagoya, Japan} \& \textsc{Department of Mathematics, Massachusetts Institute of Technology, Cambridge, Massachusetts 02139, USA}\par\nopagebreak
  \textit{E-mail address}, A.~Parzygnat: \texttt{parzygnat@nagoya-u.jp}, \texttt{arthurjp@mit.edu}

}}

\begin{document}
\emergencystretch 2em

\maketitle

\vspace{-7mm}
\tableofcontents

\begin{abstract}
In this work, we use the theory of quantum states over time to define an entropy $S(\rho,\mathcal{E})$ associated with quantum processes $(\rho,\mathcal{E})$, where $\rho$ is a state and $\mathcal{E}$ is a quantum channel responsible for the dynamical evolution of $\rho$. The entropy $S(\rho,\mathcal{E})$ is a generalization of the von~Neumann entropy in the sense that $S(\rho,\id)=S(\rho)$ (where $\id$ denotes the identity channel), and is a dynamical analogue of the quantum joint entropy for bipartite states. Such an entropy is then used to define dynamical formulations of the quantum conditional entropy and quantum mutual information, and we show such information measures satisfy many desirable properties, such as a \emph{quantum entropic Bayes' rule}. 
We also use our entropy function to quantify  the information loss/gain associated with the dynamical evolution of quantum systems, which enables us to formulate a precise notion of information conservation for quantum processes.  
\end{abstract}

\section{Introduction}

In classical probability theory, there are dual perspectives one may take when considering a joint distribution $\mathbb{P}(x,y)$. On the one hand, there is the \emph{static} perspective, where $\mathbb{P}(x,y)$ is viewed as a distribution associated with a pair of random variables $(X,Y)$ whose outputs are occurring in parallel, or rather, whose outputs are \emph{spacelike} separated. On the other hand, there is the \emph{dynamical} perspective, where $\mathbb{P}(x,y)$ is viewed as a \emph{state over time}, i.e., as the distribution associated with a random variable $X$ evolving stochastically into the random variable $Y$ according to the associated family of conditional distributions $\mathbb{P}(y|x)$. As such, classical information measures associated with a pair of random variables---such as the joint entropy, conditional entropy, and mutual information---also admit dual static/dynamic interpretations.

But while the static and dynamic perspectives are equivalent for classical random variables, it is only the static perspective that generalizes to the quantum setting in a straightforward manner. In particular, the quantum analogue of a classical joint distribution is a positive operator $\rho_{\VA\VB}\in \VA\otimes \VB$ of unit trace. And while $\rho_{\VA\VB}$ has a natural interpretation as a joint state with associated marginal spacelike separated states $\rho_{\VA}=\tr_{\VB}(\rho_{\VA\VB})$ and $\rho_{\VB}=\tr_{\VA}(\rho_{\VA\VB})$, there is in general no known associated dual perspective where we may view $\rho_{\VA}$ as dynamically evolving into $\rho_{\VB}$ via a quantum channel $\mathcal{E}:\VA\to \VB$~\cites{LeSp13,PaQPL21}. Conversely, given a quantum channel $\mathcal{E}:\VA\to \VB$ with $\mathcal{E}(\rho_{\VA})=\rho_{\VB}$, there is no known canonical joint state in the tensor product $\VA\otimes \VB$ encoding temporal correlations associated with the dynamical evolution of $\rho_{\VA}$ into $\rho_{\VB}$~\cites{HHPBS17,FuPa22,FuPa22a}. Thus, the dynamical perspective of bipartite quantum states is presently fragmented according to the various approaches to circumventing such issues. 

Due to the lack of a consensus on dynamical perspectives associated with bipartite quantum states, the standard approach to quantum information measures is taken via the static perspective, i.e., in terms of the information content of a density operator $\rho_{\VA\VB}$ representing the joint state of a system on two spacelike separated regions at a single instant of time~\cites{NiCh11,Wilde2017}. For example, the quantum joint entropy, the quantum conditional entropy, and the quantum mutual information of such a bipartite quantum state $\rho_{\VA\VB}$ are given by
\begin{eqnarray*}
\text{\underline{Quantum Joint Entropy}}:\quad \, S(\VA,\VB)&=&S(\rho_{\VA\VB}) \\
\text{\underline{Quantum Conditional Entropy}}: \quad \, H(\VB|\VA)&=&S(\rho_{\VA\VB})-S(\rho_{\VA}) \\
\text{\underline{Quantum Mutual Information}}:\quad I(\VA:\VB)&=&S(\rho_{\VA})+S(\rho_{\VB})-S(\rho_{\VA\VB}), 
\end{eqnarray*}
where $S(*)$ denotes the von~Neumann entropy and $\rho_{\VA}$ and $\rho_{\VB}$ are the reduced density matrices associated with $\rho_{\VA\VB}$. But what if $\rho_{\VA}$ and $\rho_{\VB}$ are instead timelike separated? Certainly, there should exist an analogue of quantum joint entropy, quantum conditional entropy, and quantum mutual information in a scenario where $\rho_{\VA}$ dynamically evolves into $\rho_{\VB}$. While various proposals for dynamical measures of quantum information exist~\cites{Li91,NiSh96,Sc96,ChMa20,SSW14,RZF11,GoWi21,MaCh22,KPP21,JSK23}, we propose a general formalism which not only recovers some of the aforementioned proposals, but which also may be applied to hybrid classical-quantum systems, such as preparations and measurements.  

To be more precise, we define an entropy $S(\rho,\mathcal{E})$ associated with a quantum process $(\rho,\mathcal{E})$, where $\rho$ is a state and $\mathcal{E}$ is a quantum channel responsible for the dynamical evolution of $\rho$. In the case of the identity channel $\id:\VA\to \VA$, we show $S(\rho,\id)=S(\rho)$ (cf. Theorem~\ref{BTXETXFX787}), 
thus illustrating how our entropy may be viewed as an extension of von~Neumann entropy from states to processes.
Such an entropy then takes the place of $S(\rho_{\VA\VB})$ in the equations above defining quantum information measures in the static setting, and is used to define dynamical formulations of the quantum joint entropy, the quantum conditional entropy, and the quantum mutual information. We also use $S(\rho,\mathcal{E})$ to define an information measure that we refer to as \emph{information discrepancy}, which we introduce as a quantum analogue of the ``conditional information loss'' defined in Ref.~\cite{FuPa21}. We use the terminology ``information discrepancy'', as opposed to ``information loss'', since information can actually be gained in certain quantum operations. In particular, for quantum measurements (POVMs), we find that the information discrepancy is always negative, and provides a measure of disturbance associated with the measurement (cf. Remark~\ref{rmk:CQI} (c) and Theorem~\ref{prop:PVMKN56}). 

Crucial to quantifying the information flow associated with a quantum process will be the notion of a \emph{state over time function}, which provides a unified framework for incorporating various formulations of a dynamical perspective associated with quantum bipartite states~\cite{FuPa22a}. A state over time function $\psi$ takes as its input a quantum process $(\rho,\mathcal{E})$ and outputs a \emph{state over time}, i.e., a self-adjoint operator $\psi(\rho,\mathcal{E})\in \VA\otimes \VB$ of unit trace such that
\[
\tr_{\VB}\big(\psi(\rho,\mathcal{E})\big)=\rho \quad \text{and} \quad \tr_{\VA}\big(\psi(\rho,\mathcal{E})\big)=\mathcal{E}(\rho).
\]
A state over time is then a single entity encompassing the dynamical evolution of a quantum state, similar to how spacetime is a single entity encompassing the dynamical evolution of a classical system. 

As emphasized in the work of Fitzsimons, Jones, and Vedral~\cite{FJV15}, if a state over time $\psi(\rho,\mathcal{E})$ is to encode not only the states $\rho$ and $\mathcal{E}(\rho)$, but also \emph{temporal} correlations between $\rho$ and $\mathcal{E}(\rho)$, then the operator $\psi(\rho,\mathcal{E})$ must admit negative eigenvalues. Thus, a state over time is a hermitian operator of unit trace which is not positive in general. We then refer to hermitian operators of unit trace as \emph{quasi-states}, which we view as a quantum analogue of quasi-probability distributions. While density operators are mathematical objects which encode spatial correlations, a quasi-state has the capacity to encode \emph{both} spatial and temporal correlations, which is a characteristic property of states over time. 

In Ref.~\cite{FuPa22}, we constructed a state over time function satisfying a list of axioms put forth in Ref.~\cite{HHPBS17}, and at present, our construction is the only known state over time function satisfying such axioms. However, there are still various approaches to states over time, such as the causal states of Leifer and Spekkens \cites{Le06,Le07,LeSp13}, the pseudo-density operators  of Fitzsimons, Jones, and Vedral~\cite{FJV15}, the two-state vector formalism~\cites{Wat55,ABL64,ReAh95} of Watanabe, Aharonov et alia, the Wigner-function approach of Wootters~\cites{Woot87,HHPBS17}, 
and the compound states of Ohya~\cite{Ohya1983}. As such, we formulate our results \emph{with respect to} the choice of a state over time function, thus incorporating the various approaches to states over time and their associated measures of information~\cites{Li91,SSW14,ChMa20,MaCh22,JSK23}.  

For a precise formulation of the entropy of a state over time, we extend the von~Neumann entropy to quasi-states, and show that such an extension satisfies many of the characteristic properties of von~Neumann entropy, such as unitary invariance, additivity over product quasi-states, and a Fannes-type inequality, extending the usual Fannes inequality for von~Neumann entropy. 
Our definition of entropy for quasi-states however can be negative, which is an essential feature of quantum information \cites{CeAd97,HorOpp05,dRARDV11,ReWo14}. We then apply our extension of the von~Neumann entropy to states over time, thus yielding an entropy $S(\rho,\mathcal{E})$ associated with quantum processes $(\rho,\mathcal{E})$. 

While there are approaches to extending von~Neumann entropy to non-positive operators using analytic continuation of the logarithm~\cites{NTTTW21,TTT21}, we find that analytic continuation is not ideal for our purposes for two reasons. First, the branch cut used in Ref.~\cite{NTTTW21} is along the negative real axis, which is problematic for quasi-states, which have eigenvalues along the whole real line. Second, even if an alternative branch cut is chosen, it is unclear what the meaning of the imaginary component of the complex logarithm should be, given that its value is contingent upon the choice of such a branch cut. Nevertheless, Ref.~\cite{DHMTT22} argues for a possible interpretation of the imaginary component of a complex-valued entropy in terms of the emergence of time, analogous to how the standard entanglement entropy is argued to be related to the emergence of space in the context of holography~\cite{RyTa06}. 
In any case, in our approach we keep only the real part of the complex logarithm extended to the negative real axis (which is independent of the choice of a branch cut), which in turn yields the odd completion of the function $x\mapsto -x\log(x)$, which is given by $x\mapsto -x\log|x|$. 

In terms of the functional calculus, our entropy $S(\tau)$ for a quasi-state $\tau$ is then given by
\be \label{FPENTROPY23}
S(\tau)=-\tr\left(\tau\log|\tau|\right).
\ee
When applied to states over time, such an entropy yields direct generalizations to the quantum setting of fundamental aspects of dynamical measures of classical information, such as the vanishing of conditional entropy under deterministic evolution, which would not hold if we were to use analytic continuation (cf. Remark~\ref{rmk:pseudoentropy}). Interestingly, while the entropy functional given by \eqref{FPENTROPY23} does not in general satisfy subadditivity, it does seem to satisfy subadditivity when applied to states over time (cf.\ Remark~\ref{rmk:NNPMI}). In particular, in all known examples we find
\[
S(\rho)+S(\mathcal{E}(\rho))\geq S(\psi(\rho,\mathcal{E})),
\]
leading us to conjecture that states over time form a subclass of quasi-states for which the subadditivity property associated with entropy functional \eqref{FPENTROPY23} indeed holds. 

Yet another approach to extending von Neumann entropy to quasi-states that appeared recently is to use the functional calculus associated with the even completion $x\mapsto -|x|\log|x|$ of the function $x\mapsto -x\log(x)$, which only uses the singular values of a self-adjoint matrix, rather than its eigenvalues~\cite{JSK23}. We will also see that this even extension of entropy applied to states over time does not satisfy certain properties that we expect from the classical theory, yet which our definition satisfies (cf. Remark~\ref{rmk:JSK}). 

For example, a fundamental symmetry for classical information measures associated with a pair of random variables $(X,Y)$ is the \emph{entropic Bayes' rule}, which states that
\be \label{EBRX17}
H(Y|X)+H(X)=H(X|Y)+H(Y),
\ee
where $H(*|*)$ is the conditional entropy and $H(*)$ is the Shannon entropy~\cites{Sh48,Kh57}. In this work, we employ a notion of quantum Bayesian inversion with respect to the choice of a state over time function $\psi$~\cite{FuPa22a}, and show that a \emph{quantum entropic Bayes' rule} generalizing equation \eqref{EBRX17} holds for the dynamical quantum conditional entropy associated with $\psi$ (cf. Proposition~\ref{QEXTPXBR91}). Such a notion of Bayesian inversion associated with a state over time function $\psi$ specializes to other cases of retrodiction maps appearing in the literature, and a more in-depth analysis of this construction and its more general relation to time-reversal symmetry in quantum theory is the content of Ref.~\cite{FuPa22a}.

The present paper is organized as follows. In Section~\ref{S2}, we provide the necessary definitions for a precise mathematical formulation of our results. In Section~\ref{S3}, we recall the basic theory of states over time associated with quantum processes and give some fundamental examples, such as the Leifer and Spekkens state over time, the symmetric bloom, and the compound state. In order to be unbiased with regards to the different approaches to states over time, in Section ~\ref{S4}, we introduce a parametric family of state over time functions that interpolate between most of the aforementioned examples. In Section~\ref{S5}, we motivate our eventual definition of entropy for quantum processes by showing that there is a unique functional on hermitian matrices that, when evaluated on a state over time $\psi(\rho,\id)$ associated with an identity process $(\rho,\id)$, specializes to the von~Neumann entropy $S(\rho)$. In Section~\ref{S6}, we use this functional to define an extension of von~Neumann entropy to hermitian matrices, and we prove that such an extension satisfies many desirable properties. In Section~\ref{S7}, we use our extension of von~Neumann entropy to define an entropy $S_{\psi}(\rho,\mathcal{E})$ for quantum processes $(\rho,\mathcal{E})$, which we view as a dynamical formulation of quantum joint entropy. We note that the subscript $\psi$ appears on the entropy of a process as it is defined \emph{with respect to} a choice of a state over time function $\psi$. We then use this dynamical joint entropy to define dynamical analogues of quantum conditional entropy, quantum mutual information, and a quantum analogue of the information loss associated with a stochastic map \cite{FuPa21}. In Section~\ref{S81}, we go over some explicit examples in detail, such as the bit-flip channel, the amplitude damping channel, the partial trace, and also a projection-valued measure. In Section~\ref{S82}, we consider a general CPTP map corresponding to a positive operator-valued measure (POVM), and show that in such a case the dynamical mutual information defined in Section~\ref{S7} may be interpreted as a measure of disturbance for an associated quantum instrument. In Section~\ref{S9}, we prove a quantum entropic Bayes' rule associated with the dynamical conditional entropy associated with any state over time function. In Section~\ref{S8}, we prove some general results about the entropy of deterministic processes, such as unitary evolution and partial trace. We then make some concluding remarks in Section~\ref{S10}.

\section{Preliminaries} \label{S2}
In this section, we provide the basic definitions, notation and terminology that will be used throughout~\cites{Fa01,GdHJ89,Pa03,NiCh11,FuPa22a}.

\bd
Let $X$ be a finite set. A function $p:X\to \R$ will be referred to as a \define{quasi-probability distribution} if and only if $\sum_{x\in X}p(x)=1$. In such a case, $p(x)\in \R$ will be denoted by $p_x$ for all $x\in X$. If $p_x\in [0,1]$ for all $x\in X$, then $p$ will be referred to as a \define{probability distribution}.  
\ed

\bd
Let $X$ and $Y$ be finite sets. A \define{stochastic} map $f:X\to Y$ consists of the assignment of a probability distribution $f_x:Y\to [0,1]$ for every $x\in X$. In such a case, $(f_x)_y$ will be denoted by $f_{yx}$ for all $x\in X$ and $y\in Y$, which is interpreted as the conditional probability of $y$ given $x$. A stochastic map $f:X\to Y$ together with a prior distribution $p$ on its set of inputs $X$ will be denoted by $(p,f)$, which is referred to as a \define{classical process}.
\ed

\bd
Given a natural number $m\in \N$, the set of $m\times m$ matrices with complex entries will be denoted by $\M_m$, and will be referred to as a \define{matrix algebra}. As the matrix algebra $\M_1$ is simply the complex numbers, it will be denoted by $\C$. The matrix units in $\M_m$ will be denoted by $E_{ij}^{(m)}$ (or simply $E_{ij}$ if $m$ is clear from the context), and for every $\rho\in \M_m$, $\rho^{\dag}\in \M_m$ denotes the conjugate transpose of $\rho$. Given a finite set $X$, a direct sum $\bigoplus_{x\in X}\M_{m_x}$ will be referred to as a \define{multi-matrix algebra}, whose multiplication and addition are defined component-wise. If $\VA$ and $\VB$ are multi-matrix algebras, then the vector space of all linear maps from $\VA$ to $\VB$ will be denoted by $\hom(\VA,\VB)$. The \define{trace} of an element $A=\bigoplus_{x\in X}A_{x}\in \bigoplus_{x\in X}\M_{m_x}$ is the complex number $\tr(A)=\sum_{x\in X}\tr(A_{x})$, where $\tr(*)$ is the usual trace on matrices. The \define{adjoint} of $A$ is the element $A^{\dag}\in \bigoplus_{x\in X}\M_{m_x}$ given by $A^{\dag}=\bigoplus_{x\in X}A_{x}^{\dag}$, where $A^{\dag}_{x}$ is the usual conjugate transpose. As a conjugate-linear map, the adjoint operation will be denoted by $\dag$.
Given $\mathcal{E}\in \hom(\VA,\VB)$, we let $\mathcal{E}^*\in \hom(\VB,\VA)$ denote the Hilbert--Schmidt adjoint of $\mathcal{E}$, which is uniquely determined by the condition
\[
\tr\left(\mathcal{E}(A)^{\dag}B\right)=\tr\left(A^{\dag}\mathcal{E}^*(B)\right)
\] 
for all $A\in \VA$ and $B\in \VB$. The identity map between algebras will be denoted by $\id$, while the unit element in an algebra will be denoted by $\mathds{1}$ (subscripts, such as in $\id_{\mA}$ and $\mathds{1}_{\mA}$, will be used if deemed necessary).
\ed

\bd
Given a multi-matrix algebra $\VA$, the \define{multiplication map} is the map $\mu_{\VA}:\VA\otimes \VA\to \VA$ corresponding to the linear extension of the assignment $A_1\otimes A_2\mapsto A_1A_2$.
\ed

\bnot
Given a finite set $X$,  the multi-matrix algebra $\bigoplus_{x\in X}\C$ is canonically isomorphic to the algebra $\C^X$ of complex-valued functions on $X$, and as such, $\bigoplus_{x\in X}\C$  will be denote simply by $\C^X$. 
\enot

\bd \label{STXDFX99}
Let $X$ be a finite set and let $\VA=\bigoplus_{x\in X}\M_{m_x}$ be a multi-matrix algebra. An element $A=\bigoplus_{x\in X} A_x\in \VA$ is said to be 
\begin{itemize}
\item
\define{self-adjoint} if and only if $A^{\dag}=A$.
\item
\define{positive} if and only if $A_x\in \M_{m_x}$ is self-adjoint and has non-negative eigenvalues for all $x\in X$.
\item
a \define{state} if and only if $A$ is positive and of unit trace. If $X$ contains only one element, then $A$ will often be referred to as a \define{density matrix}. 
\item
a \define{quasi-state} if and only if $A$ is self-adjoint and of unit trace. If $X$ contains only one element, then $A$ will often be referred to as a \define{quasi-density matrix}. 
\end{itemize}
\ed

\bnot
The real vector space of all self-adjoint matrices in $\M_n$ will be denoted by $\H_n$. 
If $\VA$ is a multi-matrix algebra, the set of all states in $\VA$ will be denoted by $\mathcal{S}(\VA)$, while the set of all quasi-states in $\VA$ will be denoted by $\mathcal{Q}(\VA)$.
\enot

\bd
\label{defn:mspec}
The \define{multi-spectrum} of a matrix $\rho\in \M_n$ is the multiset $\mathfrak{mspec}(\rho)$ corresponding to the eigenvalues of $\rho$ together with their (algebraic) multiplicities. 
If $\rho$ is self-adjoint, then there is a natural ordering on the multispectrum of $\rho$ induced by the total order $\leq$ on $\R$. In such a case, the notation $\{\lambda_i\}$ will be used to denote the multi-spectrum of $\rho$, where the index $i$ corresponds to to the ordering $\lambda_1\leq \lambda_2\leq \cdots \leq \lambda_n$. 
\ed

\bnot
Given a finite set $X$ and an element $\rho\in \C^X$, we let $\rho_x\in \C$ for all $x\in X$ denote the complex numbers such that $\rho=\bigoplus_{x\in X}\rho_x$. In such a case $\rho_x$ will be referred to as the \define{$x$-component} of $\rho$. For each $x\in X$, the \define{Dirac-delta} at $x$ is the state $\delta_x\in \mathcal{S}(\C^X)$, the $x$-component of which is 1. 
\enot 

\br
If $X$ is a finite set and $\VA=\C^X$, then a state on $\VA$ may be identified with a probability distribution on $X$.
\er

\bd
Let $\VA$ and $\VB$ be multi-matrix algebras. A map $\mathcal{E}\in \hom(\VA,\VB)$ is said to be
\begin{itemize}
\item
\define{$\dag$-preserving} if and only if $\dag\circ\mathcal{E}=\mathcal{E}\circ\dag$, i.e., $\mathcal{E}(A)^{\dag}=\mathcal{E}(A^{\dag})$ for all $A\in \VA$.
\item
\define{trace-preserving} (\define{TP}) if and only if $\tr\circ\mathcal{E}=\tr$, i.e., $\tr(\mathcal{E}(A))=\tr(A)$ for all $A\in \VA$.
\item
\define{positive} if and only if $\mathcal{E}(A)$ is positive whenever $A\in \VA$ is positive.
\item
\define{completely positive} (\define{CP}) if and only if $\mathcal{E}\otimes \id_{\VC}:\VA\otimes \VC\to \VB\otimes \VC$ is positive for every matrix algebra $\VC$.
\item
\define{CPTP} if and only if $\mathcal{E}$ is completely positive and trace-preserving. The convex space of all $CPTP$ maps from $\VA$ to $\VB$ will be denoted by $\CPTP(\VA,\VB)$.
\end{itemize} 
\ed

\bd
Let $X$ and $Y$ be finite sets, let $\VA=\bigoplus_{x\in X}\M_{m_x}$, let $\VB=\bigoplus_{y\in Y}\M_{n_y}$ and let $\mathcal{E}\in \CPTP(\VA,\VB)$. 
\begin{itemize}
\item
If $m_x=n_y=1$ for all $x\in X$ and $y\in Y$, then $\mathcal{E}$ is said to be a \define{classical channel}.   
\item
If $\# X=1$ and $n_y=1$ for all $y\in Y$, then $\mathcal{E}$ is said to be a \define{measurement}.
\item
If $\# Y=1$ and $m_x=1$ for all $x\in X$, then $\mathcal{E}$ is said to be an \define{ensemble preparation}.
\item
If $\# X=1$ and $n_y=n>1$ for all $y\in Y$, then $\mathcal{E}$ is said to be a \define{quantum instrument}.
\end{itemize}
Here, $\# X$ denotes the cardinality of $X$. 
\ed

\br
The definitions of classical channel, measurement, preparation, and quantum instrument in terms of CPTP maps between multi-matrix algebras are equivalent to the usual notions from quantum information theory~\cite{FuPa22a}. As such, the formalism of CPTP maps between multi-matrix algebras incorporates many fundamental notions from quantum information theory in a single mathematical formalism.
\er

\bnot
Given a pair $(X,Y)$ of finite sets and a classical channel $\mathcal{E}\in \CPTP(\C^X,\C^Y)$, then for all $(x,y)\in X\times Y$, we let $\mathcal{E}_{yx}\in [0,1]$ denote the elements such that
\[
\mathcal{E}(\delta_x)=\bigoplus_{y\in Y}\mathcal{E}_{yx}.
\]
In such a case, the elements $\mathcal{E}_{yx}$ will be referred to as the \define{conditional probabilities} associated with $\mathcal{E}$.
\enot 

\bd
Given a pair $(\VA,\VB)$ of multi-matrix algebras, an element $(\rho,\mathcal{E})\in \mathcal{S}(\VA)\times \CPTP(\VA,\VB)$ will be referred to as a \define{process}, and the set of processes $\mathcal{S}(\VA)\times \CPTP(\VA,\VB)$ will be denoted by $\mathscr{P}(\VA,\VB)$. When $\VA=\C^X$ and $\VB=\C^Y$ for finite sets $X$ and $Y$, then $(\rho,\mathcal{E})\in \mathscr{P}(\VA,\VB)$ will be referred to as a \define{classical process}.
\ed

\section{States over time} \label{S3}
In this section we introduce the notion of a \emph{state over time function}, which will play a fundamental role moving forward. 
A state over time is essentially a quantum generalization of a classical joint probability distribution associated with the stochastic evolution of a random variable $X$ into a random variable $Y$. To incorporate more examples and also various approaches to states over time, the definition given here is less restrictive than the definition of a state over time function given in Ref.~\cite{FuPa22} (see also Ref.~\cite{FuPa22a}). 

\bd\label{DXS81}
A \define{state over time function} associates every pair $(\VA,\VB)$ of multi-matrix algebras with a map $\psi_{\VA,\VB}:\mathscr{P}(\VA,\VB)\lra \VA\otimes \VB$ such that 
\be \label{SOTX77}
\tr_{\VB}\big(\psi_{\VA\VB}(\rho,\mathcal{E})\big)=\rho \qquad\text{ and }\qquad
\tr_{\VA}\big(\psi_{\VA\VB}(\rho,\mathcal{E})\big)=\mathcal{E}(\rho).
\ee
for all $(\rho,\mathcal{E})\in \mathscr{P}(\VA,\VB)$. In such a case, the element $\psi_{\VA\VB}(\rho,\mathcal{E})\in \VA\otimes \VB$ will be referred to as the \define{state over time} associated with the process $(\rho,\mathcal{E})$.
\ed

\bnot
While a state over time function is actually a \emph{family} of functions, with each member $\psi_{\VA\VB}:\mathscr{P}(\VA,\VB)\lra \mA\otimes\mB$ corresponding to an ordered pair $(\VA,\VB)$ of multi-matrix algebras, the input process $(\rho,\mathcal{E})$ uniquely determines the function $\psi_{\VA\VB}$ in the family which is to be evaluated at $(\rho,\mathcal{E})$. As such, the state over time $\psi_{\VA\VB}(\rho,\mathcal{E})$ will be denoted simply by $\psi(\rho,\mathcal{E})$ for all processes $(\rho,\mathcal{E})\in \mathscr{P}(\VA,\VB)$.
\enot

\br A state over time function was denoted by $\star$ in Ref.~\cite{FuPa22a} so that $\psi(\rho,\mathcal{E})=\mathcal{E}\star\rho$.
\er

\br
Either of the conditions in \eqref{SOTX77} ensures that a state over time function $\psi$ preserves the trace of its input state, i.e., 
\be \label{TRXST71}
\tr\big(\psi(\rho,\mathcal{E})\big)=\tr(\rho)=1
\ee
for all processes $(\rho,\mathcal{E})$. However, there is nothing in the definition of a state over time function ensuring that $\psi(\rho,\mathcal{E})$ is an actual state, i.e., an element of $\mathcal{S}(\VA\otimes \VB)$. As such, a state over time should be viewed as a generalized quantum state, the significance of which will be further addressed in due course.
\er

To give explicit examples of state over time functions we make use of the \emph{channel state}, which we now define.

\bd
Let $\VA$ and $\VB$ be multi-matrix algebras, and let $\mathcal{E}\in \hom(\VA,\VB)$. The \define{channel state} of $\mathcal{E}$ is the element $\mathscr{J}[\mathcal{E}]\in \VA\otimes \VB$ given by $\mathscr{J}[\mathcal{E}]=(\id_{\VA}\otimes \mathcal{E})(\mu_{\VA}^{*}(\mathds{1}_{\VA}))$.
\ed

We now use the channel state to define several examples of state over time functions~\cite{FuPa22a}. 

\bd
\label{ex:rbloom}
The \define{right bloom} state over time function is the map $\rbloom$ given by $\rbloom(\rho,\mathcal{E})=(\rho\otimes\mathds{1})\Jamiol[\mathcal{E}]$. 
\ed

\bd
The \define{left bloom} state over time function is the map $\lbloom$ given by $\lbloom(\rho,\mathcal{E})=\Jamiol[\mathcal{E}](\rho\otimes\mathds{1})$.
\ed

\bd
The \define{symmetric bloom} state over time function is the map $\sbloom$ given by $\sbloom(\rho,\mathcal{E})=\frac{1}{2}\big\{\rho\otimes\mathds{1},\Jamiol[\mathcal{E}]\big\}$, where $\{A,B\}=AB+BA$ denotes the \emph{anti-commutator}, also known as the \emph{Jordan product}. 
\ed

\bd
The \define{Leifer--Spekkens} (or simply \define{LS}) state over time function is the map $\psi_{\LS}$ given by $\psi_{\LS}(\rho,\mathcal{E})=\left(\sqrt{\rho}\otimes \mathds{1}\right)\Jamiol[\mathcal{E}]\left(\sqrt{\rho}\otimes \mathds{1}\right)$.
\ed

The next state over time function is an extension of Ohya's compound state to arbitrary processes, and which does not make use of the channel state~\cites{Ohya1983,Oh83b,OhPe93}.

\bd
Given a state $\rho=\bigoplus_{x\in X}\rho_{x}$ in $\VA=\bigoplus_{x\in X}\matr_{m_{x}}$, write $\rho_{x}=\sum_{\alpha_{x}}\lambda_{x\alpha_{x}}P_{x\alpha_{x}}$ as the spectral decomposition, where each $P_{x\alpha_{x}}$ is the projection onto the $\lambda_{x\alpha_{x}}$ eigenspace of $\rho_{x}$. Then Ohya's \define{compound state over time} is given by~\cites{Ohya1983,Oh83b,OhPe93,FuPa22a}
\[
\psi_{\mathrm{co}}(\rho,\mathcal{E})=\bigoplus_{x\in X}\sum_{\alpha_{x}}\lambda_{x\alpha_{x}}P_{x\alpha_{x}}\otimes\mathcal{E}\left(\frac{P_{x\alpha_{x}}}{\tr(P_{x\alpha_{x}})}\right).
\]
\ed

For explicit calculations of states over time, we make use of the following result.

\bn  \label{CSXS1971}
Let $\VA$ and $\VB$ be multi-matrix algebras and suppose $\mathcal{E}\in \hom(\VA,\VB)$. Then the following statements hold.
\begin{enumerate}[i.] 
\item \label{CSXS1}
The channel state $\mathscr{J}[\mathcal{E}]$ is self-adjoint if and only if $\mathcal{E}$ is $\dag$-preserving.
\item  \label{CSXS2}
If $\VA$ is a matrix algebra, then 
\be \label{ANXNA101}
\mathscr{J}[\mathcal{E}]=\sum_{i,j} E_{ij}^{\VA}\otimes \mathcal{E}(E_{ji}^{\VA})
\ee
\item  \label{CSXS3}
If $\VA$ and $\VB$ are both matrix algebras and if $\mathcal{E}$ is $\dag$-preserving, then
\be \label{ANXNA111}
\mathscr{J}[\mathcal{E}]=\sum_{k,l}\mathcal{E}^*(E_{kl}^{\VB})\otimes E_{lk}^{\VB}
\ee
\end{enumerate}
\en

\bprf
{\color{white}You found me!}

\noindent
\underline{Item \ref{CSXS1}}: The claim follows from Lemma~3.4 in Ref.~\cite{FuPa22} (see also Ref.~\cite{Jam72}).

\noindent
\underline{Item \ref{CSXS2}}: Indeed,
\[
\mathscr{J}[\mathcal{E}]
=(\id_{\VA}\otimes \mathcal{E})\left(\mu_{\VA}^{*}(\mathds{1})\right)
\overset{\eqref{eq:musEij}}=(\id_{\VA}\otimes \mathcal{E})\left(\sum_{i,j}E^{\VA}_{ij}\otimes E^{\VA}_{ji}\right) \\
=\sum_{i,j}E^{\VA}_{ij}\otimes \mathcal{E}(E^{\VA}_{ji}),
\]
as desired.

\noindent
\underline{Item \ref{CSXS3}}: Let $\mathcal{E}=\sum_{\alpha}\Ad_{V_{\alpha}}-\sum_{\beta}\Ad_{W_{\beta}}$ be a Kraus decomposition of $\mathcal{E}$~\cites{dePi67,Ch75,Kr83,Wa18}. We then have
\begin{eqnarray*}
\mathscr{J}[\mathcal{E}]&=&\sum_{i,j}E_{ij}^{\VA}\otimes \mathcal{E}\left(E_{ji}^{\VA}\right) =\sum_{i,j}E_{ij}^{\VA}\otimes \left(\sum_{\alpha}V_{\alpha}E_{ji}^{\VA}V_{\alpha}^{\dag}-\sum_{\beta}W_{\beta}E_{ji}^{\VA}W_{\beta}^{\dag}\right) \\
&\overset{\eqref{ESX17}}=&\sum_{k,l}\left(\sum_{\alpha} V_{\alpha}^{\dag}E_{kl}^{\VB}V_{\alpha}-\sum_{\beta} W_{\beta}^{\dag}E_{kl}^{\VB}W_{\beta}\right)\otimes E_{lk}^{\VB}=\sum_{k,l} \hspace{0.35mm}  \mathcal{E}^*(E_{kl}^{\VB})\otimes E_{lk}^{\VB},
\end{eqnarray*}
as desired.
\eprf

\br
The map $\mathscr{J}:\hom(\VA,\VB)\to \VA\otimes \VB$ given by $\mathcal{E}\mapsto \mathscr{J}[\mathcal{E}]$ is a linear isomorphism, which we refer to as the \define{Jamio{\l}kowski isomorphism} \cite{Jam72}. Note that in the case of matrix algebras, the formula from item~\ref{CSXS2} in Proposition~\ref{CSXS1971} for the channel state $\Jamiol[\mathcal{E}]$ is
different from 
the formula which defines that Choi matrix
$\mathscr{C}[\mathcal{E}]=\sum_{i,j}E_{ij}\otimes \mathcal{E}(E_{ij})$~\cite{Ch75}.
It then follows that the matrices $\Jamiol[\mathcal{E}]$ and $\mathscr{C}[\mathcal{E}]$ differ by a partial transpose~\cites{FuPa22a,FrCa22}, and as such, are generically distinct. Moreover, while the \define{Choi--Jamio{\l}kowski isomorphism} $\mathcal{E}\mapsto \mathscr{C}[\mathcal{E}]$ is basis-\emph{dependent}, the Jamio{\l}kowski isomorphism $\mathcal{E}\mapsto \Jamiol[\mathcal{E}]$ is basis-\emph{independent}. 
\er

The following example shows how the use of the Jamio{\l}kowsi ismorphism is motivated by the case of classical channels.

\bx[Classical states over time] \label{CXST177}
Let $\mathcal{E}:\C^X\to \C^Y$ be a classical channel, let $\rho$ be a state on $\C^X$, and let $\mathcal{E}_{yx}$ be the associated conditional probabilities, so that
\[
\mathcal{E}(\delta_x)=\bigoplus_{y\in Y}\mathcal{E}_{yx}
\]
for all $x\in X$. We then have $\rbloom(\rho,\mathcal{E})=\lbloom(\rho,\mathcal{E})=\sbloom(\rho,\mathcal{E})=\psi_{\LS}(\rho,\mathcal{E})=(\rho\otimes \mathds{1})\mathscr{J}[\mathcal{E}]$, and 
\[
(\rho\otimes \mathds{1})\mathscr{J}[\mathcal{E}]=
\bigoplus_{(x,y)\in X\times Y}\rho_x\mathcal{E}_{yx}.
\]
As such, the right bloom, left bloom, symmetric bloom, and the Leifer--Spekkens state over time functions all yield the same result, which we refer to as the \define{classical state over time}. If we associate $\mathcal{E}$ with the stochastic map $f:X\to Y$ given by $f_x=\mathcal{E}(\delta_x)$, and $\rho$ is associated with a probability distribution $p$ on $X$, then the classical state over time is the distribution $\vartheta(p,f)$ on $X\times Y$ given by $\vartheta(p,f)_{(x,y)}=p_xf_{yx}$, which is the usual joint distribution associated with the pair of random variables $X$ and $Y$.
\ex

\bx[Partial trace of an EPR pair] \label{EPRX87}
Let $\mathcal{E}:\M_2\otimes \M_2\to \M_2$ be the partial trace over the first factor, and let $\rho_{\EPR}$ be the density matrix given by
\be \label{EPRXS}
\rho_{\EPR}=\frac{1}{2}\left( 
\begin{array}{cccc}
0 & 0 & 0 & 0 \\
0 & 1 & \matminus{1} & 0 \\
0 & \matminus{1} & 1 & 0 \\
0 & 0 & 0 & 0 \\
\end{array}
\right),
\ee
so that $\rho_{\EPR}$ is the projection associated with an EPR pair, or Bell state, often written in Dirac notation as $\frac{1}{\sqrt{2}}\big(|\!\up\up\>-|\!\dn\dn\>\big)$~\cites{Bo51,EPR}. Then 
\[
\rbloom(\rho_{\EPR},\mathcal{E})=
\frac{1}{2}
\left(\begin{smallmatrix}
0 & 0 & 0 & 0 & 0 & 0 & 0 & 0 \\
0 & 0 & 0 & 0 & 0 & 0 & 0 & 0 \\
0 & 1 & 0 & 0 & -1 & 0 & 0 & 0 \\
0 & 0 & 0 & 1 & 0 & 0 & -1 & 0 \\
0 & -1 & 0 & 0 & 1 & 0 & 0 & 0 \\
0 & 0 & 0 & -1 & 0 & 0 & 1 & 0 \\
0 & 0 & 0 & 0 & 0 & 0 & 0 & 0 \\
0 & 0 & 0 & 0 & 0 & 0 & 0 & 0 \\
\end{smallmatrix}\right)=\lbloom(\rho_{\EPR},\mathcal{E})^{\dag}
,
\;\;
\sbloom(\rho_{\EPR},\mathcal{E})=\frac{1}{4}
\left(\begin{smallmatrix}
0 & 0 & 0 & 0 & 0 & 0 & 0 & 0 \\
0 & 0 & 1 & 0 & -1 & 0 & 0 & 0 \\
0 & 1 & 0 & 0 & -1 & 0 & 0 & 0 \\
0 & 0 & 0 & 2 & 0 & -1 & -1 & 0 \\
0 & -1 & -1 & 0 & 2 & 0 & 0 & 0 \\
0 & 0 & 0 & -1 & 0 & 0 & 1 & 0 \\
0 & 0 & 0 & -1 & 0 & 1 & 0 & 0 \\
0 & 0 & 0 & 0 & 0 & 0 & 0 & 0 \\
\end{smallmatrix}\right),
\]
and
\[
\psi_{\LS}(\rho_{\EPR},\mathcal{E})=
\frac{1}{4}
\left(\begin{smallmatrix}
0 & 0 & 0 & 0 & 0 & 0 & 0 & 0 \\
0 & 0 & 0 & 0 & 0 & 0 & 0 & 0 \\
0 & 0 & 1 & 0 & -1 & 0 & 0 & 0 \\
0 & 0 & 0 & 1 & 0 & -1 & 0 & 0 \\
0 & 0 & -1 & 0 & 1 & 0 & 0 & 0 \\
0 & 0 & 0 & -1 & 0 & 1 & 0 & 0 \\
0 & 0 & 0 & 0 & 0 & 0 & 0 & 0 \\
0 & 0 & 0 & 0 & 0 & 0 & 0 & 0 \\
\end{smallmatrix}\right)=\rho_{\EPR}\otimes \frac{\mathds{1}_{2}}{2}.
\]
As such, the right bloom, left bloom, symmetric bloom, and the Leifer--Spekkens state over time all differ in this example. Furthermore, Ohya's compound state over time is given by $\rho_{\EPR}\otimes\frac{\mathds{1}_{2}}{2}$, which agrees with the Leifer--Spekkens state over time in this example. We will revisit this example in the context of dynamical information measures in Section~\ref{CIMX47} and Section~\ref{S81}.
\ex

We now introduce the following definition to distinguish state over time functions not by their formulas, but by the properties they satisfy.

\bd
A state over time function $\psi$ is said to be
\begin{itemize}
\item
\define{hermitian} if and only if $\psi(\rho,\mathcal{E})$ is self-adjoint for all processes $(\rho,\mathcal{E})$.
\item
\define{locally positive} if and only if for all processes $(\rho,\mathcal{E})$ we have $\tr\big(\psi(\rho,\mathcal{E})(A\otimes B)\big)\geq 0$ whenever $A\in \VA$ and $B\in \VB$ are both positive.
\item
\define{channel-linear} if and only if 
\[
\psi\Big(\rho,\lambda \mathcal{E}+(1-\lambda)\mathcal{F}\Big)=\lambda\psi(\rho,\mathcal{E})+(1-\lambda)\psi(\rho,\mathcal{F})
\]
for all $\lambda\in [0,1]$ and for all processes $(\rho,\mathcal{E})$ and $(\rho,\mathcal{F})$.
\item
\define{state-linear} if and only if 
\[
\psi\Big(\lambda\rho+(1-\lambda)\sigma,\mathcal{E}\Big)=\lambda\psi(\rho,\mathcal{E})+(1-\lambda)\psi(\sigma,\mathcal{E})
\]
for all $\lambda\in [0,1]$ for all processes $(\rho,\mathcal{E})$ and $(\sigma,\mathcal{E})$.
\item
\define{bilinear} if and only if it is both channel-linear and state-linear.
\item
\define{classically reducible} if and only if 
\[
\big[(\rho\otimes \mathds{1}),\mathscr{J}[\mathcal{E}]\big]=0\implies \psi(\rho,\mathcal{E})=(\rho\otimes \mathds{1})\mathscr{J}[\mathcal{E}].
\]
\end{itemize}
\ed

\br \label{RMXPDMXDX781}
If $\psi$ is a hermitian state over time function, then it follows from \eqref{SOTX77} that $\psi(\rho,\mathcal{E})$ is a self-adjoint operator of unit trace for all input processes $(\rho,\mathcal{E})$, i.e., $\psi$ maps processes to quasi-states (cf. Definition~\ref{STXDFX99}). In particular, if $(\rho,\mathcal{E})\in \mathscr{P}(\VA,\VB)$ with $\VA$ and $\VB$ both matrix algebras, then $\psi(\rho,\mathcal{E})$ is a quasi-density matrix. 
\er

The following properties of the left/right bloom, the symmetric bloom, and the Leifer--Spekkens state over time function are well-known (see e.g.\ Refs.~\cites{HHPBS17,FuPa22,FuPa22a}).

\bn \label{STXPXT87}
The following statements hold.
\begin{enumerate}[i.]
\item \label{STXPXT1}
The right bloom $\rbloom$ and the left bloom $\lbloom$ are both bilinear and classically reducible.
\item \label{STXPXT2}
The symmetric bloom $\sbloom$ is hermitian, bilinear, and classically reducible.
\item \label{STXPXT3}
The Leifer--Spekkens state over time function $\psi_{\emph{LS}}$ is hermitian, locally positive, channel-linear, and classically reducible.
\end{enumerate}
\en

\br
At present, the symmetric bloom is the only known state over time function that is hermitian, bilinear, and classically reducible, while the Leifer--Spekkens state over time function is the only known state over time function that is hermitian, locally positive, channel-linear, and classically reducible. Proofs (or counter-examples) that these properties characterize these states over time are currently under investigation.
\er 

\br[The Causality Monotone] \label{CSTXMTX67}
In Ref.~\cite{FJV15}, Fitzsimons, Jones, and Vedral argue that if a self-adjoint state over time is to encode causal correlations, then it must necessarily admit negative eigenvalues. In particular, they define a \emph{causality monotone} $\mathfrak{C}:\mathcal{Q}(\VA)\to \R$, with $\VA=\matr_{n}$, given by
\[
\mathfrak{C}(A)=\norm{A}_{1}-1.
\]
Here, $\mathcal{Q}(\VA)$ denotes the real affine space of all quasi-states in $\VA$ and $\norm{*}_1$ denotes the trace-norm, i.e., $\norm{A}_{1}:=\tr\big(\sqrt{A^{\dag}A}\big)$, which is the sum of the singular values of $A$. The causality monotone $\mathfrak{C}$ satisfies the following properties.
\begin{enumerate}[(C1)]
\item
\label{item:C1}
$\mathfrak{C}(A)\geq 0$ for all $A\in \mathcal{Q}(\VA)$.
\item
\label{item:C2}
$\mathfrak{C}(A)=\mathfrak{C}\big(\Ad_U(A)\big)$ for all unitary $U\in \VA$ and all $A\in \mathcal{Q}(\VA)$.
\item
\label{item:C3}
$\mathfrak{C}$ is non-increasing under CPTP maps.
\item
\label{item:C4}
$\mathfrak{C}\left(\sum_{x\in X}p_xA_x\right)\leq \sum_{x\in X}p_x\mathfrak{C}(A_x)$ for all convex combinations $\sum_{x\in X}p_xA_x$ of quasi-states.
\end{enumerate}
Condition~(C\ref{item:C1}) follows from the fact that $A^{\dag}=A$ (so that the singular values equal the absolute value of the eigenvalues) and $\tr(A)=1$ for all $A\in\mathcal{Q}(\VA)$. Condition~(C\ref{item:C2}) holds because the singular values do not change under the adjoint action by a unitary.
Given a CPTP map $\mathcal{E}:\VA\to\VB$, with $\VB$ another matrix algebra, condition~(C\ref{item:C3}) follows from
\be
\label{eq:CMNICPTP}
\mathfrak{C}\big(\mathcal{E}(A)\big)
=\big\lVert\mathcal{E}(A)\big\rVert_{1}-1
\le\lVert\mathcal{E}\rVert_{1-1}\lVert A\rVert_{1}-1
\le\lVert A\rVert_{1}-1
=\mathfrak{C}(A), 
\ee
where $\lVert\mathcal{E}\rVert_{1-1}:=\sup_{A\in\VA}\norm{\mathcal{E}(A)}_{1}/\norm{A}_{1}$ denotes the operator norm, where both $\VA$ and $\VB$ are equipped with the trace norm. Note that the first inequality in~\eqref{eq:CMNICPTP} is a property of any operator norm \cite[Section~5.1]{Fo07}, while the second inequality follows from the contractive property of CPTP maps under the trace norm, i.e.,  $\lVert\mathcal{E}\rVert_{1-1}\le1$~\cite[Theorem~2.1]{PGWPR06}.
Finally, condition~(C\ref{item:C4}) follows from the triangle inequality for norms. 

Moreover, it follows directly from the definition of $\mathfrak{C}$ that a quasi-state $A\in \mathcal{Q}(\M_n)$ admits negative eigenvalues if and only if $\mathfrak{C}(A)>0$. As such, if $A$ is a state over time, then $\mathfrak{C}(A)$ is a measure of deviation of $A$ from being an actual quantum state, which is then interpreted as a measure of the extent to which $A$ is encoding causal correlations. In particular, in Example~\ref{EPRX87} we have 
\be \label{CXTYMTX863}
\mathfrak{C}\big(\sbloom(\rho_{\EPR},\mathcal{E})\big)=1>0 \quad \text{and} \quad \mathfrak{C}\big(\psi_{\LS}(\rho_{\EPR},\mathcal{E})\big)=0.
\ee
Thus, $\sbloom(\rho_{\EPR},\mathcal{E})$ is encoding causal correlations associated with the process $(\rho_{\EPR},\mathcal{E})$, while $\psi_{\LS}(\rho_{\EPR},\mathcal{E})$ and $\psi_{\mathrm{co}}(\rho_{EPR},\mathcal{E})$ are not. While the Leifer--Spekkens state over time function seems to be the most natural in the context of quantum Bayesian inference and retrodiction of quantum states~\cites{LeSp13,FuPa22a,PaBu22}, the observation in \eqref{CXTYMTX863} lends support in favor of the symmetric bloom for capturing both temporal and causal correlations that exist within a quantum process $(\rho,\mathcal{E})$. This viewpoint is further supported in the context of minimizing the mean-square error in quantum estimation theory, where the symmetric bloom provides an optimal estimator for retrodicting hidden quantum observables~\cites{Pe71,Ohki15,Ohki17,Ts22,Ts22b}.
\er

In what follows, we will associate an entropy with quantum processes $(\rho,\mathcal{E})$ by extending the von~Neumann entropy to self-adjoint matrices, and then evaluating this extended von~Neumann entropy on the associated state over time $\psi(\rho,\mathcal{E})$. As our extension of von Neumann entropy is to self-adjoint matrices, hermitian state over time functions will be needed to define such an entropy, and this entropy will then be defined \emph{with respect to} the choice of a hermitian state over time function. Before doing so, however, in the next section, we first define a parametric family of state over time functions that interpolates between the symmetric bloom and Leifer--Spekkens state over time functions. We then motivate the definition of our extension of von~Neumann entropy to self-adjoint matrices by identifying certain charateristic properties such an extension should satisfy.

\section{A parametric family} \label{S4}
We now define a parametric family of state over time functions that interpolates between the right bloom and the left bloom, and whose symmetrization interpolates between the symmetric bloom and Leifer--Spekkens state over time function. 

\bd
\label{defn:SpB}
For every $p\in [0,1]$, the \define{$p$-bloom} is the state over time function $\pbloom{p}$ given by
\[
\pbloom{p}(\rho,\mathcal{E})=(\rho^p\otimes \mathds{1})\mathscr{J}[\mathcal{E}](\rho^{1-p}\otimes \mathds{1}),
\]
where $\rho^{0}$ is set equal to $\mathds{1}$ for every state $\rho$. 
The \define{symmetric $p$-bloom} is the state over time function $\spbloom{p}$ given by
\[
\spbloom{p}=\frac{1}{2}\left(\pbloom{p}+\pbloom{1-p}\right).
\]
\ed

\bn \label{PNXNP1999}
Let $\pbloom{p}$ and $\spbloom{p}$ be the parametric families of state over time functions as in Definition~\ref{defn:SpB}. Then the following statements hold. 
\begin{enumerate}[i.]
\item \label{PNXNP501}
$\pbloom{1}=\rbloom$ .
\item \label{PNXNP0}
$\pbloom{0}=\lbloom$ .
\item \label{PNXNP5}
$\pbloom{1-p}=\dag\circ \pbloom{p}$ \; for all $p\in [0,1]$.
\item \label{PNXNP1}
$\spbloom{p}$ is a hermitian state over time function for all $p\in [0,1]$.
\item \label{PNXNP2}
$\spbloom{p}=\spbloom{1-p}$ \; for all $p\in [0,1]$.
\item \label{PNXNP3}
$\pbloom{\frac{1}{2}}=\spbloom{\frac{1}{2}}=\psi_{\emph{LS}}$ .
\item \label{PNXNP4}
$\spbloom{1}=\spbloom{0}=\sbloom$ .
\item \label{PNXNP7}
$r\hspace{0.5mm}\pbloom{p}+(1-r)\pbloom{q}$ is a state over time function for all $p,q,r\in [0,1]$. 
\end{enumerate}
\en

\bprf 
{\color{white}You found me!}

\noindent
\underline{Item \ref{PNXNP501}}: For every process $(\rho,\mathcal{E})$, we have
\[
\pbloom{1}(\rho,\mathcal{E})=(\rho^1\otimes \mathds{1})\mathscr{J}[\mathcal{E}](\rho^{0}\otimes \mathds{1})=(\rho\otimes \mathds{1})\mathscr{J}[\mathcal{E}]=\rbloom(\rho,\mathcal{E}),
\]
as desired.

\noindent
\underline{Item \ref{PNXNP0}}: The proof is similar to that of item \ref{PNXNP501}.

\noindent
\underline{Item \ref{PNXNP5}}: For all $p\in [0,1]$ and for every process $(\rho,\mathcal{E})$, we have
\begin{eqnarray*}
(\dag\circ \pbloom{p})(\rho,\mathcal{E})&=&\big(\pbloom{p}(\rho,\mathcal{E})\big)^{\dag}=\left((\rho^p\otimes \mathds{1})\mathscr{J}[\mathcal{E}](\rho^{1-p}\otimes \mathds{1})\right)^{\dag} \\
&=&(\rho^{1-p}\otimes \mathds{1})^{\dag}\mathscr{J}[\mathcal{E}]^{\dag}(\rho^p\otimes \mathds{1})^{\dag}=(\rho^{1-p}\otimes \mathds{1})\mathscr{J}[\mathcal{E}](\rho^{p}\otimes \mathds{1}) \\
&=&\pbloom{1-p}(\rho,\mathcal{E}), 
\end{eqnarray*}
where the fourth equality follows from the fact that $\rho$, $\mathds{1}$, and $\Jamiol[\mathcal{E}]$ are all self-adjoint, the latter of which follows from item \ref{CSXS1} of Proposition~\ref{CSXS1971}.

\noindent
\underline{Item \ref{PNXNP1}}: The statement follows directly from item \ref{PNXNP5}.

\noindent
\underline{Item \ref{PNXNP2}}: The statement follows directly from the definition of $\spbloom{p}$.

\noindent
\underline{Item \ref{PNXNP3}}: The statement follows directly from the definition of  
$\psi_{\LS}$. 

\noindent
\underline{Item \ref{PNXNP4}}: 
This follows directly from the definitions of $\spbloom{p}$ and $\sbloom$.

\noindent
\underline{Item \ref{PNXNP7}}: The statement follows from the linearity of partial trace.
\eprf

\bd
The set $\{r\hspace{0.5mm}\pbloom{p}+(1-r)\pbloom{q}\hspace{1mm}|\hspace{1mm} p,q,r\in [0,1]\}$ will be referred to as the \define{$(p,q,r)$-family} of state over time functions.
\ed

In the next section, we prove that if $\psi$ is any state over time function in the $(p,q,r)$-family, then the multi-spectrum of $\psi(\id,\rho)$ 
generically 
contains negative eigenvalues. As such, for even the simplest of processes, states over time are in general not positive. In light of such non-positivity, if we are to define information measures associated with states over time, we need to extend the von~Neumann entropy to non-positive hermitian matrices. We will define such an extension in Section~\ref{S6}, while in the next section, we first motivate our definition.

\section{Extending von~Neumann entropy to quasi-density matrices} \label{S5}

Let 
$f:X\to Y$
be a stochastic map corresponding to a classical channel, and let 
$p$
be a prior distribution on $X$, the inputs of $f$. The associated \emph{entropy} of the classical process $(p,f)$ is the non-negative real number $S(p,f)$ corresponding to the Shannon entropy of the associated joint distribution 
$\vartheta(p,f)$ on $X\times Y$
given by $\vartheta(p,f)_{(x,y)}=p_xf_{yx}$, which, in our language, is the \emph{state over time} associated with the classical process $(p,f)$. If $f$ is a bijection (and so completely deterministic), then the entropy $S(p,f)$ is just the Shannon entropy $H(p)$ of the initial state $p$. In particular, $S(p,\id)=H(p)$ for any classical state $p$.

In the purely quantum domain, bijections correspond to unitary evolution. Thus, if we are to associate an entropy $S(\rho,\mathcal{E})$ with quantum processes $(\rho,\mathcal{E})$ in such a way that retains the essential features of the classical case, then we must do so in such a way so that $S(\rho,\Ad_U)=S(\rho)$, where $U$ is a unitary matrix and $S(\rho)$ is the von~Neumann entropy of the state $\rho$. In particular, we would require $S(\rho,\id)=S(\rho)$, which we view as an essential property for extending entropy from states to processes. But while in the classical case the entropy of $(p,f)$ is simply the Shannon entropy of the state over time $\vartheta(p,f)$, in the quantum domain our states over time $\psi(\rho,\mathcal{E})$ are not necessarily density matrices. Thus, they do not in general have a von~Neumann entropy. As such, if we wish to define entropy in a way such that the equation $S(\rho,\Ad_U)=S(\rho)$ holds for all unitary processes $(\rho,\Ad_U)$, then we need to extend the von~Neumann entropy $S(*)$ to quasi-density matrices in such a way that $S(\psi(\rho,\Ad_U))=S(\rho)$. The following theorem shows that this indeed can be achieved.

\bt \label{BTXETXFX787}
Let $\psi$ be an element of the $(p,q,r)$-family of state over time functions. Then
\be \label{LTBXENTX001}
S(\rho)=-\emph{tr}\Big(\psi(\rho,\Ad_U)\log\big|\psi(\rho,\Ad_U)\big|\Big)
\ee
for every unitary process $(\rho,\Ad_U)\in \mathscr{P}(\M_m,\M_m)$. In particular,
\[
S(\rho)=-\tr\Big(\psi(\rho,\id)\log\big|\psi(\rho,\id)\big|\Big)
\]
for every state $\rho\in\mS(\matr_{m})$.
\et

\begin{figure}
\centering
\includegraphics[width=12cm]{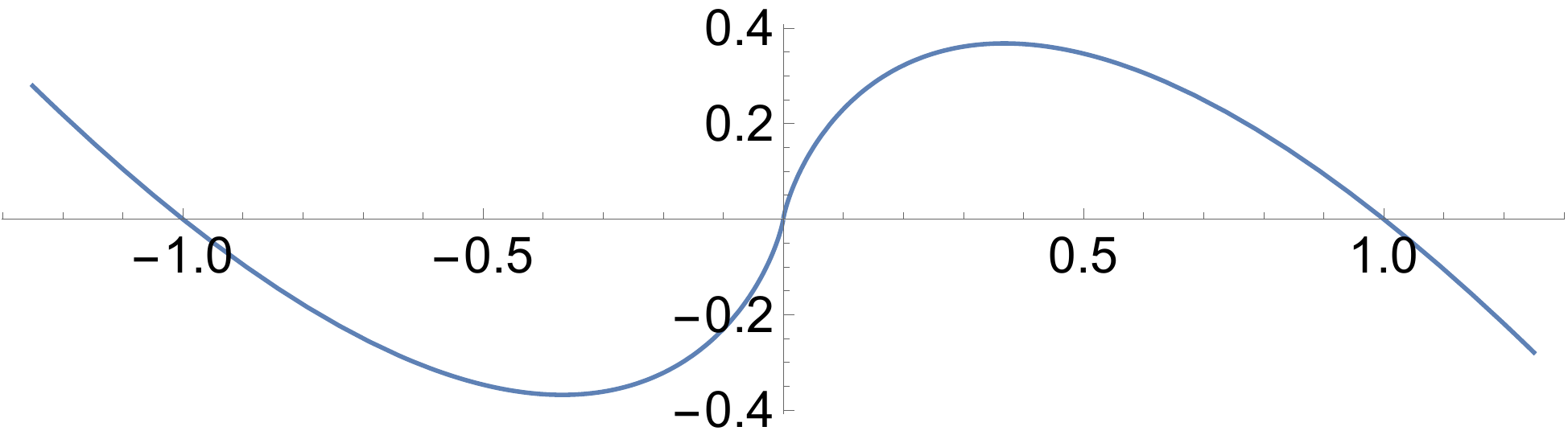}
\caption{A graph of the function $\R\ni x\mapsto\widetilde{\eta}(x)=-x\log|x|$ on part of its domain. This function is continuous on all of $\R$ (with the convention $0\log(0)=0$), and it is infinitely differentiable on $\R\setminus\{0\}$. The function $\widetilde{\eta}$ is the odd completion of the function $[0,\infty)\ni x\mapsto\eta(x)=-x\log (x)$.}
\label{fig:mxlogAbx}
\end{figure}

Before proving Theorem~\ref{BTXETXFX787}, we first explain the meaning of equation~\eqref{LTBXENTX001}, and then we briefly explain how this allows us to extend the von~Neumann entropy to self-adjoint matrices.

\br
\label{rmk:definingEntviaFC}
In the statement of Theorem~\ref{BTXETXFX787}, $|A|:=\sqrt{A^{\dag}A}$ denotes the unique positive squareroot of $A^{\dag}A$, and the convention $0\log0=0$ is used. To make sense of the RHS of equation~\eqref{LTBXENTX001}, one uses the functional calculus for matrices that are not necessarily self-adjoint (see Section 1.2 of Ref.~\cite{Hi08} for more details). 
In the process of proving this theorem, we will see that $\psi(\rho,\Ad_U)$ is actually diagonalizable, even though it need not be hermitian (nor even normal). This provides an alternative method for computing the RHS of~\eqref{LTBXENTX001}, which will be used throughout this work. 
Namely, since the function $f:\R\to\R$ given by $f(x)=-x\log|x|$ is continuous (see Figure~\ref{fig:mxlogAbx} and the next section for more details), if $A=P\Lambda P^{-1}$ is a diagonalization of $A\in\matr_{m}$, with $\Lambda=\diag(\lambda_{1},\dots,\lambda_{m})$ a diagonal matrix of eigenvalues of $A$, and $P$ an (invertible) matrix of corresponding eigenvectors, then $f(A)=Pf(\Lambda)P^{-1}$, where $f(\Lambda)=\diag\big(f(\lambda_{1}),\dots,f(\lambda_{m})\big)$. In this case, this shows that
\be
\label{eqn:FCDENT}
-\tr\big(A\log|A|\big)
=-\tr\Big(P\big(\Lambda\log|\Lambda|\big)P^{-1}\Big)
=-\tr\big(\Lambda\log|\Lambda|\big)
=-\sum_{\lambda\in\mathfrak{mspec}(A)}\lambda\log|\lambda|,
\ee
where $\mathfrak{mspec}(A)$ denotes the multispectrum of $A$ (cf. Definition~\ref{defn:mspec}) and the cyclicity property of trace was used to cancel $P$ with $P^{-1}$.
\er

\br
An immediate consequence of Theorem~\ref{BTXETXFX787}, is that if we define an extension of the von~Neumann entropy to a function $S$ on all self-adjoint matrices given by 
\be \label{EXTENT1681}
S(A)=-\tr\big(A\log|A|\big),
\ee
then for every hermitian state over time function $\psi$, $S(\rho,\mathcal{E}):=S(\psi(\rho,\mathcal{E}))$ would yield an entropy for quantum processes such that $S(\rho,\Ad_U)=S(\rho)$ for every \emph{unitary} process $(\rho,\Ad_U)$. In particular, such an entropy yields the characteristic property
\[
S(\rho,\id)=S(\rho)
\]
for all states $\rho$. Moreover, in the next section, we prove that the extended entropy function given by \eqref{EXTENT1681} satisfies many desirable properties, such as unitary invariance, additivity over product states, and  additivity over orthogonal convex conbinations (cf.\ Proposition~\ref{PTXENT81}). As such, we take the extended entropy function as given by \eqref{EXTENT1681} as the foundation for defining information measures associated with quantum processes. We provide further justification for our use of the extended entropy function, as opposed to alternative proposals, in Remark~\ref{rmk:pseudoentropy} and Remark~\ref{rmk:JSK}.
\er

\blem \label{UEVXTX1973}
Let $\rho\in \mathcal{S}(\M_m)$, with $\mathfrak{mspec}(\rho)=\{\lambda_{1},\dots,\lambda_m\}$, and let $U\in \M_m$ be unitary. Then for all $p,q,r\in [0,1]$, the state over time $r\,\pbloom{p}(\rho,\Ad_U)+(1-r)\pbloom{q}(\rho,\Ad_U)$ is diagonalizable and
\[
\mathfrak{mspec}\Big(r\hspace{0.5mm}\pbloom{p}\left(\rho,\Ad_U\right)+(1-r)\pbloom{q}\left(\rho,\Ad_U\right)\Big)=\mathfrak{mspec}(\rho)\cup  \left\{\left.\pm\sqrt{a_{ij}a_{ji}} \hspace{1mm} \right| \hspace{1mm} 1\leq i<j\leq m \right\} ,
\] 
where $a_{ij}=r\lambda_i^p\lambda_j^{1-p}+(1-r)\lambda_i^q\lambda_j^{1-q}$ . 
\elem

The following proof of Lemma~\ref{UEVXTX1973} relies on several technical results that can be found in Appendix~\ref{BLXCXCS1971}. 

\bprf[Proof of Lemma~\ref{UEVXTX1973}]
Let $\chi_r=r\hspace{0.5mm}\pbloom{p}\left(\rho,\Ad_U\right)+(1-r)\pbloom{q}\left(\rho,\Ad_U\right)$ and let $\rho_{d}$ and $V$ be as in Lemma~\ref{PHIADXU81}. Then 
\begin{eqnarray*}
\chi_r
&\overset{\eqref{DGXRSXP87}}=&r \Ad_{V\otimes UV}\big(\pbloom{p}(\rho_d,\id)\big)+(1-r)\Ad_{V\otimes UV}\big(\pbloom{q}(\rho_d,\id)\big) \\
&=&\Ad_{V\otimes UV}\big(r \pbloom{p}(\rho_d,\id)+(1-r)\pbloom{q}(\rho_d,\id)\big) \\
&\overset{\eqref{DIAGX77}}=&\Ad_{V\otimes UV}\left(r\sum_{i,j}^{m}E_{ij}\otimes \left(\lambda_{i}^{p}\lambda_{j}^{1-p}\right)E_{ji}+(1-r)\sum_{i,j}^{m}E_{ij}\otimes \left(\lambda_{i}^{q}\lambda_{j}^{1-q}\right)E_{ji}\right) \\
&=&\Ad_{V\otimes UV}\left(\sum_{i,j}^{m}E_{ij}\otimes \left(r\lambda_{i}^{p}\lambda_{j}^{1-p}+(1-r)\lambda_{i}^{q}\lambda_{j}^{1-q}\right)E_{ji}\right)
\end{eqnarray*}
for all $r\in [0,1]$. If $a_{ij}>0$ for all distinct $i,j\in\{1,\dots,m\}$, then Lemma~\ref{LXSTAR77} implies the statement. Now suppose there exist distinct $i,j$ such that $a_{ij}=0$. Then, using the fact that $a_{ij}$ is a convex combination of $\lambda_{i}^{p}\lambda_{j}^{1-p}$ and $\lambda_{i}^{q}\lambda_{j}^{1-q}$, at least one of 
\[
0\le \lambda_{i}^{p}\lambda_{j}^{1-p}\le a_{ij} \le \lambda_{i}^{q}\lambda_{j}^{1-q}\le1
\quad\text{ or }\quad
0\le \lambda_{i}^{q}\lambda_{j}^{1-q}\le a_{ij} \le \lambda_{i}^{p}\lambda_{j}^{1-p}\le1
\]
must hold. 
In either case, $a_{ij}=0$ implies that at least one of $\lambda_{i}$ or $\lambda_{j}$ is zero. Hence, 
\[
a_{ji}
=r\lambda_{j}^{p}\lambda_{i}^{1-p}+(1-r)\lambda_{j}^{q}\lambda_{i}^{1-q}
=0.
\]
The fact that the state over time $\chi_{r}$ is diagonalizable then follows from Lemma~\ref{LXSTAR77}.
\eprf

\bprf[Proof of Theorem~\ref{BTXETXFX787}]
Let $(\rho,\Ad_U)\in \mathscr{P}(\M_m,\M_m)$ be a unitary process with $\mathfrak{mspec}(\rho)=\{\lambda_1,\dots,\lambda_m\}$, and let $\psi$ be a state over time function in the $(p,q,r)$-family. By Lemma~\ref{UEVXTX1973}, we have that $\psi(\rho,\Ad_U)$ is diagonalizable, and
\[
\mathfrak{mspec}\big(\psi(\rho,\Ad_U)\big)=\mathfrak{mspec}(\rho)\cup  \left\{\left.\pm\sqrt{a_{ij}a_{ji}} \hspace{1mm} \right| \hspace{1mm} 1\leq i<j\leq m \right\} ,
\]
where $a_{ij}=r\lambda_i^p\lambda_j^{1-p}+(1-r)\lambda_i^q\lambda_j^{1-q}$. As such, if we set $\aleph=\psi(\rho,\Ad_U)$ and $\chi=-\tr(\aleph\log|\aleph|)$, we then have
\begingroup
\allowdisplaybreaks
\begin{eqnarray*}
\chi
&=&-\tr(\aleph\log|\aleph|) \\
&=&-\sum_{\lambda\in \mathfrak{mspec}(\aleph)}\lambda\log|\lambda| \\
&=&-\sum_{i=1}^{m}\lambda_{i}\log\left|\lambda_i\right|-\sum_{1\leq i<j\leq m}\sqrt{a_{ij}a_{ji}}\log\left|\sqrt{a_{ij}a_{ji}}\right|-\sum_{1\leq i<j\leq m}\left(-\sqrt{a_{ij}a_{ji}}\right)\log\left|-\sqrt{a_{ij}a_{ji}}\right| \\
&=&-\sum_{i=1}^{m}\lambda_{i}\log(\lambda_i)-\sum_{1\leq i<j\leq m}\sqrt{a_{ij}a_{ji}}\log\left(\sqrt{a_{ij}a_{ji}}\right)+\sum_{1\leq i<j\leq m}\left(\sqrt{a_{ij}a_{ji}}\right)\log\left(\sqrt{a_{ij}a_{ji}}\right) \\
&=&-\sum_{i=1}^{m}\lambda_{i}\log(\lambda_i) \\
&=&S(\rho),
\end{eqnarray*}
\endgroup
where the second equality follows from~\eqref{eqn:FCDENT}, which is applicable since $\psi(\rho,\Ad_U)$ is diagonalizable by Lemma~\ref{UEVXTX1973} (see also Remark~\ref{rmk:DSOTRMK}), and the third equality follows from Lemma~\ref{UEVXTX1973}.
\eprf

\br[Extending von~Neumann entropy via analytic continuation]
\label{rmk:pseudoentropy}
While there are other approaches to extending the von~Neumann entropy using analytic continuation of the logarithm~\cite{TTT21}, such extensions would not yield analogues of Theorem~\ref{BTXETXFX787}. In particular, if $(\rho,\Ad_{U})$ is a unitary process and $\psi$ is a state over time function which is a member of the $(p,q,r)$-family, then we know by Lemma~\ref{UEVXTX1973} that the multispectrum of $\psi(\rho,\Ad_{U})$ is of the form
\[
\mathfrak{mspec}(\rho)\cup \{\pm \mu_{ij} \hspace{1mm}| \hspace{1mm} 1\leq i<j\leq m\},
\]
where $m$ is such that $\rho\in \mathcal{S}(\M_m)$. As such, the fact that $f(x)=-x\log|x|$ is an odd function ensures that the RHS of \eqref{LTBXENTX001} from Theorem~\ref{BTXETXFX787} coincides with $S(\rho)$. On the other hand, if $\log(z)$ denotes an analytic continuation of the logarithm containing the negative real axis, then $g(z)=-z\log(z)$ is \emph{not} an odd function when restricted to real values of $z$. Therefore, the quantity
\[
-\tr\Big(\psi(\rho,\Ad_U)\log\big(\psi(\rho,\Ad_U)\big)\Big)
\]
does not coincide with $S(\rho)$. For example, if $\rho\in \mathcal{S}(\M_2)$ represents a qubit in a pure state and $\psi$ is the symmetric bloom state over time function, then 
\[
\mathfrak{mspec}\left(\psi(\rho,\id)\right)=\left\{-\frac{1}{2},0,\frac{1}{2},1\right\},
\]
where $\id:\M_2\to \M_2$ denotes the identity map. It then follows that if $\log(z)$ denotes any analytic continuation of the logarithm containing the negative real axis, then
\begin{eqnarray*}
-\tr\Big(\psi(\rho,\id)\log\big(\psi(\rho,\id)\big)\Big)&=&-1\log(1)-0\log(0)-\frac{1}{2}\log\left(\frac{1}{2}\right)-\left(-\frac{1}{2}\right)\log\left(-\frac{1}{2}\right) \\
&=&-\frac{1}{2}\left(\log\left|\frac{1}{2}\right|+i\text{Arg}\left(\frac{1}{2}\right)\right)+\frac{1}{2}\left(\log\left|-\frac{1}{2}\right|+i\text{Arg}\left(-\frac{1}{2}\right)\right) \\
&=&\frac{i}{2}\left(\text{Arg}\left(-\frac{1}{2}\right)-\text{Arg}\left(\frac{1}{2}\right)\right) \\
&=&\frac{i \pi}{2}\\
&\neq&0,
\end{eqnarray*}
while $S(\rho)=0$. Thus, even in the simplest of examples, an analogue of Theorem~\ref{BTXETXFX787} does not hold when using analytic continuation for extending the von~Neumann entropy.
\er 

\br[Extending von Neumann entropy using $-|x|\log|x|$]
\label{rmk:JSK}
In Ref.~\cite{JSK23}, the von~Neumann entropy is extended to self-adjoint matrices using the formula
\be \label{NAIVE81}
S(A)=-\tr\big(|A|\log|A|\big)\equiv -\sum_{i}|\lambda_i|\log|\lambda_i|,
\ee
where $\{\lambda_i\}=\mathfrak{mspec}(A)$. But if $\psi$ is the symmetric bloom and $S(\rho,\mathcal{E}):=S(\psi(\rho,\mathcal{E}))$ with $S(*)$ as given by \eqref{NAIVE81}, then such an extension of entropy from states to processes does not yield the characteristic property $S(\rho,\id)=S(\rho)$ for all states $\rho$. In particular, for the case of a single qubit in a pure state we have
\[
S(\rho,\id)=S(\rho)+\log(2)\neq S(\rho).
\]
Therefore, the characteristic property $S(\rho,\id)=S(\rho)$ fails for this form of entropy. In addition, the functional~\eqref{NAIVE81} does not satisfy other properties, such as additivity over product quasi-states, as compared to our extension of entropy from~\eqref{EXTENT1681} (cf.\ Proposition~\ref{PTXENT81}). However,
the entropy function~\eqref{NAIVE81} is more closely related to the causality monotone from Remark~\ref{CSTXMTX67}, as discussed in Ref.~\cite{JSK23}. Yet another recent extension of the von~Neumann entropy is given by a normalized version of~\eqref{NAIVE81}, which is called the SVD entropy~\cite{PTTW23}.
\er

We conclude this section with a generalization of Theorem~\ref{BTXETXFX787} to arbitrary multi-matrix algebras and $*$-isomorphisms. 

\bc
\label{cor:TEMMA}
Let $\psi$ be an element of the $(p,q,r)$-family of state over time functions. Then 
\[
S(\rho)=-\emph{tr}\Big(\psi(\rho,\mathcal{E})\log\big|\psi(\rho,\mathcal{E})\big|\Big)
\]
for every process $(\rho,\mathcal{E})\in\mathscr{P}(\VA,\VB)$, where $\VA,\VB$ are multi-matrix algebras and $\mathcal{E}$ is a $*$-isomorphism. 
\ec

\bprf
Write $\VA=\bigoplus_{x\in X}\matr_{m_{x}}$ and $\VB=\bigoplus_{y\in Y}\matr_{n_{y}}$. If $\mathcal{E}$ is a $*$-isomorphism, then there exists a bijection $f:X\to Y$ and unitaries $U_{x}\in\matr_{m_{x}}$ such that $n_{f(x)}=m_{x}$ for all $x\in X$ and 
\[
\mathcal{E}_{yx}(A_{x})=
\begin{cases}
U_{x}A_{x}U_{x}^{\dag}&\mbox{ if $y=f(x)$}\\
0 &\mbox{ otherwise}
\end{cases}
\]
for all $x\in X$ and $y\in Y$.
This implies that the channel state takes the particularly simple form
\[
\Jamiol[\mathcal{E}]=\bigoplus_{x\in X}\sum_{i_{x},j_{x}}E^{(m_{x})}_{i_{x}j_{x}}\otimes \Ad_{U_{x}}\big(E^{(m_{x})}_{j_{x}i_{x}}\big).
\]
Therefore, if $\rho\in\mathcal{S}(\VA)$ is decomposed as $\rho=\bigoplus_{x\in X}p_{x}\rho^{x}$, with each $\rho^{x}$ a density matrix so that $\{p_{x}\}_{x\in X}$ defines a probability measure on $X$, then 
\[
\begin{split}
\pbloom{p}(\rho,\mathcal{E})&=(\rho^{p}\otimes\mathds{1}_{\VB})\Jamiol[\mathcal{E}](\rho^{1-p}\otimes\mathds{1}_{\VB})
\\
&=\bigoplus_{x\in X}p_{x}\sum_{i_{x},j_{x}}(\rho^{x})^p E^{(m_{x})}_{i_{x}j_{x}}(\rho^{x})^{1-p}\otimes \Ad_{U_{x}}\big(E^{(m_{x})}_{j_{x}i_{x}}\big)\\
&=\bigoplus_{x\in X}p_{x}\pbloom{p}\big(\rho^{x},\Ad_{U_{x}}\big),
\end{split}
\]
where $p$ without a subscript refers to the $p$ inside a $(p,q,r)$-family (which is not to be confused with the $p_{x}$ using subscripts, the latter of which refer to the probability measure associated with $\rho$ on the direct sum factors). Hence, 
\[
\psi(\rho,\mathcal{E})=\bigoplus_{x\in X}p_{x}\psi\big(\rho^{x},\Ad_{U_{x}}\big),
\]
which itself implies 
\[
\big|\psi(\rho,\mathcal{E})\big|=\bigoplus_{x\in X}p_{x}\big|\psi(\rho^{x},\Ad_{U_{x}})\big|.
\]
Putting this all together gives
\[
\begin{split}
-\tr\Big(\psi(\rho,\mathcal{E})\log\big|\psi(\rho,\mathcal{E})\big|\Big)
&=-\sum_{x\in X}p_{x}\tr\bigg(\psi(\rho^{x},\Ad_{U_{x}})\log\Big(p_{x}\big|\psi(\rho^{x},\Ad_{U_{x}})\big|\Big)\bigg)\\
&=H(p)+\sum_{x\in X}p_{x}\bigg(-\tr\Big(\psi(\rho^{x},\Ad_{U_{x}})\log\big|\psi(\rho^{x},\Ad_{U_{x}})\big|\Big)\bigg)\\
&=H(p)+\sum_{x}p_{x}S(\rho^{x})\\
&=S(\rho).
\end{split}
\]
The third equality follows from Theorem~\ref{BTXETXFX787}, while the remaining equalities follow from the previously computed identities as well as basic properties of the logarithm and the von~Neumann entropy~\cite{AJP22}. 
\eprf

\section{Properties of the extended entropy function} \label{S6}
In this section we investigate mathematical aspects of our extension of the von~Neumann entropy as given by \eqref{EXTENT1681}. We recall that $\H_n$ denotes the set of all self-adjoint matrices in $\M_n$.

\begin{figure}
\begin{tabular}{m{5.5cm} m{5.5cm} m{5.5cm}}
\includegraphics[width=5.5cm]{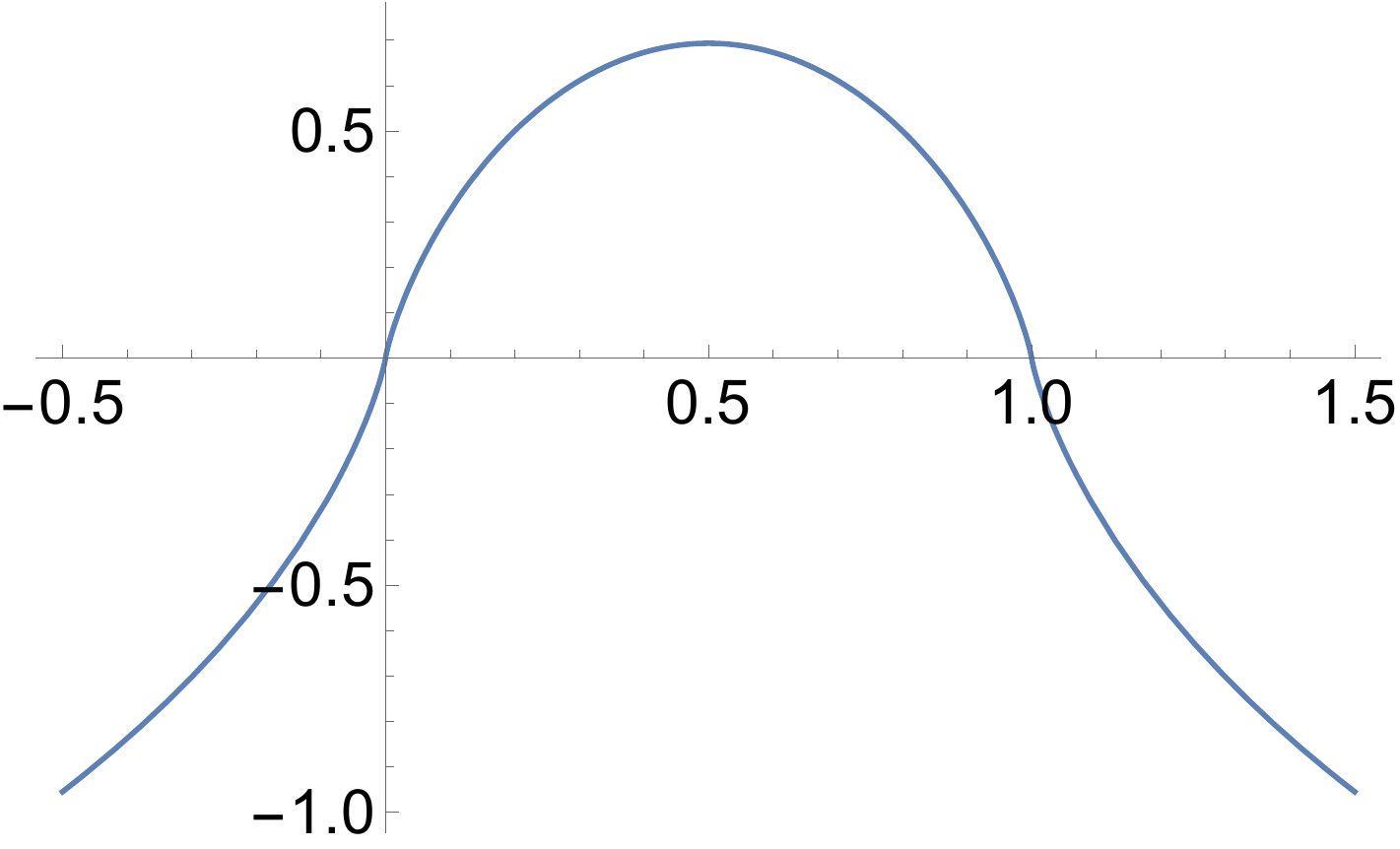}
&
\includegraphics[width=5.5cm,trim={2cm 0cm 1.9cm 0cm},clip]{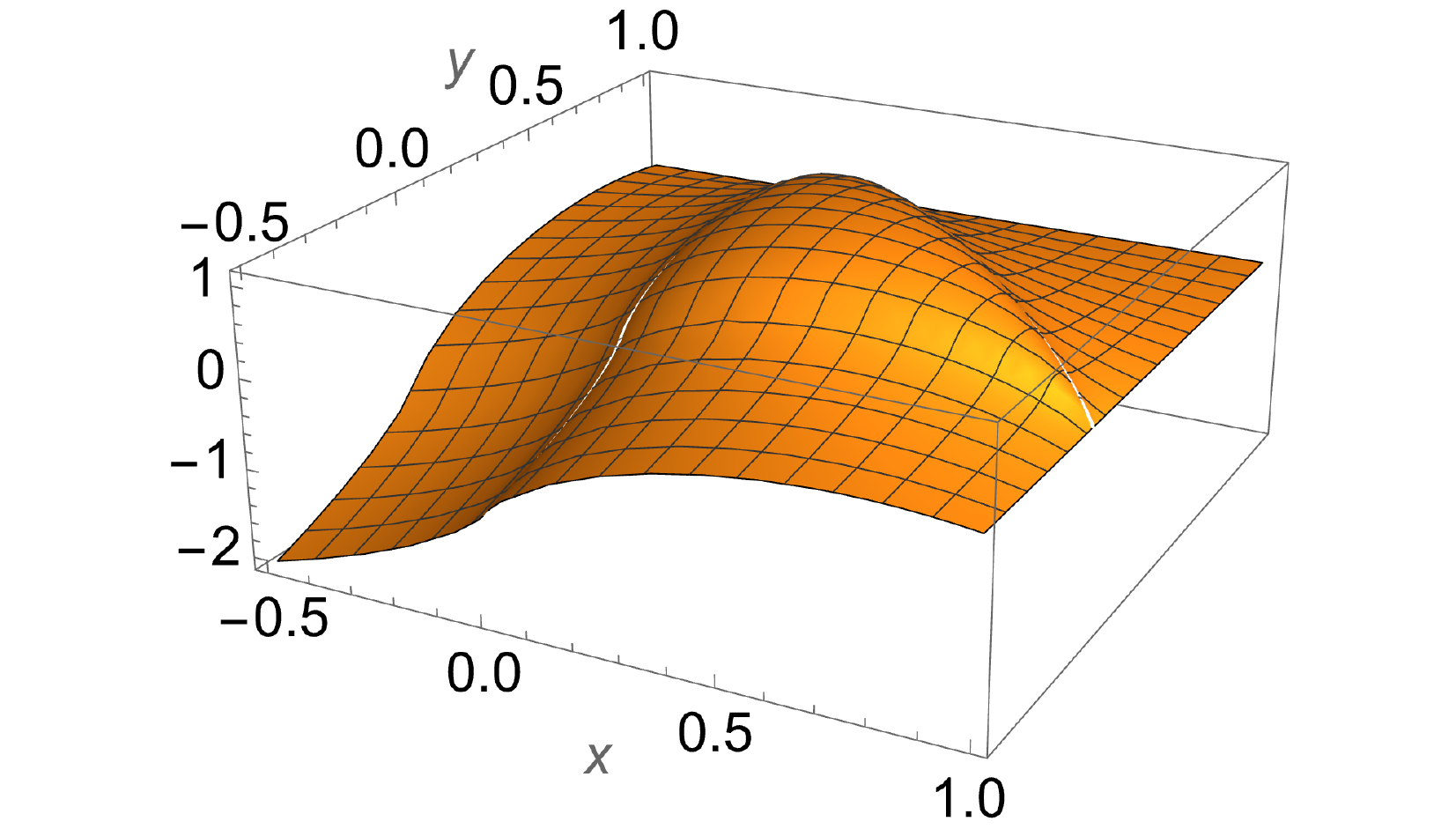}
&
\includegraphics[width=5.5cm]{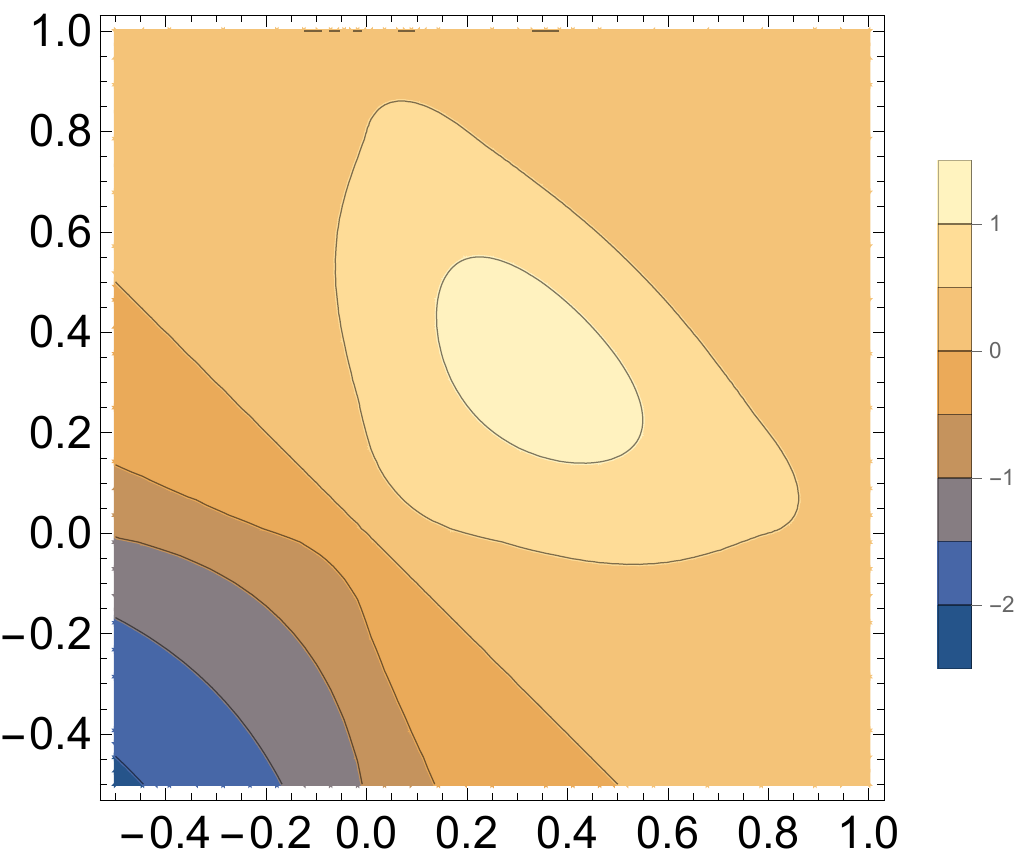}
\end{tabular}
\caption{The left plot shows the graph of the function $-x\log|x|-(1-x)\log|1-x|$ for $x\in[-1/2,3/2]$, which is what the entropy from Definition~\ref{defn:extEnt} looks like for quasi-probability distributions whose entries are contained in the interval $[-1/2,3/2]$. The middle plot shows the graph of the function $-x\log|x|-y\log|y|-(1-x-y)\log|1-x-y|$ for $(x,y)\in[-1/2,1]\times[-1/2,1]$. The right plot shows the same information of the middle plot drawn as a contour plot.}
\label{fig:newshannon}
\end{figure}

\bd
\label{defn:extEnt}
Let $p:X\to \R$ be a quasi-probability distribution on a finite set $X$. The \define{entropy} of $p$ is the real number $H(p)$ given by 
\[
H(p)=-\sum_{x\in X}p_x\log|p_x|.
\]
If $p$ is a probability distribution, then $H(p)$ will be referred to as the \define{Shannon entropy} of $p$ (see Figure~\ref{fig:newshannon} for examples).
\ed

\bd
\label{defn:JEnt}
The \define{entropy} of a self-adjoint matrix $A\in \M_n$  is the real number $S(A)$ given by
\be\label{ENTDFX181}
S(A)= -\tr\big(A\log\left|A\right|\big)\equiv -\sum_{\lambda\in \mathfrak{mspec}(A)}\lambda\log\left|\lambda\right|.
\ee
If $A$ is a density matrix, then $S(A)$ will be referred to as the \define{von~Neumann entropy} of $A$.
\ed

Besides the consistency condition of Theorem~\ref{BTXETXFX787} for unitary processes, the entropy function of Definition~\ref{defn:JEnt} satisfies many important properties that are extensions of the usual properties for the von~Neumann entropy function. 

\bn\label{PTXENT81}
The entropy function $S:\bigcup_{n=1}^{\infty}\mathbb{H}_n\to \R$ given by \eqref{ENTDFX181} satisfies the following properties.
\begin{enumerate}[i.]
\item\label{ETX1}
\underline{\emph{Extension}}: $S(\rho)$ is the von~Neumann entropy for every density matrix $\rho$.
\item\label{ETX2}
\underline{\emph{Isometric Invariance}}: $S(\rho)=S(U\rho U^{\dag})$ for every unitary $U$ and self-adjoint $\rho$.
\item\label{ETX3}
\underline{\emph{Additivity}}: $S(\rho\otimes \xi)=S(\rho)+S(\xi)$ for all self-adjoint matrices $\rho$ and $\xi$.
\item\label{ETX4}
\underline{\emph{Orthogonal Affinity}}: If $p:X\to \R$ is a quasi-probability distribution on a finite set $X$ and $\rho^x\in \mathbb{H}_n$ is a collection of mutually orthogonal, unit-trace self-adjoint matrices indexed by $X$, then
\be\label{EZNTXAD97}
S\left(\sum_{x\in X}p_x\rho^x\right)=H(p)+\sum_{x\in X}p_xS(\rho^x).
\ee
\item \label{ETX6}
\underline{\emph{Continuity}}: The entropy function $S$ is continuous. Moreover, if $\rho,\xi\in \mathbb{H}_n$ are quasi-states with multispectrums $\{\lambda_i\}$ and $\{\mu_i\}$ such that $\lambda_i \mu_i\geq 0$ for all $i\in \{1,\dots,n\}$, and if $\norm{\rho-\xi}_1\leq 1/e$, then
\be \label{FANNES17}
\big|S(\rho)-S(\xi)\big|\leq \norm{\rho-\xi}_1\log(n)+\eta\big(\norm{\rho-\xi}_1\big),
\ee
where $\eta(x)=-x\log(x)$.
\end{enumerate}
\en

\bprf
{\color{white}You found me!}

\noindent
\underline{Item \ref{ETX1}}: The statement follows from the fact that a density matrix $\rho$ is positive and therefore satisfies $\rho=|\rho|$.

\noindent
\underline{Item \ref{ETX2}}: The statement follows from the cyclicity of the trace and the functional calculus for matrices, or equivalently, from the fact that $U\rho U^{\dag}$ has the same eigenvalues as $\rho$.

\noindent
\underline{Item \ref{ETX3}}: Suppose $\mathfrak{mspec}(\rho)=\{\lambda_i\}$ and $\mathfrak{mspec}(\xi)=\{\mu_j\}$, so that $\mathfrak{mspec}(\rho\otimes \sigma)=\{\lambda_i\mu_j\}$. Then
\begin{eqnarray*}
S(\rho\otimes \xi)&=&-\sum_{i,j}\lambda_i\mu_j\log\left|\lambda_i\mu_j\right|=-\sum_{i,j}\lambda_i\mu_j\left(\log\left|\lambda_i\right|+\log\left|\mu_j\right|\right) \\
&=&-\sum_{i,j}\lambda_i\mu_j\log\left|\lambda_i\right|-\sum_{i,j}\lambda_i\mu_j\log\left|\mu_j\right|=-\sum_{i}\lambda_i\log\left|\lambda_i\right|-\sum_{j}\mu_j\log\left|\mu_j\right| \\
&=&S(\rho)+S(\xi),
\end{eqnarray*}
as desired.

\noindent
\underline{Item \ref{ETX4}}: Let $\{\lambda_i^x\}$ denote the multispectrum of $\rho^x$ for all $x\in X$, so that the multispectrum of $\sum_{x\in X}p_x\rho^x$ is $\{p_x\lambda_i^x\}$. Then
\begin{eqnarray*}
S\left(\sum_{x\in X}p_x\rho^x\right)&=&-\sum_{x\in X}\sum_{i=1}^{n}p_x\lambda^x_i\log\left|p_x\lambda^x_i\right| \\
&=&-\sum_{x\in X}\sum_{i=1}^{n}p_x\lambda^x_i\log|p_x|-\sum_{x\in X}\sum_{i=1}^{n}p_x\lambda^x_i\log\left|\lambda^x_i\right| \\
&=&-\sum_{x\in X}p_x\log|p_x|\sum_{i=1}^{n}\lambda^x_i-\sum_{x\in X}p_x\sum_{i=1}^{n}\lambda^x_i\log\left|\lambda^x_i\right| \\
&=&H(p)+\sum_{x\in X}p_xS(\rho^x),
\end{eqnarray*}
as desired.

\noindent
\underline{Item \ref{ETX6}}: For all $n\in \N$, let $\widetilde{\eta}_n:\R^n\to \R$ be the function given by
\[
\widetilde{\eta}_n(x_1,\dots,x_n)=-\sum_{i=1}^{n}x_i\log\left|x_i\right|
\]
(where we set $0\log0=0$), and let $\mathscr{E}_n:\mathcal{Q}(\M_n)\to \R^n$ be the function given by
\[
\mathscr{E}_n(\rho)=(\lambda_1,\dots,\lambda_n),
\]
where $\mathfrak{mspec}(\rho)=\{\lambda_1,\dots,\lambda_n\}$. Recall that $\mathcal{Q}(\M_n)$ denotes the set of all quasi-states in $\M_n$. 
The identity $S(\rho)=(\widetilde{\eta}_n\circ \mathscr{E}_n)(\rho)$ for all $\rho\in \mathcal{Q}(\M_n)$ shows that $S$ is the composite of two continuous functions, from which it follows that $S$ is continuous as well (cf.\ Chapters 1 and 5 in Ref.~\cite{Wa18}).

To prove the Fannes-type inequality \eqref{FANNES17}, we adapt the standard proof for density matrices to the case at hand (cf. Theorem~11.6 of Ref. \cite{NiCh11}). Let $\eta:[0,\infty)\to \R$ be the function given by $\eta(x)=-x\log(x)$, and let $\widetilde{\eta}:\R\to \R$ be the odd completion of $\eta$, so that $\widetilde{\eta}(x)=-x\log|x|$. Suppose now that $\rho,\xi\in \mathcal{Q}(\M_n)$ are quasi-states with multispectrums $\{\lambda_i\}$ and $\{\mu_i\}$, suppose $\lambda_i \mu_i\geq 0$ for all $i\in \{1,\dots,n\}$, and also suppose $\norm{\rho-\xi}_1\leq 1/e$, so that $|\lambda_i-\mu_i|<\frac{1}{2}$. Then 
\begin{eqnarray*}
\big|S(\rho)-S(\xi)\big|
&=&\left|\sum_{i=1}^n\widetilde{\eta}(\lambda_i)-\widetilde{\eta}(\mu_i)\right|\leq \sum_{i=1}^n\big|\widetilde{\eta}(\lambda_i)-\widetilde{\eta}(\mu_i)\big|  \\
&=&\sum_{i=1}^n\Big|\eta\big(|\lambda_i|\big)-\eta\big(|\mu_i|\big)\Big|\leq \sum_{i=1}^n\eta\Big(\big|\hspace{1mm}|\lambda_i|-|\mu_i|\hspace{1mm}\big|\Big) \\
&=&\sum_{i=1}^n\eta\big(|\lambda_i-\mu_i|\big),
\end{eqnarray*}
where the second equality follows from the fact that $rs\geq 0$ implies 
\[
\big|\widetilde{\eta}(r)-\widetilde{\eta}(s)\big|=\Big|\eta\big(|r|\big)-\eta\big(|s|\big)\Big|,
\]
and the second inequality follows from the fact that $|r-s|<\frac{1}{2}$ with $r$ and $s$ non-negative implies
\[
\big|\eta(r)-\eta(s)\big|\leq \eta\big(|r-s|\big).
\]
Now set $\epsilon_i=\left|\lambda_i-\mu_i\right|$ for all $i\in \{1,\dots,n\}$, and set $\epsilon=\epsilon_1+\cdots+\epsilon_n$, so that $\epsilon\leq \norm{\rho-\xi}_1$. We then have
\begin{eqnarray*}
\left|S(\rho)-S(\xi)\right|&\leq& \sum_{i=1}^n\eta\left(\left|\lambda_i-\mu_i\right|\right)=\sum_{i=1}^n\eta\left(\epsilon_i\right)=\sum_{i=1}^n\big(\epsilon\eta(\epsilon_i/\epsilon)-\epsilon_i\log(\epsilon)\big) \\
&=&\epsilon\left(\sum_{i=1}^{n}\eta(\epsilon_i/\epsilon)\right)+\eta(\epsilon)\leq \norm{\rho-\xi}_1\log(n)+\eta\big(\norm{\rho-\xi}_1\big),
\end{eqnarray*}
where the final inequality follows from the fact that $\epsilon\leq \norm{\rho-\xi}_1$ and $\eta$ is monotone-increasing on $[0,1/e]$, thus concluding the proof.
\eprf

\br[Failure of subadditivity for quasi-states] \label{SAXT67} 
If $\rho_{\VA\VB}\in \VA\otimes \VB$ is a density matrix with marginals $\rho_{\VA}=\tr_{\VB}(\rho_{\VA\VB})$ and $\rho_{\VB}=\tr_{\VA}(\rho_{\VA\VB})$, then
\[
S(\rho_{\VA\VB})\leq S(\rho_{\VA})+S(\rho_{\VB}),
\]
which is a property of von~Neumann entropy commonly referred to as \emph{subadditivity}. This property does not hold in general for the extended entropy function for arbitrary quasi-states whose marginals are states. For a counter-example, consider the non-positive self-adjoint matrix 
\[
\rho_{\mA\mB}=\frac{1}{12}
\left(
\begin{array}{cccc}
-6&\sqrt{5}&\sqrt{5}&0\\
\sqrt{5}&8&0&\sqrt{5}\\
\sqrt{5}&0&8&\sqrt{5}\\
0&\sqrt{5}&\sqrt{5}&2
\end{array}
\right),
\]
whose marginals are given by the same reduced density matrix
\[
\rho_{\mA}=\frac{1}{6}
\left(\!
\begin{array}{cc}
1&\sqrt{5}\\\sqrt{5}&5
\end{array}
\right)
=\rho_{\mB}.
\]
In such a case, we have $\rho_{\mA}$ and $\rho_{\mB}$ correspond to pure states, so that their entropies vanish. Meanwhile, $S(\rho_{\mA\mB})\approx0.29$, illustrating that $S(\rho_{\VA\VB})>S(\rho_{\VA})+S(\rho_{\VB})$.

\begin{figure}[ht]
\centering
\begin{tabular}{m{8cm} m{0.25cm} m{8cm}}
\includegraphics[width=8cm]{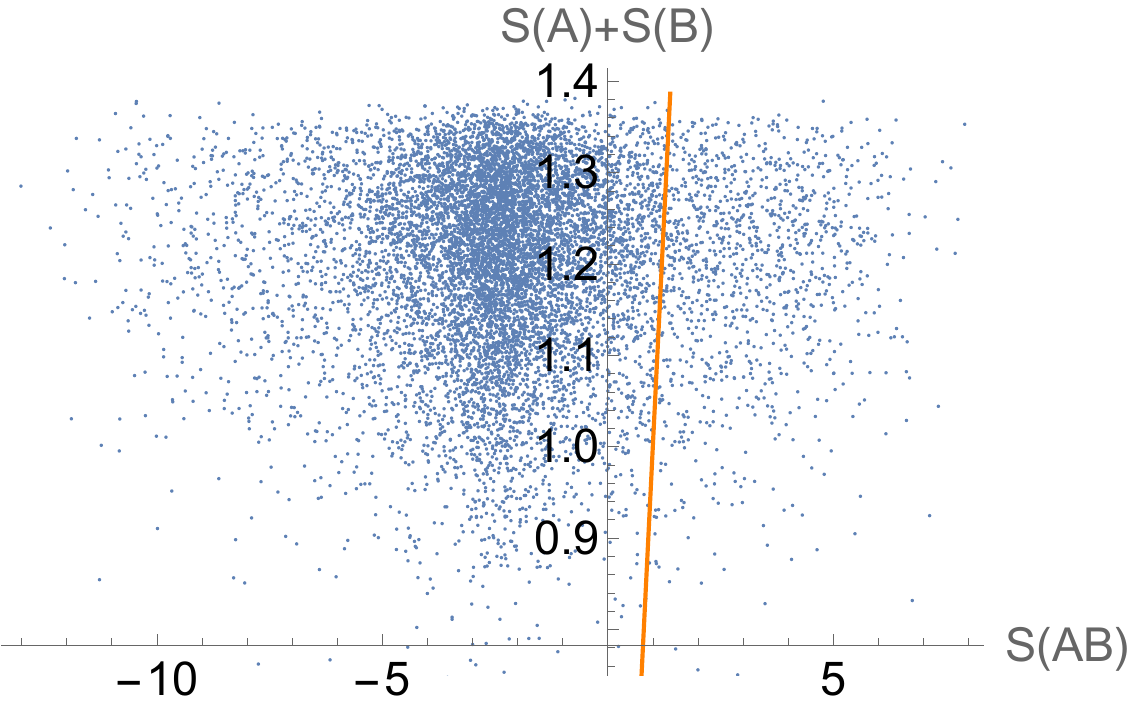}
& &
\includegraphics[width=8cm]{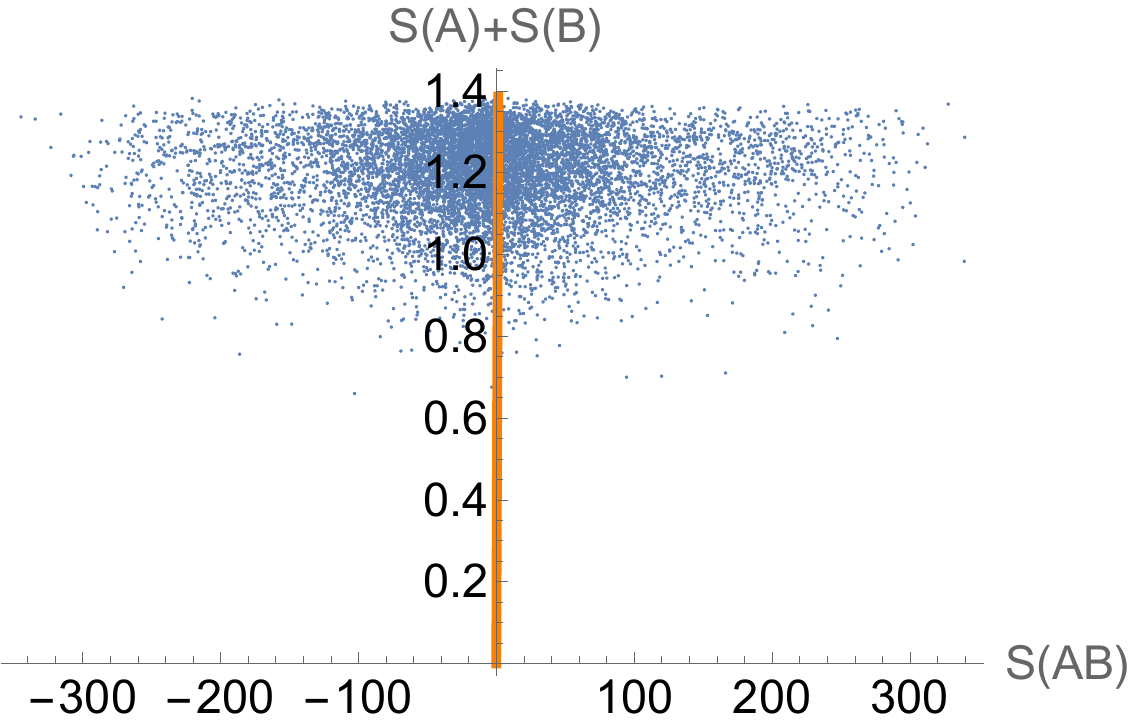}
\end{tabular}
\caption{In both scatter plots, $\mA=\matr_{2}=\mB$. A randomly selected hermitian matrix $\sigma_{\mA\mB}\in\mA\otimes\mB$ was generated with marginals $\rho_{\mA}=\tr_{\mB}(\sigma_{\mA\mB})$ and $\rho_{\mB}=\tr_{\mA}(\sigma_{\mA\mB})$ that are states. Each scatter plot consists of 10000 values of $(S(AB),S(A)+S(B)):=(S(\sigma_{\mA\mB}),S(\rho_{\mA})+S(\rho_{\mB}))$, and the line (in orange) is the graph of the identity function indicating that points above this line (in the figures, these look like points to the left of the line) satisfy $S(\sigma_{\mA\mB})\le S(\rho_{\mA})+S(\rho_{\mB})$, while points below the line satisfy $S(\sigma_{\mA\mB})\ge S(\rho_{\mA})+S(\rho_{\mB})$. The procedure used to generate one of these data points is as follows. First, create a random density matrix $\rho_{\mA\mB}$ by constructing a Haar random pure state on $\mA\otimes\mB\otimes\mC$, where $\mC=\matr_{4},$ and then partially trace out $\mC$ to obtain $\rho_{\mA\mB}$. Then, randomly generate a hermitian matrix $\tau_{\mA\mB}\in\mA\otimes\mB$ that satisfies $\tr_{\mB}(\tau_{\mA\mB})=0$ and $\tr_{\mA}(\tau_{\mA\mB})=0$. In this special case of $4$ by $4$ matrices $(a_{ij})$, the entry $a_{11}$ of $\tau_{\VA\VB}$ was selected uniformly at random from the interval $[-3,3]$ in the left plot (and the interval $[-100,100]$ in the right plot), while the entries $a_{12},a_{13},a_{14}$, and $a_{23}$ were selected uniformly at random from the square in the complex plane of side length 6 in the plot on the left (and side length 200 in the plot on the right) centered at the origin (and whose sides are parallel to the coordinate axes). Finally, $\sigma_{\mA\mB}$ is defined as $\sigma_{\mA\mB}=\rho_{\mA\mB}+\tau_{\mA\mB}$. 
}
\label{fig:SAFH}
\end{figure}

In addition, the failure of subadditivity is not restricted to some subset of measure zero, as Figure~\ref{fig:SAFH} illustrates. In fact, not only does subadditivity fail, but one sees from Figure~\ref{fig:SAFH} that the entropy of quasi-densities is unbounded for a fixed dimension, even if the marginals are states (this phenomena is somewhat reminiscent of the amplification of pseudo-entropy~\cite{IKMT22}).
We will revisit subadditivity for \emph{states over time} in the next section (cf. Remark~\ref{rmk:NNPMI}).  
\er

\bd
Let $X$ be a finite set. The \define{entropy} of a self-adjoint element $\rho=\bigoplus_{x\in X}A_x\in \bigoplus_{x\in X}\M_{m_x}$ is the real number $S(\rho)$ given by 
\be \label{ENTDFQDX151}
S(\rho)=S\big(\mathfrak{bloc}(\rho)\big),
\ee
where $\mathfrak{bloc}(\rho)$ is a block-diagonal representation of $\rho$ in the matrix algebra $\M_n$ and $n=\sum_{x\in X}m_x$.   
\ed

\bn
Let $X$ be a finite set, let $\rho=\bigoplus_{x\in X}A_x\in \bigoplus_{x\in X}\M_{m_x}$ be a quasi-state such that $\tr(A_x)\neq 0$ for all $x\in X$, let $p:X\to \R$ be the quasi-probability distribution given by $p_x=\emph{tr}(A_x)$, and let $\rho^x=\frac{1}{p_x}A_x$ for all $x\in X$. Then, $\rho=\bigoplus_{x\in X}p_x\rho^x$ and
\be \label{SENTX81}
S(\rho)=H(p)+\sum_{x\in X}p_xS(\rho^x).
\ee
\en

\bprf
The statement follows from the orthogonal affinity condition \eqref{EZNTXAD97} applied to $\mathfrak{bloc}(\rho)$.
\eprf

\br
The formula $S(A)=-\tr\big(A\log|A|\big)$ appears in Refs.~\cites{SSW14,TTC22,CiKu23} for different purposes, which we briefly discuss. 
Ref.~\cite{SSW14} uses it as an extension of entropy (called conditional entropy in Ref.~\cite{SSW14}) to two-states~\cites{Wat55,ABL64,ReAh95} (also called transition matrices~\cite{NTTTW21}). Two-states are discussed in more detail in terms of states over time in Ref.~\cite{FuPa22a}. Briefly, a two-state is a particular case of the left bloom state over time with an initial density matrix corresponding to a pure state and a quantum channel given by a projection-valued measure. And because every member of the $(p,q,r)$ family, which includes the left bloom state over time function, is diagonalizeable by Lemma~\ref{UEVXTX1973} (see also Remark~\ref{rmk:definingEntviaFC}), our formula for the entropy from Definition~\ref{defn:JEnt} extends the entropy of Ref.~\cite{SSW14} to a much larger class of state over time functions.
Meanwhile, Ref.~\cite{TTC22} (see also the earlier Ref.~\cite{CJS17}) focuses on the applications of such an entropy formula to quantifying entanglement in non-hermitian quantum systems and providing a relationship to negative central charges in non-unitary conformal field theories. 
Furthermore, Ref.~\cite{CiKu23} computes the Page curve associated with such an entropy.
However, we will not discuss the extension of entropy as in Equation~\eqref{ENTDFX181} to non-hermitian matrices in more detail in this paper, as our primary focus here is to apply our formula to hermitian states over time.
\er

Before closing this section, we prove one more fact regarding the isometric invariance of the entropy. This will be used in Section~\ref{S9} when we prove a \emph{quantum entropic Bayes' rule} (cf. Theorem~\ref{QEXTPXBR91}).

\bn
\label{prop:SEinv}
If $\mathcal{E}:\VA\to\VB$ is a $*$-isomorphism and  $\rho\in\VA$ is a quasi-state, then $S(\mathcal{E}(\rho))=S(\rho)$. 
\en

\bprf
Given multimatrix algebras $\VA=\bigoplus_{x\in X}\matr_{m_{x}}$ and $\VB=\bigoplus_{y\in Y}\matr_{n_{y}}$, if a $*$-isomorphism $\mathcal{E}:\VA\to\VB$ exists, then $m=n$, where $m=\sum_{x\in X}m_{x}$ and $n=\sum_{y\in Y}$, and there exists a unitary $U\in\matr_{m}$ such that $U\big(\mathfrak{bloc}(\rho)\big)U^{\dag}=\mathfrak{bloc}\big(\mathcal{E}(\rho)\big)$ (see the beginning of the proof of Corollary~\ref{cor:TEMMA}). The claim then follows from the isometric invariance of the entropy in Proposition~\ref{PTXENT81}.
\eprf

\section{Dynamical information measures}\label{CIMX47} \label{S7}

We now define dynamical measures of quantum information associated with a hermitian state over time $\psi(\rho,\mathcal{E})$. Before doing so, however, we first recall the dynamical measures of classical information, which serve as motivation for our definitions~\cites{Sh48,Kh57,BFL11,Fu21AXM,FuPa21}.

\bd[Dynamical measures of classical information] \label{DYXNXC19} 
Let $f:X\to Y$ be a stochastic map, let $p:X\to [0,1]$ be a prior distribution on $X$, let $q:Y\to [0,1]$ be the associated output distribution on $Y$, and let $\vartheta(p,f):X\times Y\to [0,1]$ be the associated classical state over time, so that
\[
q_y=\sum_{x\in X}p_xf_{yx} \quad \text{and} \quad \vartheta(p,f)_{(x,y)}=p_xf_{yx}.
\] 
\begin{itemize}
\item
The \define{entropy} of $(p,f)$ is the real number $S(p,f)$ given by
\[
S(p,f)=H\big(\vartheta(p,f)\big).
\]
\item
The \define{conditional entropy} of $(p,f)$ is the real number $H(p,f)$ given by
\[
H(p,f)=S(p,f)-H(p).
\]
\item
The \define{mutual information} of $(p,f)$ is the real number $I(p,f)$ given by
\[
I(p,f)=H(p)+H(q)-S(p,f).
\]
\item
The \define{information loss} of $(p,f)$ is the real number $K(p,f)$ given by
\[
K(p,f)=S(p,f)-H(q).
\]
\end{itemize}
\ed

What we call ``information loss'' above was called ``conditional information loss'' in Ref.~\cite{FuPa21}. 
While the entropy $S(p,f)$ may be viewed as the entropy of the entire process $(p,f)$, one can show 
\be
\label{eq:CCE}
H(p,f)=\sum_{x\in X}p_xH(f_x).
\ee
Thus, the conditional entropy may be viewed as a measure of the uncertainty of the outputs of $f$ weighted by the prior distribution $p$ on the inputs of $f$. And while the mutual information $I(p,f)$ may be viewed as the information that is shared in the process $(p,f)$, the information loss $K(p,f)$ may be viewed---as its name suggests---as the information that is lost during the process $(p,f)$. We illustrate this in the following two examples. 

\bx[Classical deterministic evolution] \label{CDXE771}
Let $f:X\to Y$ be a surjective function between finite sets, so that we may view $f$ as a stochastic map that associates every $x\in X$ with the point-mass distribution supported on $f(x)\in Y$. Given a prior distribution $p$ on $X$, the associated output distribution $q$ on $Y$ is then given by
\[
q_y=\sum_{x\in f^{-1}(y)}p_x.
\]
It then follows that
\[
S(p,f)=H(p), \quad H(p,f)=0, \quad I(p,f)=H(q), \quad \text{and} \quad  K(p,f)=H(p)-H(q).
\]
The equality $S(p,f)=H(p)$ can then interpreted as the fact that all the uncertainty in the process $(p,f)$ is contained in the uncertainty of its inputs, and $H(p,f)=0$ follows from the fact that there is no uncertainty in the outputs of $f$ given knowledge of its inputs. The information that is shared in the process is then $I(p,f)=H(q)\leq H(p)$, and the information that is lost in the process is captured by the entropy difference $H(p)-H(q)=K(p,f)$ (since $H(p,f)=0$). 
\ex

\bx[Vanishing of information loss and correctable codes]
Let $f:X\to Y$ be a stochastic map and let $p:X\to[0,1]$ be a probability distribution on $X$. Viewing $f$ as a noisy communication channel, one can prove that $f$ is correctable if and only if $K(p,f)=0$~\cite{FuPa21}. The precise definition of correctability is given in~\cite[Remark~A1]{FuPa21}. The interpretation
is that, although $f$ describes a noisy channel, no information is lost because there exists a deterministic procedure that is able to perfectly correct any error that can occur (within the model). 
\ex

Axiomatic characterizations of the dynamical measures of classical information given by Definition~\ref{DYXNXC19} can be found in Refs.~\cites{BFL11,FuPa21,Fu21AXM}, and a dynamical axiomatic characterization of von~Neumann entropy appears in Ref.~\cite{AJP22}. Many other axiomatic characterizations of related entropy functions can be found in Refs.~\cites{Re61,Oc75,Cs08,BaFr14,Br21,BGWG21,Leinster21} and the references therein.

In what follows we generalize the classical information measures given by Definition~\ref{DYXNXC19} to the quantum dynamics associated with a general process $(\rho,\mathcal{E})\in  \mathscr{P}(\VA,\VB)$. But while there is a canonical state over time $\vartheta(p,f)$ associated with the classical dynamics of $(p,f)$, state over time functions are not unique in the quantum setting. As such, we generalize such information measures \emph{with respect to} the choice of a hermitian state over time function $\psi$.  The hermiticity assumption on $\psi$ is needed to ensure $\psi(\rho,\mathcal{E})$ is a quasi-state for all processes $(\rho,\mathcal{E})$ (see Remark~\ref{RMXPDMXDX781}), which will be assumed for our definitions.  

\bd[Dynamical measures of quantum information] \label{DMXQIX981}
Let $\psi$ be a hermitian state over time function, let $\VA$ and $\VB$ be multi-matrix algebras, and let $(\rho,\mathcal{E})\in \mathscr{P}(\VA,\VB)$. 
\begin{itemize}
\item
The $\psi$-\define{entropy} of $(\rho,\mathcal{E})$ is the real number $S_{\psi}(\rho,\mathcal{E})$ given by
\[
S_{\psi}(\rho,\mathcal{E})=S\big(\psi(\rho,\mathcal{E})\big).
\]
\item
The $\psi$-\define{conditional entropy} of $(\rho,\mathcal{E})$ is the real number $H_{\psi}(\rho,\mathcal{E})$ given by
\[
H_{\psi}(\rho,\mathcal{E})=S_{\psi}(\rho,\mathcal{E})-S(\rho).
\]
\item
The $\psi$-\define{mutual information} of $(\rho,\mathcal{E})$ is the real number $I_{\psi}(\rho,\mathcal{E})$ given by
\[
I_{\psi}(\rho,\mathcal{E})=S(\rho)+S\big(\mathcal{E}(\rho)\big)-S_{\psi}(\rho,\mathcal{E}).
\]
\item
The $\psi$-\define{information discrepancy} of $(\rho,\mathcal{E})$ is the real number $K_{\psi}(\rho,\mathcal{E})$ given by
\[
K_{\psi}(\rho,\mathcal{E})=S_{\psi}(\rho,\mathcal{E})-S\big(\mathcal{E}(\rho)\big).
\]
\end{itemize}
\ed

\br
Let $\psi$ be a hermitian state over time function that is classically reducible and let $(\rho,\mathcal{E})\in \mathscr{P}(\C^X,\C^Y)$ be a classical process. Then all of the dynamical measures of classical information in Definition~\ref{DMXQIX981} evaluated on $(\rho,\mathcal{E})$ agree with those from Definition~\ref{DYXNXC19} after the identification of classical channels with stochastic maps.
\er

\bnot
\label{not:SHIKp}
The information measures associated with the symmetric $p$-bloom $\spbloom{p}$ will be denoted by $S_{p},H_{p},I_{p},K_{p}$, while the information measures associated with the Leifer--Spekkens state over time function $\psi_{\LS}$ will be denoted by $S_{\LS},H_{\LS},I_{\LS},K_{\LS}$.
\enot 

\br[Extending von~Neumann entropy to processes] 
A direct consequence of Theorem~\ref{BTXETXFX787} and Corollary~\ref{cor:TEMMA} is that for every state $\rho\in \mathcal{S}(\VA)$ we have
\[
S(\rho)=S_p(\rho,\id_{\VA}).
\]
As such, the entropy $S_p$ associated with the symmetric $p$-bloom may be viewed as an extension of von~Neumann entropy to processes for all $p\in [0,1]$.
\er 

\begin{figure}
\centering
\begin{subfigure}[b]{0.49\textwidth}
\includegraphics[width=8.5cm]{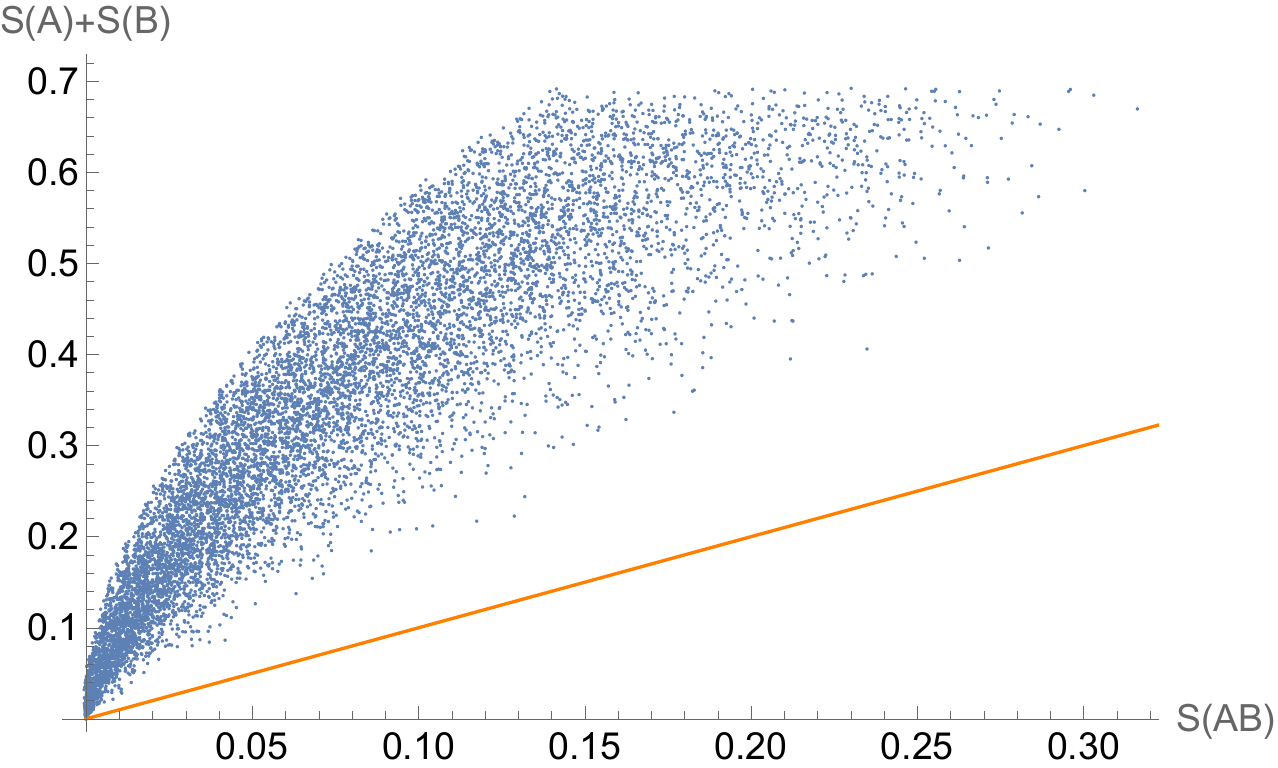}
\subcaption{$m=2,d_1=1,d_2=1,d_3=2$}
\end{subfigure}
\;
\begin{subfigure}[b]{0.49\textwidth}
\includegraphics[width=8.5cm]{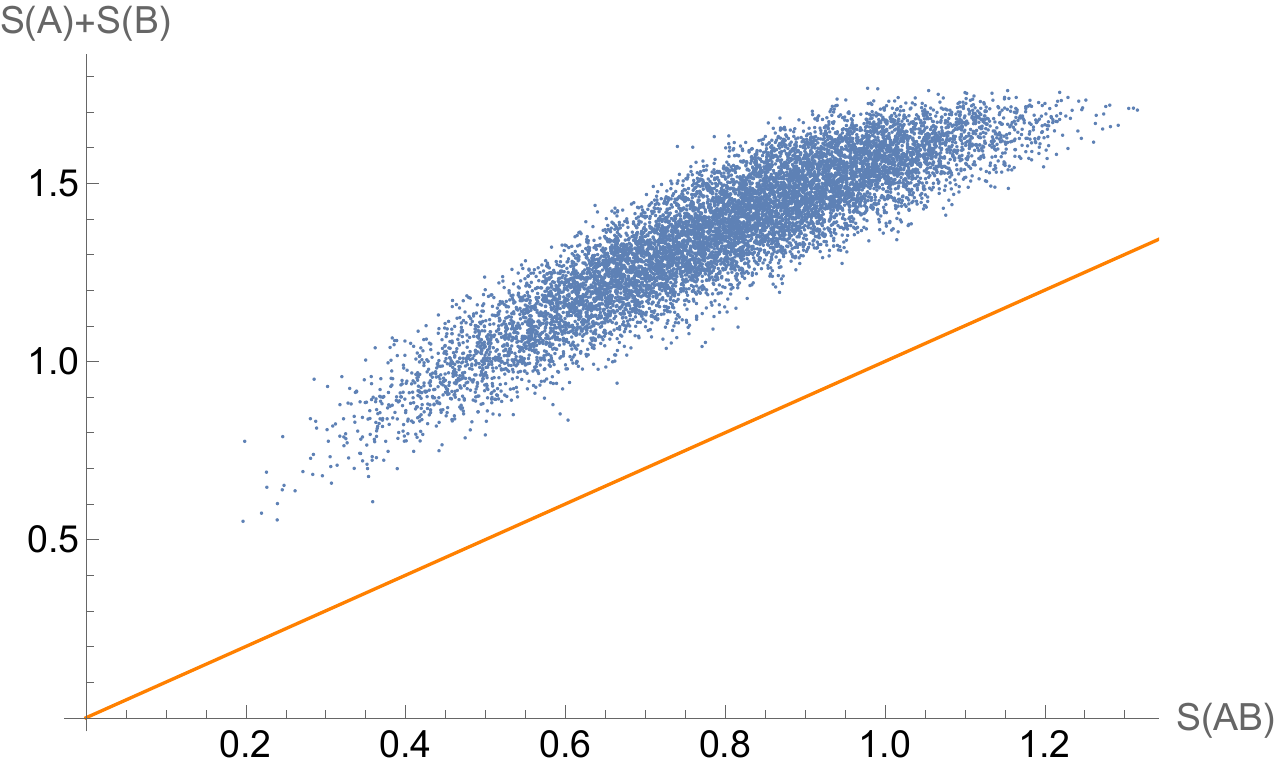}
\subcaption{$m=3,d_1=2,d_2=2,d_3=3$}
\end{subfigure}

\begin{subfigure}[b]{0.49\textwidth}
\includegraphics[width=8.5cm]{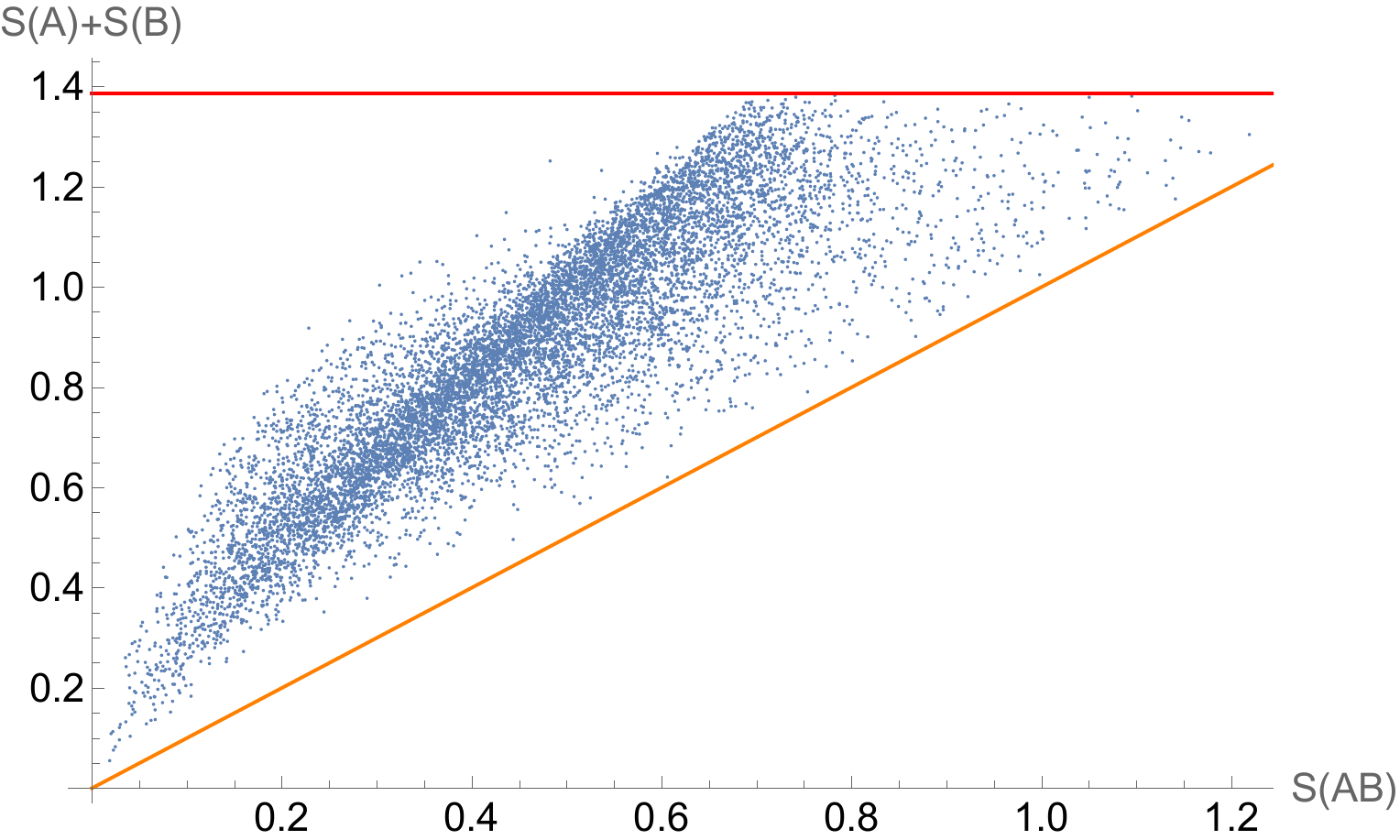}
\subcaption{$m=2,d_1=2,d_2=2,d_3=2$}
\end{subfigure}
\;
\begin{subfigure}[b]{0.49\textwidth}
\includegraphics[width=8.5cm]{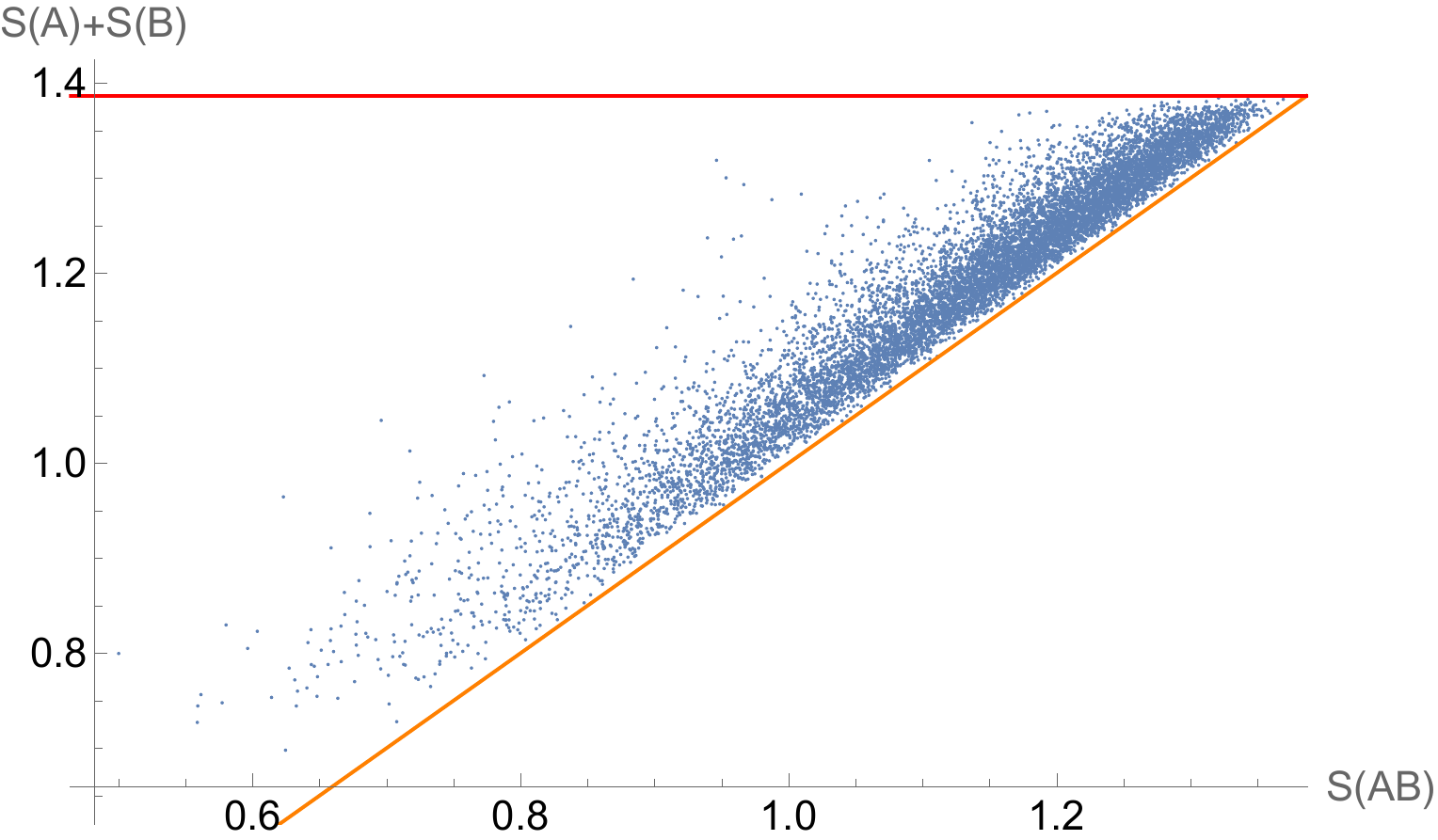}
\subcaption{$m=2,d_1=4,d_2=4,d_3=8$}
\end{subfigure}
\caption{In these scatter plots, $10000$ Haar-random random input density matrices $\rho\in\mA$ and quantum channels $\mathcal{E}:\mA\to\mB$ are randomly sampled, with $\mA=\mB=\matr_{m}$ for some positive integer value of $m$. The Haar-random density matrices $\rho$ are obtained by first generating a Haar-random pure state $|\psi\>$ in $\C^{m}\otimes\C^{d_{1}}$ and then $\rho$ is defined via the partial trace as $\rho=\tr_{\matr_{d_{1}}}\big(|\psi\>\<\psi|\big)$. The random channel $\mathcal{E}$ is constructed in the following way. First, a Haar-random environment system state $\rho_{E}$ is constructed on $\matr_{d_{3}}$ by generating a Haar-random $|\varphi\>$ in $\C^{d_{3}}\otimes\C^{d_{2}}$ and taking the partial trace $\rho_{E}=\tr_{\matr_{d_{2}}}\big(|\varphi\>\<\varphi|\big)$. Then, a Haar-random unitary $U:\C^{m}\otimes\C^{d_{3}}\to\C^{m}\otimes\C^{d_{3}}$ is sampled. Finally, $\mathcal{E}=\tr_{\matr_{d_{3}}}\circ\Ad_{U}\circ j_{E}$, where $j_{E}:\matr_{m}\to\matr_{m}\otimes\matr_{d_{3}}$ is the environment inclusion map sending an arbitrary state $\sigma$ to $\sigma\otimes\rho_{E}$. Although this is not the most general quantum channel that one can construct, it nevertheless has a direct physical interpretation in terms of coupling a system, $\mA$, to a random environment and undergoing a random unitary evolution, followed by discarding (i.e., tracing out) the environment~\cite{KNPPZ21}. 
The data points on these plots are given by $\big(S(AB),S(A)+S(B)\big):=\big(S(\psi(\rho,\mathcal{E})),S(\rho)+S(\mathcal{E}(\rho))\big)$, where $\psi$ is the symmetric bloom state over time function. The orange diagonal line depicts the line for which these two values would be equal. In other words, points above the line depict states $\rho$ and channels $\mathcal{E}$ for which sub-additivity $S(\psi(\rho,\mathcal{E}))\le S(\rho)+S(\mathcal{E}(\rho))$ holds. 
Meanwhile, the red horizontal line is given by $2\log(m)$ (this horizontal line is not shown in the scatter plots (a) and (b) because it lies well above the data points shown).
In all plots, not a single violation of $0\le I_{\psi}(\rho,\mathcal{E})\le2\min\big\{\log(\tr(1_{\VA})),\log(\tr(1_{\VB}))\big\}$ has been found. 
The dimensions $m, d_{1}, d_{2}, d_{3}$ are different in each plot and are provided under the respective figures. 
}
\label{fig:subaddSOT}
\end{figure}

\br[On the non-negativity of $\psi$-mutual information] \label{rmk:NNPMI}
As the extended entropy function does not satisfy subadditivity for quasi-densities whose marginals are states (see Remark~\ref{SAXT67}), 
it is not immediately obvious whether the mutual information $I_{\psi}$ is a non-negative information measure or not. 
However, we have yet to come across any example of a hermitian matrix that arises as the symmetric bloom state over time of some $(\rho,\mathcal{E})$ and for which $I_{\psi}(\rho,\mathcal{E})$ is  negative. 
In fact, we present numerical evidence in Figure~\ref{fig:subaddSOT} in support of the hypothesis that $I_{\psi}(\rho,\mathcal{E})\ge0$ holds. 
We do this by sampling random states $\rho$ and channels $\mathcal{E}$ according to the Haar measure for various dimensions~\cite{KNPPZ21}. This suggests that $I_{\psi}(\rho,\mathcal{E})\ge0$ may indeed hold for our entropy formula defined using states over time constructed from the symmetric bloom. If true, this would be in stark contrast to alternative proposals for dynamical forms of entropies, for which subadditivity of the entropy functions is known to fail~\cites{NTTTW21,JSK23}. 
Finally, we point out that if we assume the validity of the lower bound $0\le I_{\psi}(\rho,\mathcal{E})$, then the analogue of the upper bound $I(\VA:\VB)\le 2\min\big\{S(\rho_{\VA}),S(\rho_{\VB})\big\}$~\cites{CeAd97,AdCe97,ChMa20} must fail in our dynamical setting. This can easily be seen by taking any pure state $\rho$ as the initial state, which forces $\min\big\{S(\rho),S(\mathcal{E}(\rho))\big\}=0$, while $I_{\psi}(\rho,\mathcal{E})$ is generically positive. Nevertheless, our numerical plots suggest the possibility that the alternative weaker bound $I_{\psi}(\rho,\mathcal{E})\le 2\min\big\{\log(\tr(1_{\VA})),\log(\tr(1_{\VB}))\big\}$  might be satisfied, in analogy with the static quantum mutual information (cf.\ \cite[Exercise~11.6.3]{Wilde2017}).  

\er 

\br [The static interpretation] \label{STXPXPTV81}
If $\psi(\rho,\mathcal{E})$ is positive (i.e., if $\psi(\rho,\mathcal{E})\in \mathcal{S}(\VA\otimes \VB)$), then $\psi(\rho,\mathcal{E})$ may also be viewed as the quantum state of two space-like separated regions whose associated marginals are $\rho$ and $\mathcal{E}(\rho)$.
In such a case, $S_{\psi}(\rho,\mathcal{E})$ is the usual quantum joint entropy, $H_{\psi}(\rho,\mathcal{E})$ is the usual quantum conditional entropy, and $I_{\psi}(\rho,\mathcal{E})$ is the usual quantum mutual information of the joint state $\psi(\rho,\mathcal{E})$. Such states over time then satisfy a quantum analogue of the static-dynamic duality for classical states as discussed in the introduction. 

As an example, let $\mathcal{E}:\C^{X}\to\matr_{m}$ be an ensemble preparation~\cite{FuPa22a}, let $p$ be a probability distribution on $X$ with value on $x$ given by $p_{x}$, and set $\rho^{x}=\mathcal{E}(\delta_{x})$. Then $\psi(p,\mathcal{E})=\sum_{x\in X}p_{x}\delta_{x}\otimes\rho^{x}$ for any classically reducible  $\psi$. Such a state over time $\psi(p,\mathcal{E})$ is positive and is sometimes called a \emph{classical-quantum state}~\cite{Wilde2017}. In this case, our information measures take on familiar forms:
\begin{align*}
S_{\psi}(p,\psi)&=H(p)+\sum_{x\in X}p_{x}S(\rho^{x}),
&
H_{\psi}(p,\mathcal{E})&=\sum_{x\in X}p_{x}S(\rho^{x}),
\\
I_{\psi}(p,\mathcal{E})&=S(\rho)-\sum_{x\in X}p_{x}S(\rho^{x}),
&
K_{\psi}(p,\mathcal{E})&=H(p)-S(\rho)+\sum_{x\in X}p_{x}S(\rho^{x}),
\end{align*}
where $\rho$ is the expected density matrix $\rho=\mathcal{E}(p)=\sum_{x\in X}p_{x}\rho^{x}$. In particular, note that the conditional entropy appears just as in the classical case (cf.\ equation~\eqref{eq:CCE}) and that the mutual information $I_{\psi}(p,\mathcal{E})$ is equal to the Holevo information of the ensemble (cf.\ \cite[Section~11.6.1]{Wilde2017}). 
\er

\br 
\label{rmk:PSOTIM}
If $\rho$ is a pure state then $S(\rho)=0$. It then follows that for any process $(\rho,\mathcal{E})$ with $\rho$ pure we have
\be \label{PSTXTX181}
S_{\psi}(\rho,\mathcal{E})=H_{\psi}(\rho,\mathcal{E}) \quad \text{and} \quad I_{\psi}(\rho,\mathcal{E})=-K_{\psi}(\rho,\mathcal{E}).
\ee 
Thus, the information measures $H_{\psi}(\rho,\mathcal{E})$ and $K_{\psi}(\rho,\mathcal{E})$ are redundant in this case.
\er

\br[Conservation of Quantum Information]
\label{rmk:CQI}
Note that for every process $(\rho,\mathcal{E})$ and any hermitian state over time function $\psi$, we have
\[
I_{\psi}(\rho,\mathcal{E})+K_{\psi}(\rho,\mathcal{E})=S(\rho),
\]
which we view as a ``conservation of information'' for quantum processes $(\rho,\mathcal{E})$. Here, $I_{\psi}(\rho,\mathcal{E})$ is viewed as the information shared between the initial state $\rho$ and the final state $\mathcal{E}(\rho)$ through the channel $\mathcal{E}$, while $K_{\psi}(\rho,\mathcal{E})$ is the discrepancy between this shared information and the information content of the initial state $\rho$. There are several justifications for this interpretation; here we sketch three extreme cases assuming that the state over time function $\psi$ is also classically reducible.
\begin{enumerate}[(a)]
\item
Let $\sigma\in\mathcal{S}(\VB)$ be an arbitrary state and let $\mathcal{E}:\VA\to\VB$ be the \emph{discard-and-prepare} channel, which is given by $\mathcal{E}(A)=\tr(A)\sigma$ for all $A\in \VA$. It then follows that $\Jamiol[\mathcal{E}]=\mathds{1}_{\VA}\otimes\sigma$, so that $[\rho\otimes\mathds{1}_{\VB},\Jamiol[\mathcal{E}]]=0$ for all $\rho\in\mathcal{S}(\VA)$. Therefore, $\psi(\rho,\mathcal{E})=(\rho\otimes \mathds{1}_{\VA})\Jamiol[\mathcal{E}]=\rho\otimes\sigma$ since $\psi$ is classically reducible. Hence, 
\[
I_{\psi}(\rho,\mathcal{E})=S(\rho)+S(\sigma)-S(\rho\otimes\sigma)=0
\]
by the additive property of the von~Neumann entropy. 
This is to be expected from a measure that quantifies shared information, as the final state $\mathcal{E}(\rho)=\sigma$ does not depend on the initial state $\rho$. 
Furthermore, 
\[
K_{\psi}(\rho,\mathcal{E})=S(\rho\otimes\sigma)-S(\sigma)=S(\rho),
\]
which says that the information discrepancy equals the initial entropy $S(\rho)$. This is also to be expected as the process $(\rho,\mathcal{E})$ discards all of the initial information contained in the initial state $\rho$. 
\item
As another extreme case, let $\mathcal{E}=\Ad_{U}$ for some unitary $U\in\VA$. Then, 
\[
I_{\psi}(\rho,\mathcal{E})
=S(\rho)+S(U\rho U^{\dag})-S_{\psi}(\rho,\Ad_{U})
=S(\rho)
\]
by unitary invariance of the von~Neumann entropy and by Theorem~\ref{BTXETXFX787}. 
Hence, the shared information is exactly the entropy of $\rho$, the initial state, as expected for a completely reversible evolution (this provides another justification for the importance of Theorem~\ref{BTXETXFX787} and our definition of entropy for states over time). Meanwhile, $K_{\psi}(\rho,\mathcal{E})=0$ says there is no information discrepancy, which is also expected for unitary evolution.
\item
Finally, we consider a class of examples for which $K_{\psi}(\rho,\mathcal{E})$ is negative. 
In this regard, let $\mathcal{E}:\matr_{n}\to\C^{Y}$ be a positive operator-valued measure (POVM) with associated positive operators $\{E_{y}\}$ satisfying $\mathcal{E}(A)=\sum_{y\in Y}\tr(E_{y}A)\delta_{y}$, let $\rho\in\mathcal{S}(\matr_{m})$ be a pure state, and set $q=\mathcal{E}(\rho)$ to be the probability distribution associated with the measurement outcomes. 
We will later prove in Theorem~\ref{prop:PVMKN} that $K_{1}(\rho,\mathcal{E})\le0$ for the symmetric bloom (cf.\ Notation~\ref{not:SHIKp}). By Remark~\ref{rmk:PSOTIM} $I_{1}(\rho,\mathcal{E})=-K_{1}(\rho,\mathcal{E})$, thus yielding a proof that $I_{1}(\rho,\mathcal{E})\ge0$ for such a class of processes, in support of Remark~\ref{rmk:NNPMI}. 
This class of examples maintains the interpretation of $I_{1}(\rho,\mathcal{E})$ as shared information, but in a way that is uniquely quantum mechanical. Namely, the state $\rho$ is disturbed by the action of the measurement and this disturbance can be viewed as altering the information about the initial state with respect to the measurement. The mutual information $I_{1}(\rho,\mathcal{E})$ is then a measure of the disturbance of $\rho$ due to the measurement. In particular, only when the state is undisturbed, i.e., when the POVM $\mathcal{E}$ contains a projection that is compatible with the initial state $\rho$, does $I_{1}(\rho,\mathcal{E})=0$. Somewhat provocatively, one might interpret the positivity of mutual information $I_{1}(\rho,\mathcal{E})$---or equivalently, the negativity of the information discrepancy $K_{\psi}(\rho,\mathcal{E})$---as saying that information (entropy) has been created in the observed system due to the interaction of an agent with the system. Indeed, this creation of entropy is closely related to the fact that the probability measure $q$ need not be a Dirac delta for an arbitrary POVM, or even for a projection-valued measure (see Theorem~\ref{prop:PVMKN} for details).
\end{enumerate}
\er

We now prove convex linearity properties for the dynamical information measures introduced in this section. 

\bn \label{CVXL81}
Let $p:X\to \R$ be a quasi-probability distribution on a finite set $X$, let $\rho^x\in \M_m$ be a collection of mutually orthogonal density matrices indexed by $X$, let $\psi$ be a state over time function that is hermitian and state-linear, and suppose $\mathcal{E}\in \CPTP(\M_m,\M_n)$ is such that $\psi(\rho^x,\mathcal{E})$ is a collection of mutually orthogonal pseudo-densities indexed by $X$. Then the following statements hold.
\begin{enumerate}[i.]
\item \label{CVX1}
$S_{\psi}\left(\displaystyle \sum_{x\in X}p_x\rho^x,\mathcal{E}\right)=H(p)+\displaystyle \sum_{x\in X}p_xS_{\psi}(\rho^x,\mathcal{E})$ .
\item \label{CVX2}
$H_{\psi}\left(\displaystyle \sum_{x\in X}p_x\rho^x,\mathcal{E}\right)=\displaystyle \sum_{x\in X}p_xH_{\psi}(\rho^x,\mathcal{E})$ .
\item \label{CVX3}
$I_{\psi}\left(\displaystyle \sum_{x\in X}p_x\rho^x,\mathcal{E}\right)=H(p)+\displaystyle \sum_{x\in X}p_xI_{\psi}(\rho^x,\mathcal{E})$ .
\item \label{CVX4}
$K_{\psi}\left(\displaystyle \sum_{x\in X}p_x\rho^x,\mathcal{E}\right)=\displaystyle \sum_{x\in X}p_xK_{\psi}(\rho^x,\mathcal{E})$ .
\end{enumerate}
\en

\bprf
For item \ref{CVX1}, we have
\begin{eqnarray*}
S_{\psi}\left(\displaystyle \sum_{x\in X}p_x\rho^x,\mathcal{E}\right)&=&S\left(\psi\left(\sum_{x\in X}p_x \rho^x,\mathcal{E}\right)\right)=S\left(\sum_{x\in X}p_x\psi\left(\rho^x,\mathcal{E}\right)\right) \\
&\overset{\eqref{EZNTXAD97}}=&H(p)+\sum_{x\in X}p_xS\big(\psi \left(\rho^x,\mathcal{E}\right)\big)=H(p)+\sum_{x\in X}p_xS_{\psi}\left(\rho^x,\mathcal{E}\right), \\
\end{eqnarray*}
where the second equality follows from the assumption that $\psi$ is state-linear. The proofs of items \ref{CVX2}-\ref{CVX4} are similar.
\eprf

\br
Note that while the symmetric bloom satisfies the hypotheses of Proposition~\ref{CVXL81}, the Leifer--Spekkens state over time function is not state-linear. Thus, the information measures $S_{\LS}$, $H_{\LS}$, $I_{\LS}$, and $K_{\LS}$ do not in general satisfy the formulas in items \ref{CVX1}-\ref{CVX4} from Proposition~\ref{CVXL81}. 
\er

\section{Examples} \label{S81}
In this section we compute the information measures introduced in the previous section for some fundamental examples. This serves several purposes. First, these examples illustrate some important differences between the Leifer--Spekkens and symmetric bloom state over time functions through their corresponding information measures. Second, the examples show how the information measures we introduced can be computed concretely. Third, the examples provide explicit computations quantifying temporal correlations for states over time.

\bx[Partial trace] \label{MXEXPXS771}
Let $\mathcal{E}:\M_2\otimes \M_2\to \M_2$ be the partial trace over the first factor, and let $\rho_{\EPR}$ be the 
EPR density matrix from Equation~(\ref{EPRXS}).
As for the symmetric bloom state over time function $\sbloom$, it follows from Example~\ref{EPRX87} that the multi-spectrum of $\sbloom(\rho_{\EPR},\mathcal{E})$ is given by
\[
\mathfrak{mspec}\left(\sbloom(\rho_{\EPR},\mathcal{E})\right)=\left\{-1/4,-1/4,0,0,0,0,3/4,3/4\right\}.
\]
We then have (cf.\, Notation~\ref{not:SHIKp})
\[
S_{1}(\rho_{\EPR},\mathcal{E})\overset{\eqref{PSTXTX181}}=H_{1}(\rho_{\EPR},\mathcal{E})=\log\left(\frac{4}{3\sqrt{3}}\right)\approx-0.3774,
\]
while
\[
I_{1}(\rho_{\EPR},\mathcal{E})\overset{\eqref{PSTXTX181}}=-K_{1}(\rho_{\EPR},\mathcal{E})\approx 0.6226.
\]

As for the Leifer--Spekkens state over time $\psi_{\LS}$, it follows from Example~\ref{EPRX87} that $\psi_{\LS}(\rho_{\EPR},\mathcal{E})=\rho_{\EPR}\otimes \mathds{1}/2$. Thus, 
\[
\mathfrak{mspec}\big(\psi_{\LS}(\rho_{\EPR},\mathcal{E})\big)=\left\{0,0,0,0,0,0,1/2,1/2\right\}.
\]
We then have
\[
S_{\LS}(\rho_{\EPR},\mathcal{E})=H_{\LS}(\rho_{\EPR},\mathcal{E})=\log(2),
\quad
\text{ while }
\quad
I_{\LS}(\rho_{\EPR},\mathcal{E})=K_{\LS}(\rho_{\EPR},\mathcal{E})=0.
\]

Now consider the separable pure state 
\[
\rho=
\frac{1}{4}\left(
\begin{array}{cccc}
1&1&1&1\\
1&1&1&1\\
1&1&1&1\\
1&1&1&1\\
\end{array}
\right)
=
\frac{1}{2}\begin{pmatrix}1&1\\1&1\end{pmatrix}\otimes\frac{1}{2}\begin{pmatrix}1&1\\1&1\end{pmatrix}.
\]
We then have
\[
\sbloom(\rho,\mathcal{E})=
\frac{1}{8}
\left(
\begin{smallmatrix}
2&1&1&0&2&1&1&0 \\
1&0&2&1&1&0&2&1 \\
1&2&0&1&1&2&0&1 \\
0&1&1&2&0&1&1&2 \\
2&1&1&0&2&1&1&0 \\
1&0&2&1&1&0&2&1 \\
1&2&0&1&1&2&0&1 \\
0&1&1&2&0&1&1&2 \\
\end{smallmatrix}
\right),
\quad
\text{ while }
\quad
\psi_{\LS}(\rho,\mathcal{E})=
\frac{1}{8}\left(
\begin{smallmatrix}
1&1&1&1&1&1&1&1 \\
1&1&1&1&1&1&1&1 \\
1&1&1&1&1&1&1&1 \\
1&1&1&1&1&1&1&1 \\
1&1&1&1&1&1&1&1 \\
1&1&1&1&1&1&1&1 \\
1&1&1&1&1&1&1&1 \\
1&1&1&1&1&1&1&1 \\
\end{smallmatrix}
\right)
=
\rho\otimes\frac{1}{2}\begin{pmatrix}1&1\\1&1\end{pmatrix}.
\]
It then follows that 
\[
\mathfrak{mspec}\left(\sbloom(\rho,\mathcal{E})\right)=\{0,0,0,0,0,-1/2,1/2,1\},
\]
and
\[
\mathfrak{mspec}\big(\psi_{\LS}(\rho,\mathcal{E})\big)=\{0,0,0,0,0,0,0,1\}.
\]
Thus, all information measures associated with $\sbloom$ and $\psi_{\LS}$ are vanishing in this case, which is consistent with the intuitive idea that the separation of two uncorrelated particles is essentially an information-less process.
\ex

\br
In Example~\ref{MXEXPXS771}, we have
\[
\mathfrak{C}\big(\sbloom(\rho_{\EPR},\mathcal{E})\big)>0 \quad \text{and} \quad \mathfrak{C}\big(\sbloom(\rho,\mathcal{E})\big)>0,
\]
while
\[
\mathfrak{C}\big(\psi_{\LS}(\rho_{\EPR},\mathcal{E})\big)=\mathfrak{C}\big(\psi_{\LS}(\rho,\mathcal{E})\big)=0,
\]
where $\mathfrak{C}$ is the causality monotone as defined in Remark~\ref{CSTXMTX67}. It then follows that for both the entangled state $\rho_{\EPR}$ and the separable state $\rho$, the symmetric bloom state over time for the partial trace channel encodes causal correlations between the input and the output of the channel, while the Leifer--Spekkens state over time does not. In particular, when $\rho_{\EPR}$ is the initial state, the Leifer--Spekkens state over time is $\rho_{\EPR}\otimes (\mathds{1}/2)$, and is therefore completely uncorrelated. This is why the information measures associated the symmetric bloom and the Leifer--Spekkens state over time functions differ so drastically in the case of $\rho_{\EPR}$. 
\er

\begin{figure}
\centering
\begin{tabular}{m{7cm} m{8cm}}
\includegraphics[width=7cm,trim={0 0.5cm 0 1cm},clip]{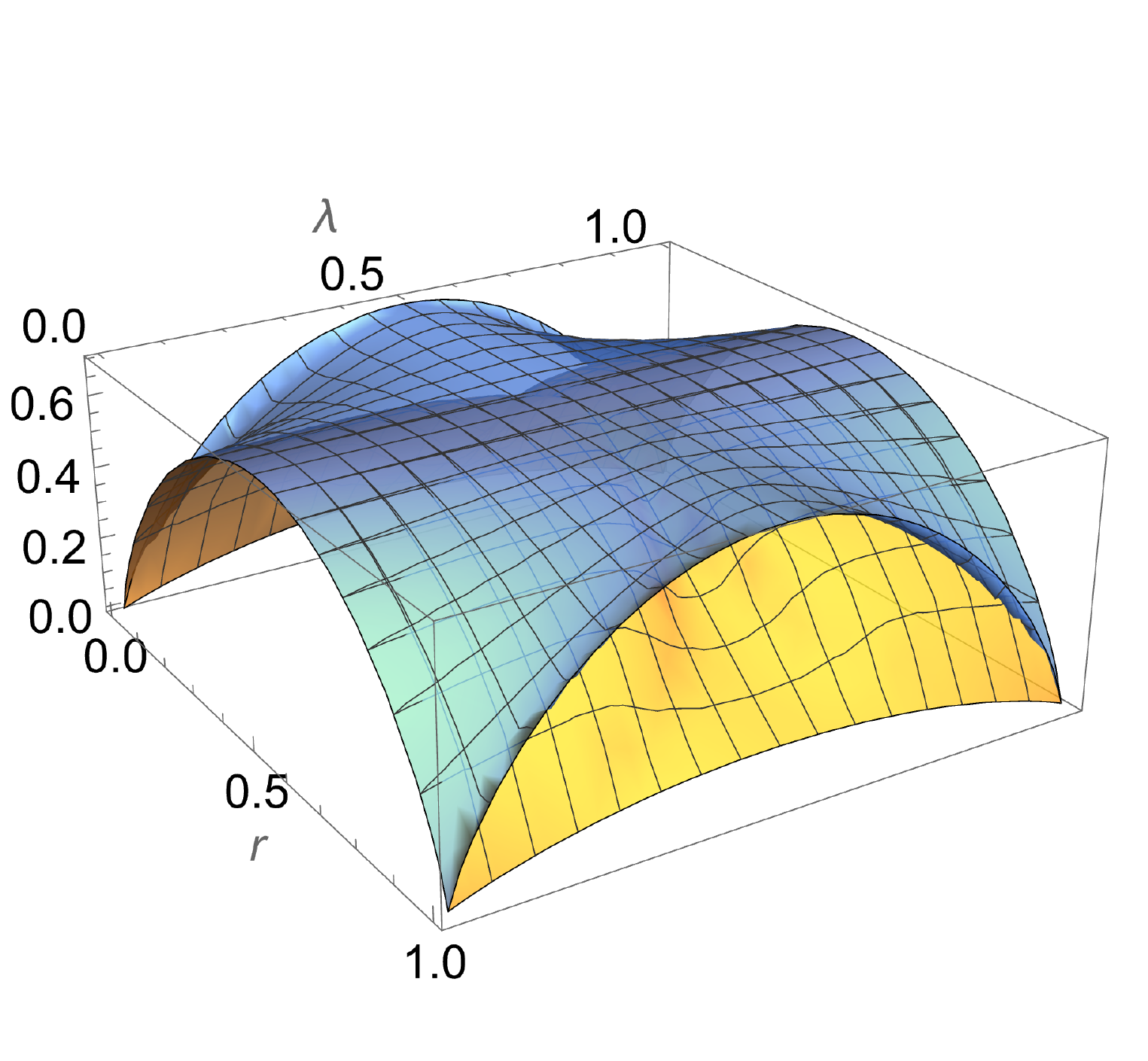}&\includegraphics[width=8cm]{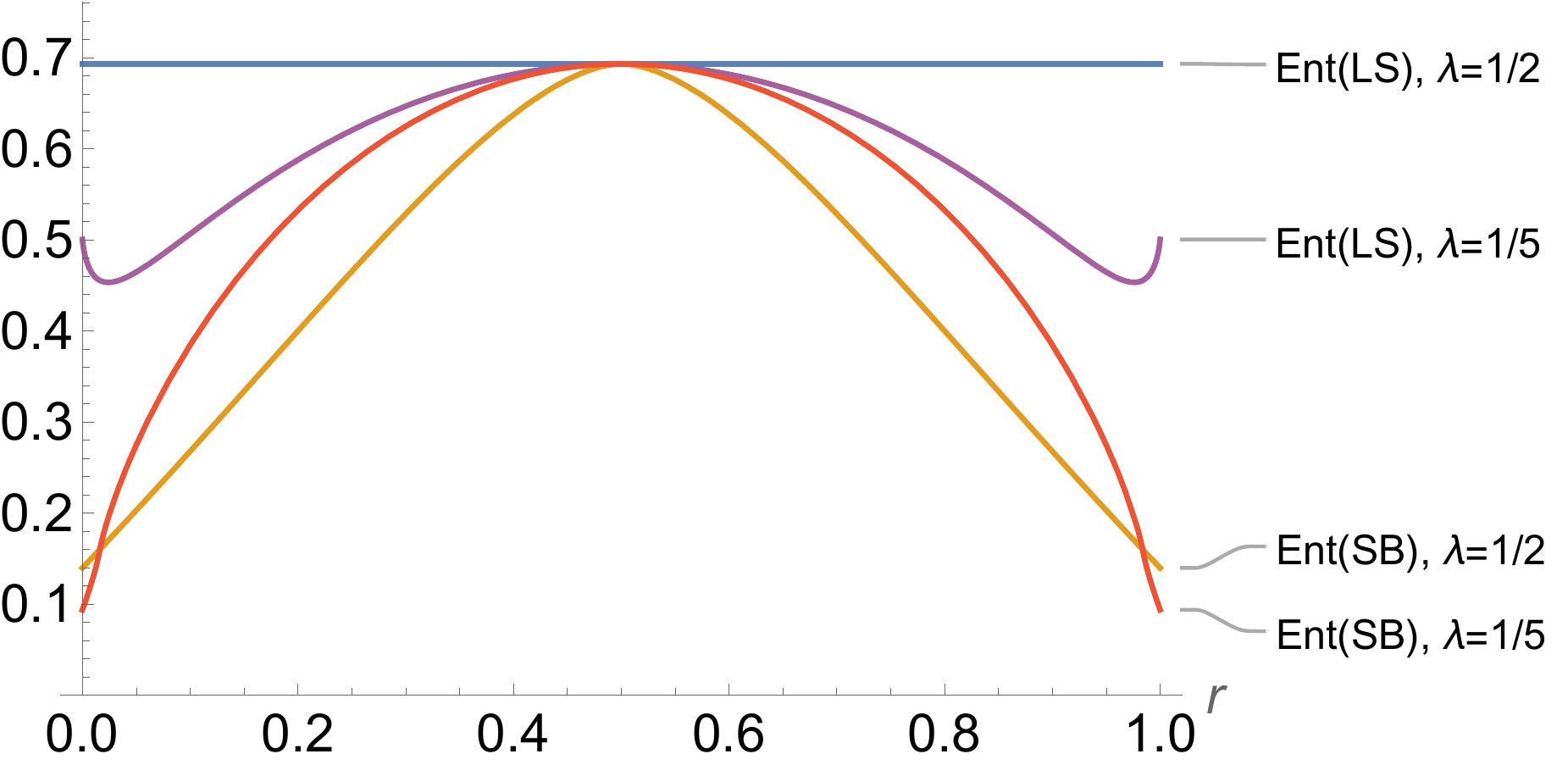}\\
\includegraphics[width=7cm,trim={0 2cm 0 3cm},clip]{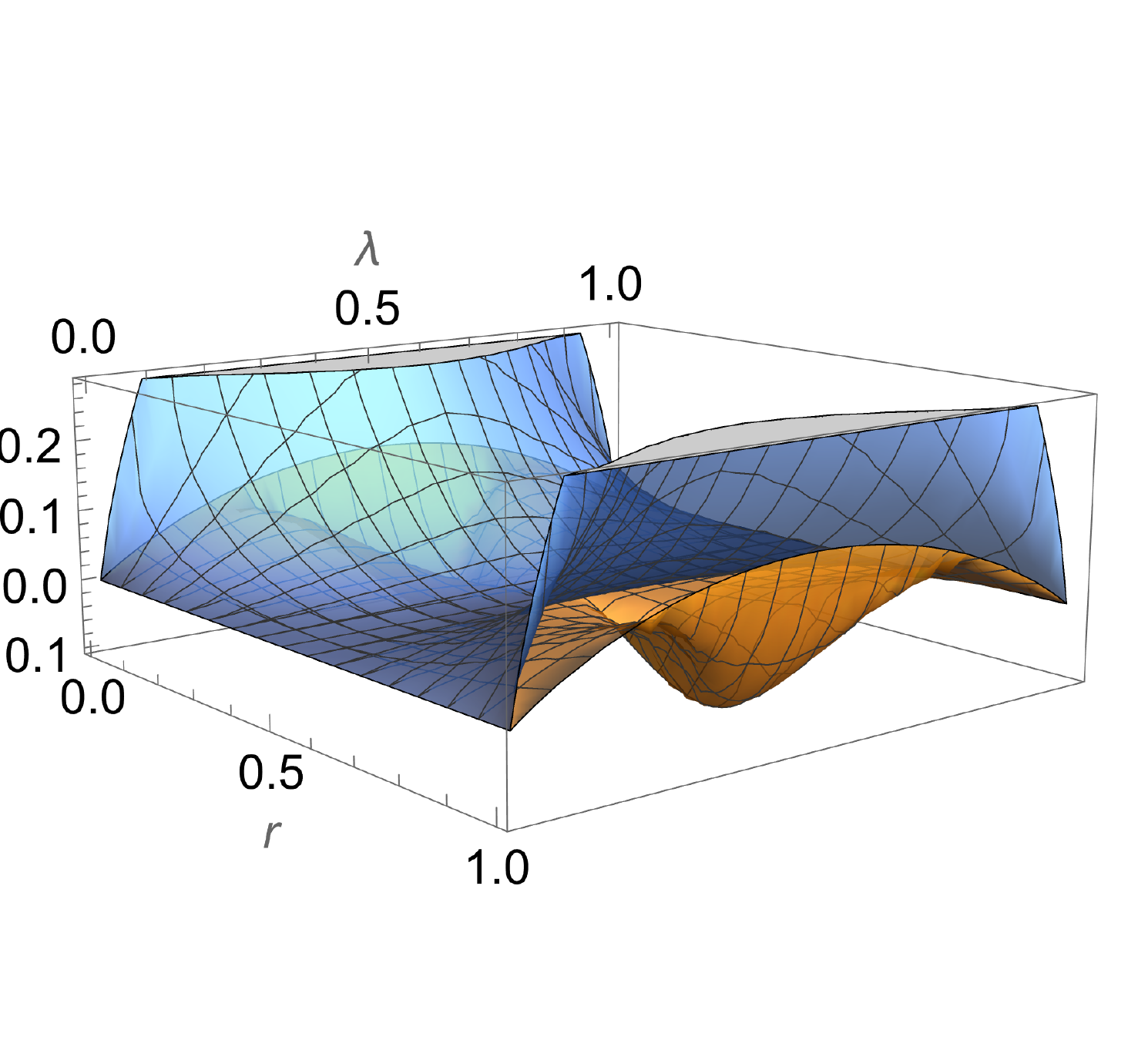}&\includegraphics[width=8cm]{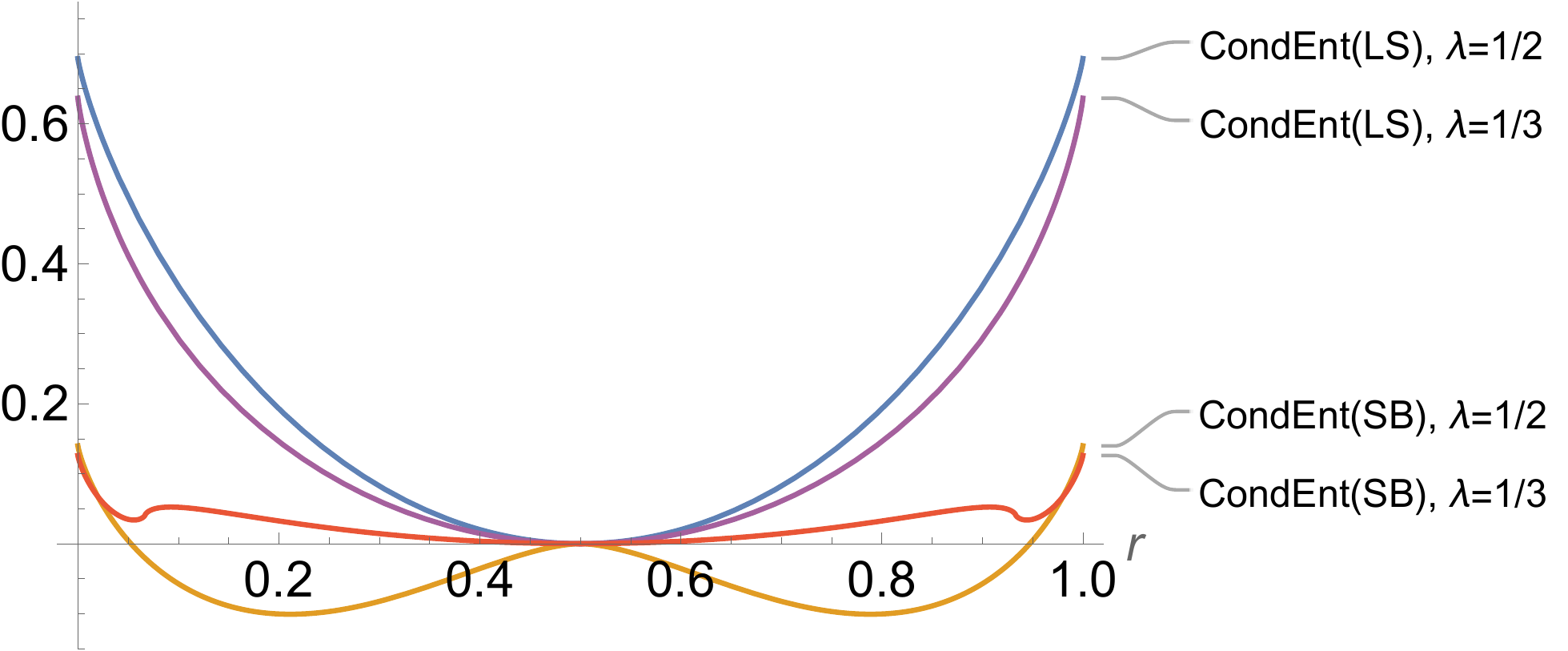}\\
\includegraphics[width=7cm,trim={0 1cm 0 2cm},clip]{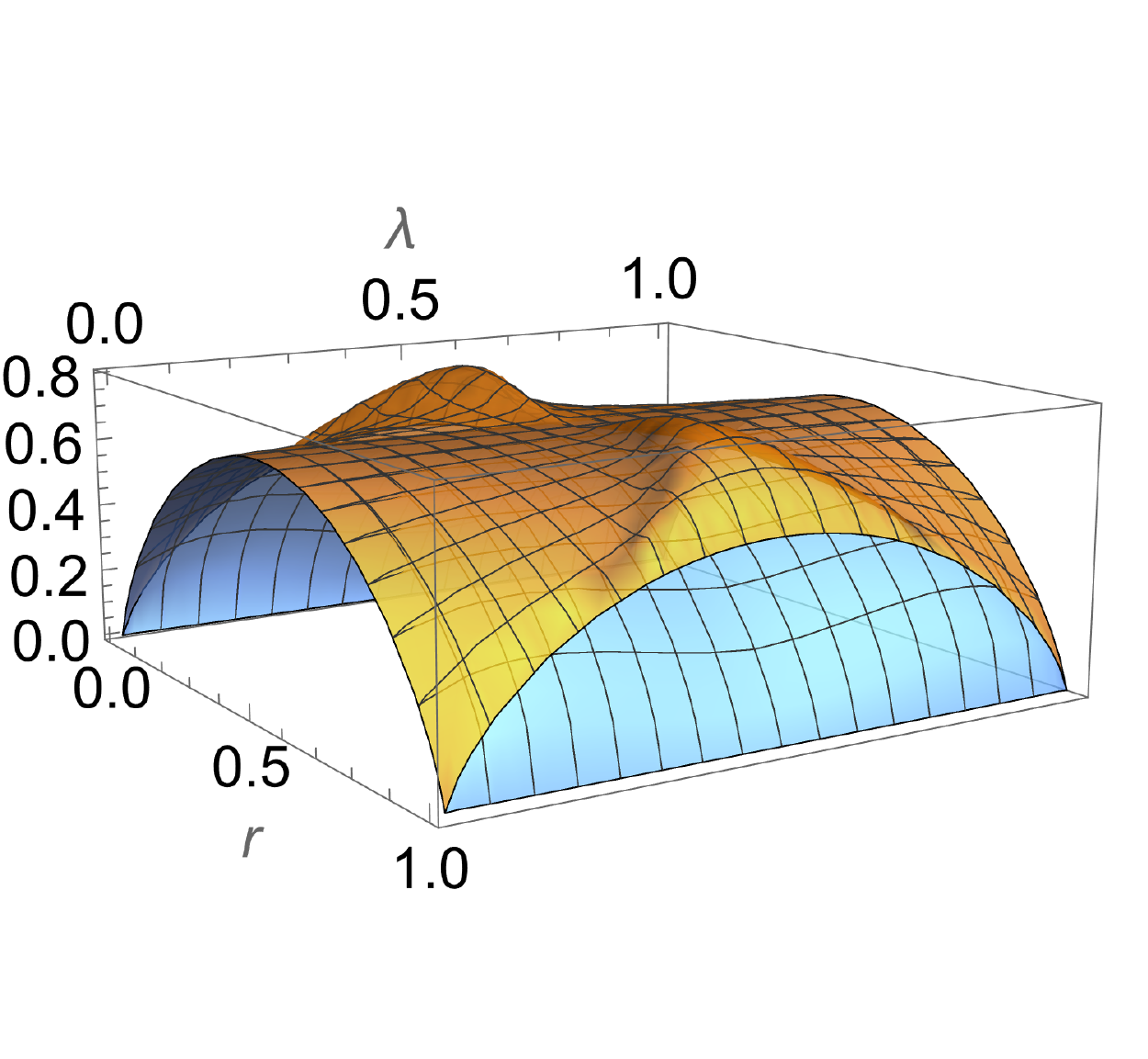}&\includegraphics[width=8cm]{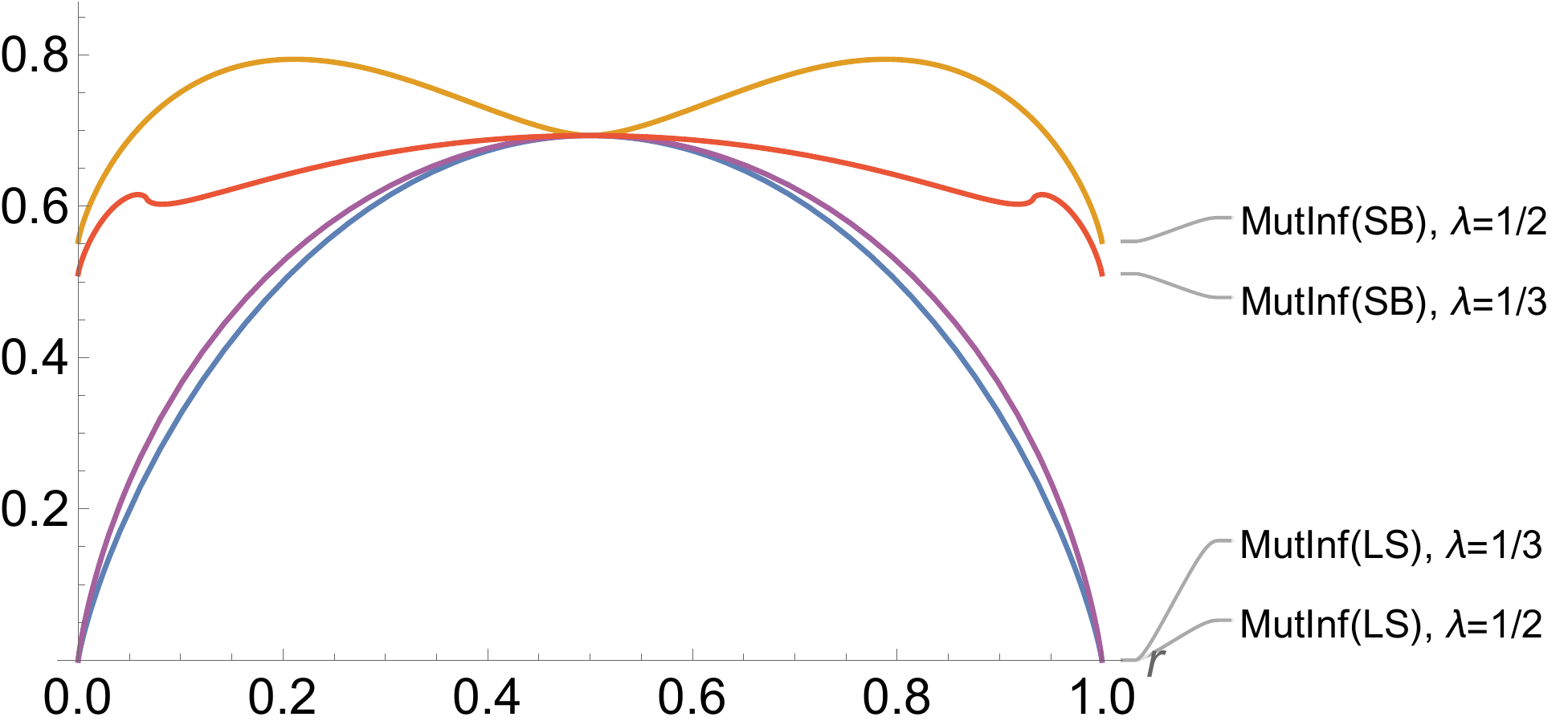}\\
\includegraphics[width=7cm,trim={0 2cm 0 2.5cm},clip]{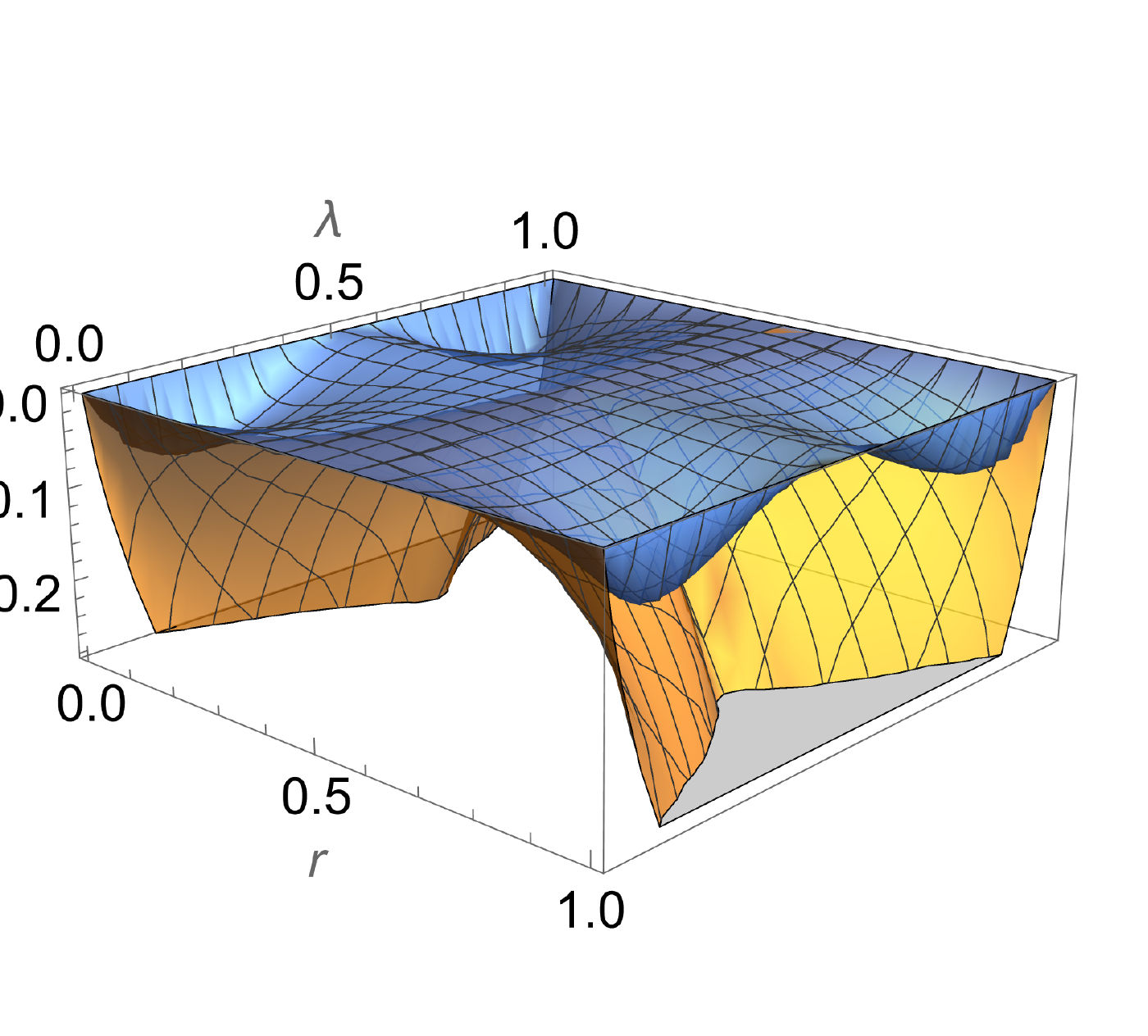}&\includegraphics[width=8cm]{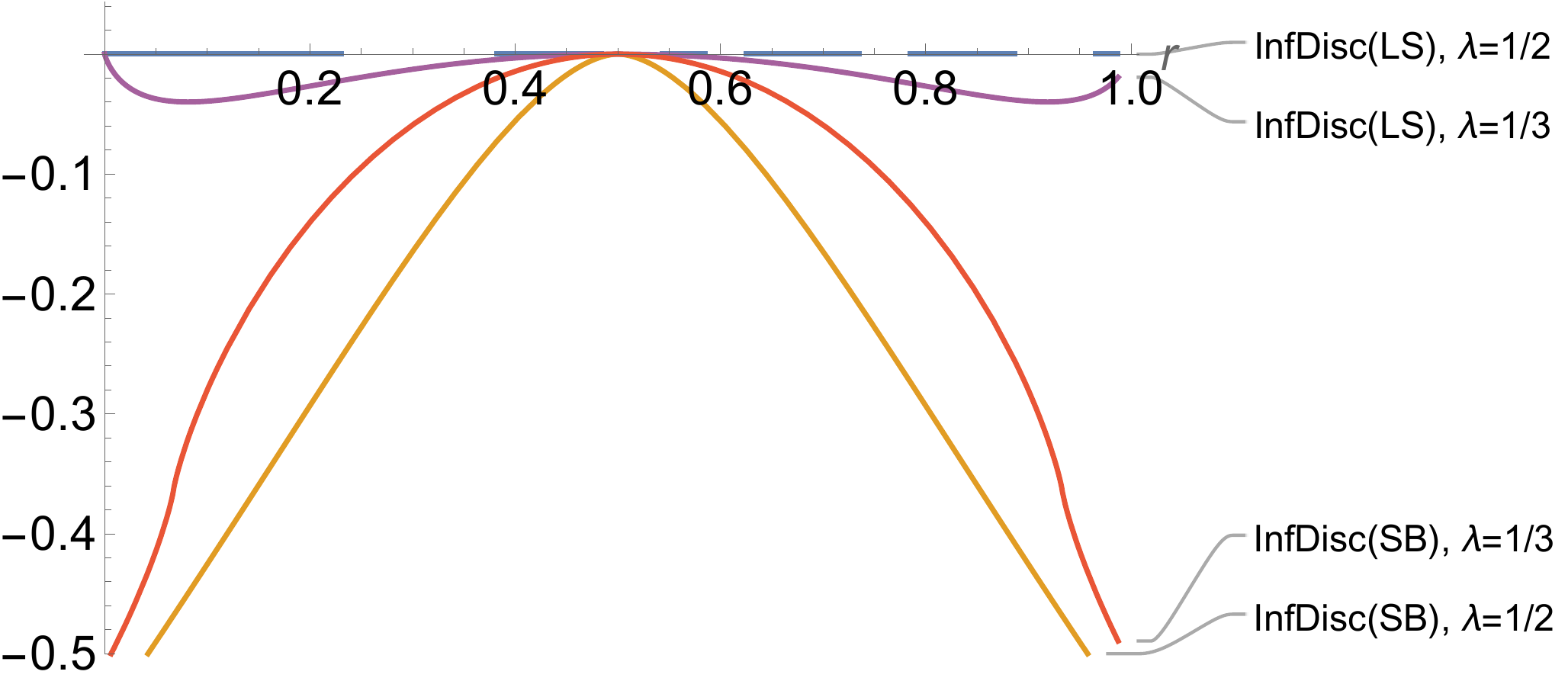}
\end{tabular}
\caption{On the left column, reading from top to bottom, graphs of the $\psi$-entropy (Ent), $\psi$-conditional entropy (CondEnt), $\psi$-mutual information (MutInf), and $\psi$-information discrepancy (InfDisc) for the bit-flip channel (Example~\ref{ex:bitflip}) are depicted as functions of $(r,\lambda)\in[0,1]^{2}$ for $\psi$ the Leifer--Spekkens (LS) and symmetric bloom (SB) state over time functions. On the right column, certain slices for fixed values of $\lambda$ are drawn as functions of $r\in[0,1]$.}
\label{fig:BFENT}
\end{figure}

\bx[The bit-flip channel]
\label{ex:bitflip}
Let $r,\lambda\in (0,1)$, let $\mathcal{E}_{\lambda}:\M_2\to \M_2$ be the map given by 
\[
\mathcal{E}_{\lambda}=\lambda\,\id_{\M_2}+(1-\lambda)\Ad_U,
\]
where 
$U=\left( 
\begin{smallmatrix}
0 & 1 \\
1 & 0 \\
\end{smallmatrix}
\right)$, 
and let  
$\rho_r=\left(
\begin{smallmatrix}
r & 0 \\
0 & 1-r \\
\end{smallmatrix}
\right)$. 
Figure~\ref{fig:BFENT} then illustrates graphs of the four information measures as functions of $(r,\lambda)$ for the Leifer--Spekkens and symmetric bloom state over time functions.

\ex

\bx[The amplitude damping channel]
Let $\mathcal{E}:\M_2\to \M_2$ be the quantum channel given by
\[
\rho\mapsto A_1\rho A_1^{\dag}+A_2\rho A_2^{\dag}, 
\]
where
\[
A_1=
\left(
\begin{array}{cc}
1 & 0 \\
0 & \sqrt{1-p}
\end{array}
\right)
\quad 
\text{ and }
\quad
A_2=
\left(
\!\!
\begin{array}{cc}
0 & 0 \\
\sqrt{p} & 0 \\
\end{array}
\right).
\]
If 
$\rho=\left(
\begin{smallmatrix}
a & b \\
c & d \\
\end{smallmatrix}
\right)$,
then
\[
\rbloom(\rho,\mathcal{E})=
\left(
\begin{array}{cccc}
a & b\sqrt{1-p} & bp & 0 \\
0 & 0 & a\sqrt{1-p} & b(1-p) \\
c & d\sqrt{1-p} & dp & 0 \\
0 & 0 & c\sqrt{1-p} & d(1-p) \\
\end{array}
\right).
\]
Thus, for $\rho_1=\frac{1}{2}\mathds{1}$, 
$\rho_2=
\frac{1}{2}
\left(
\begin{smallmatrix}
1&-1\\
-1&1\\
\end{smallmatrix}
\right)$,
and $p=1/2$, we have
\[
\sbloom(\rho_1,\mathcal{E})=
\frac{1}{4}
\left(
\begin{array}{cccc}
2&0&0&0\\
0&0&\sqrt{2}&0\\
0&\sqrt{2}&1&0\\
0&0&0&1\\
\end{array}
\right)
\quad\text{ and }\quad
\sbloom(\rho_2,\mathcal{E})=
\frac{1}{8}
\left(
\begin{array}{cccc}
4&-\sqrt{2}&-3&0\\
-\sqrt{2}&0&2\sqrt{2}&-1\\
-3&2\sqrt{2}&2&-\sqrt{2}\\
0&-1&-\sqrt{2}&2\\
\end{array}
\right).
\]
It then follows that
\[
\mathfrak{mspec}\big(\sbloom(\rho_1,\mathcal{E})\big)=\left\{\frac{1}{4},\frac{1}{2},\frac{1\pm\sqrt{17}}{8}\right\} \quad \text{and} \quad \mathfrak{mspec}\big(\sbloom(\rho_2,\mathcal{E})\big)\approx \{0.318,0.945,-0.260,-0.003\}.
\]
Thus,
\[
S\left(\sbloom(\rho_1,\mathcal{E})\right)\approx 0.8823 \quad \text{and} \quad S\left(\sbloom(\rho_2,\mathcal{E})\right)\approx 0.0723.
\]
Moreover, since 
\[
\mathcal{E}(\rho_1)=\rho_1 \quad \text{and} \quad \mathcal{E}(\rho_2)=
\frac{1}{2\sqrt{2}}
\left(\!
\begin{array}{cc}
\sqrt{2} & -1 \\
-1 & \sqrt{2} \\
\end{array}
\right),
\]
we have
\[
S(\rho_1)=S(\mathcal{E}(\rho_1))=\log(2) \quad \text{and} \quad S(\mathcal{E}(\rho_2))\approx 0.6009.
\]
Thus,
\begin{eqnarray*}
S_1(\rho_1,\mathcal{E})&\approx& 0.8823 \\
H_1(\rho_1,\mathcal{E})&\approx& 0.8823-1=-0.1177 \\
I_1(\rho_1,\mathcal{E})&\approx& 2-0.8823=1.1177 \\
K_1(\rho_1,\mathcal{E})&\approx& -0.1177 ,
\end{eqnarray*}
while
\begin{eqnarray*}
S_1(\rho_2,\mathcal{E})&=&H_1(\rho_2,\mathcal{E})\approx 0.0723 \\
I_1(\rho_2,\mathcal{E})&=&-K_1(\rho_2,\mathcal{E})\approx 0.6009-0.0723=0.5286.
\end{eqnarray*}
\ex

\bx[A projection-valued measure] \label{PVM6787}
Let $Y=\{y_1,y_2\}$, let $N_{y_1},N_{y_2}\in \M_2$ be the orthogonal projection operators given by
\[
N_{y_1}=\left(
\begin{array}{cc}
1 & 0 \\
0 & 0 \\
\end{array}
\right)
 \quad \text{and} \quad 
 N_{y_2}=\left(
\begin{array}{cc}
0 & 0 \\
0 & 1 \\
\end{array}
\right),
\]
and let $\mathcal{N}:\M_2\to \C^Y$ be the associated projection-valued measure (PVM), so that \[
\mathcal{N}(\rho)(y)=\tr(\rho P_y)
\]
for all $y\in Y$. Now let $\sigma_1=\frac{1}{2}\mathbbm{1}$ and let 
$\sigma_2
=\frac{1}{2}\left(
\begin{smallmatrix}
1 & -1 \\
-1 & 1 \\
\end{smallmatrix}
\right)$, 
so that $\sigma_1$ is the maximally mixed state and $\sigma_2$ is a pure state. We then have
\[ 
\sbloom(\mathcal{N},\sigma_1)=\bigoplus_{y\in Y}\frac{1}{2}\left(\sigma_1N_{y}+N_y\sigma_1\right)=\frac{1}{2}\left(
\begin{array}{cc}
1 & 0 \\
0 & 0 \\
\end{array}
\right)\oplus 
\frac{1}{2}\left(
\begin{array}{cc}
0 & 0 \\
0 & 1 \\
\end{array}
\right)
\]
and
\[
\sbloom\left(\mathcal{N},
\sigma_2
\right)=\bigoplus_{y\in Y}\frac{1}{2}\left(\sigma_2N_{y}+N_y\sigma_2\right)=\frac{1}{2}
\left(
\begin{array}{cc}
1 & -1/2 \\
-1/2 & 0 \\
\end{array}
\right)
\oplus
\frac{1}{2}\left(
\begin{array}{cc}
0 & -1/2 \\
-1/2 & 1 \\
\end{array}
\right).
\]
Thus, by \eqref{SENTX81}, we have
\[
S\big(\sbloom\left(\mathcal{N},
\sigma_1\right)\big)=\log(2),
\]
while
\begin{eqnarray*}
S\big(\sbloom\left(\mathcal{N},
\sigma_2\right)\big)&=&\log(2)+\frac{1}{2}S\left(
\begin{array}{cc}
1 & -1/2 \\
-1/2 & 0 \\
\end{array}
\right)+\frac{1}{2}S\left(
\begin{array}{cc}
0 & -1/2 \\
-1/2 & 1 \\
\end{array}
\right) \\
&=&\log(2)+S\left(
\begin{array}{cc}
1 & -1/2 \\
-1/2 & 0 \\
\end{array}
\right). \\
\end{eqnarray*}
And since the eigenvalues of 
$\frac{1}{2}\left(
\begin{smallmatrix}
2 & -1 \\
-1 & 0 \\
\end{smallmatrix}
\right)$
are $\lambda_{\pm}=\frac{1\pm\sqrt{2}}{2}$, it follows that 
\[
S\left(
\begin{array}{cc}
1 & -1/2 \\
-1/2 & 0 \\
\end{array}
\right)=\frac{\sqrt{2}}{2}\log\left(\frac{\sqrt{2}-1}{\sqrt{2}+1}\right)+\log(2)<0.
\]
Thus,
\[
S\left(\sbloom\left(\mathcal{N},
\sigma_2\right)\right)=2\log(2)+\frac{\sqrt{2}}{2}\log\left(\frac{\sqrt{2}-1}{\sqrt{2}+1}\right).
\]
We then have
\begin{eqnarray*}
S_1\left(\mathcal{N},\sigma_1\right)&=&I_1\left(\mathcal{N},\sigma_1\right)=\log(2) \\
H_1\left(\mathcal{N},\sigma_1\right)&=&K_1\left(\mathcal{N},\sigma_1\right)=0, 
\end{eqnarray*}
while
\begin{eqnarray*}
S_1\left(\mathcal{N},\sigma_2\right)&=&H_1\left(\mathcal{N},\sigma_2\right)\approx 0.2019 \\
I_1\left(\mathcal{N},\sigma_2\right)&=&-K_1\left(\mathcal{N},\sigma_2\right)\approx 0.7983.
\end{eqnarray*}
\ex

\section{Positive operator-valued measures} \label{S82}

In this section we derive general formulas for the dynamical information measures associated with a positive operator-valued measure (POVM), i.e., a CPTP map of the form $\mathcal{E}:\M_n\to \C^Y$ for some finite set $Y$. We note that in such a case there exists a collection of potive operators $\{E_{y}\}$ summing to the identity such that
\be
\label{eq:povmelts}
\mathcal{E}(A)=\sum_{y\in Y}\tr(AE_y)\delta_y
\ee
for all $A\in \M_n$. The operators $E_y$ will be referred to as the \define{POVM elements} associated with $\mathcal{E}$. We then show that when $\rho\in \M_n$ is a pure state, we have $K_1(\rho,\mathcal{E})\leq 0$ and $I_1(\rho,\mathcal{E})\geq 0$. Thus, Example~\ref{PVM6787} illustrates a general phenomenon for arbitrary positive operator-valued measures (of which projection-valued measures are a special case) that was alluded to in Remark~\ref{rmk:CQI}. 

\bt
\label{prop:PVMKN}
Let $(\rho,\mathcal{E})\in \mathscr{P}(\M_n,\C^Y)$, let $q=\mathcal{E}(\rho)$, let $N=\{y\in Y\;|\;q_{y}=0\}$, and for all $y\in Y\setminus N$, let $\rho^{y}_{p}\in \M_n$ be given by
\[
\rho^{y}_{p}=\frac{1}{2q_{y}}\left(\rho^{p}E_{y}\rho^{1-p}+\rho^{1-p}E_{y}\rho^{p}\right),
\]
where the $E_y$ are the POVM elements associated with $\mathcal{E}$ as in~\eqref{eq:povmelts}. Then 
\begin{align*}
S_p(\mathcal{E},\rho)&=H(q)+\sum_{y\in Y\setminus N}q_y S(\rho_p^y),&
H_p(\mathcal{E},\rho)&=H(q)-S(\rho)+\sum_{y\in Y\setminus N}q_y S(\rho_p^y),\\
I_p(\mathcal{E},\rho)&=S(\rho)-\sum_{y\in Y\setminus N}q_y S(\rho_p^y),&
K_p(\mathcal{E},\rho)&=\sum_{y\in Y\setminus N}q_y S(\rho_p^y).
\end{align*}
\et

\bprf
The state over time $\spbloom{p}(\rho,\mathcal{E})$ is given by 
\[
\spbloom{p}(\rho,\mathcal{E})=\sum_{y\in Y}\frac{1}{2}\left(\rho^{p}E_{y}\rho^{1-p}+\rho^{1-p}E_{y}\rho^{p}\right)\otimes\delta_{y}
=\sum_{y\in Y\setminus N}q_{y}\rho^{y}_{p}\otimes\delta_{y}
+\sum_{y\in N}\tilde{\rho}^{y}_{p}\otimes\delta_{y},
\]
where 
\[
\tilde{\rho}^{y}_{p}:=\frac{1}{2}\left(\rho^{p}E_{y}\rho^{1-p}+\rho^{1-p}E_{y}\rho^{p}\right)
\]
for all $y\in N$. The reason for splitting the summation into two parts is because only the first set of terms will make a contribution to the entropy identities to be proved, as we will now show that $\tilde{\rho}^{y}_{p}=0$ for all $y\in N$. 
We will prove this by examining three exhaustive cases. 
\begin{enumerate}[i.]
\item
Suppose $p=\frac{1}{2}$. Then $\tilde{\rho}^{y}_{p}=\sqrt{\rho}\,E_{y}\sqrt{\rho}$ is a positive operator. Since its trace is assumed to be $0$, it must be that $\sqrt{\rho}\,E_{y}\sqrt{\rho}=0$ by the faithfulness of the trace. 
\item
Suppose $p\in(0,\frac{1}{2})\cup(\frac{1}{2},1)$. Then, since $\tr(\tilde{\rho}^{y}_{p})=0$ still implies $\sqrt{\rho}\,E_{y}\sqrt{\rho}=0$, multiplying $\sqrt{\rho}\,E_{y}\sqrt{\rho}$ on the left by $\rho^{p-1/2}$ and on the right by $\rho^{1-p-1/2}$ (where $\rho$ raised to a negative power is meant in terms of the pseudo-inverse~\cites{Mo1920,Pe55}) gives $\rho^{p}E_{y}\rho^{1-p}=0$. Similarly,  $\rho^{1-p}E_{y}\rho^{p}=0$ so that $\tilde{\rho}^{y}_{p}=0$. 
\item
Suppose $p\in\{0,1\}$ so that $\tilde{\rho}^{y}_{p}=\frac{1}{2}\{\rho,E_{y}\}$.
Write $\rho=\sum_{i=1}^{m}p_{i}|v_{i}\>\<v_{i}|$ as an eigenvector decomposition in terms of an orthonormal basis $\{|v_{i}\>\}$ (note that some of the $p_{i}$ may vanish). Then $q_{y}=0$ reads
\[
0=q_{y}
=\sum_{j=1}^{m}\<v_{j}|\tilde{\rho}^{y}_{p}|v_{j}\>
=\sum_{i,j}^{m}\frac{p_{i}}{2}\Big(\underbrace{\<v_{j}|v_{i}\>}_{\delta_{ij}}\<v_{i}|E_{y}|v_{j}\>+\<v_{j}|E_{y}|v_{i}\>\underbrace{\<v_{i}|v_{j}\>}_{\delta_{ij}}\Big)
=\sum_{i=1}^{m}p_{i}\<v_{i}|E_{y}|v_{i}\>
\]
since the trace can be computed with respect to any orthonormal basis. Since  $p_{i}\ge0$ for all $i\in\{1,\dots,m\}$, we can conclude $\<v_{i}|E_{y}|v_{i}\>=0$ for all $i$ such that $p_{i}>0$. Hence, since $E_{y}$ is a positive operator, this implies $E_{y}|v_{i}\>=0$ (this is because $\<v_{i}|E_{y}|v_{i}\>=\<v_{i}|\sqrt{E_{y}}^{\dag}\sqrt{E_{y}}|v_{i}\>=\lVert \sqrt{E_{y}}v_{i}\rVert^{2}=0$ implies $\sqrt{E_{y}}|v_{i}\>=0$ implies $E_{y}|v_{i}\>=0$) and hence also $\<v_{i}|E_{y}=0$ for all $i$ such that $p_{i}>0$. Therefore, 
\[
\tilde{\rho}^{y}_{p}=\sum_{i=1}^{m}\frac{p_{i}}{2}\Big(|v_{i}\>\<v_{i}|E_{y}+E_{y}|v_{i}\>\<v_{i}|\Big)
=\sum_{\substack{i=1\\p_{i}\ne0}}^{m}\frac{p_{i}}{2}\Big(|v_{i}\>\<v_{i}|E_{y}+E_{y}|v_{i}\>\<v_{i}|\Big)
=0,
\]
as needed. 
\end{enumerate}
Therefore, since $\tilde{\rho}^{y}_{p}=0$ for all $y\in N$, the state over time is given by
\[
\spbloom{p}(\rho,\mathcal{E})
=\sum_{y\in Y\setminus N}q_{y}\rho^{y}_{p}\otimes\delta_{y}.
\]
Hence, 
\[
S_{p}(\rho,\mathcal{E})
=-\sum_{y\in Y\setminus N}q_{y}\tr\Big(\rho^{y}_{p}\log\big|q_{y}\rho^{y}_{p}\big|\Big)
=-\sum_{y\in Y\setminus N}q_{y}\log(q_{y})-\sum_{y\in Y\setminus N}q_{y}\tr\big(\rho^{y}_{p}\log|\rho^{y}_{p}|\big), 
\]
which proves the first entropy relation $S_p(\mathcal{E},\rho)=H(q)+\sum_{y\in Y\setminus N}q_y S(\rho_p^y)$. The rest of the first four identities follow from the definitions of the information measures.
\eprf

\bd
\label{defn:NDPDP}
A process $(\rho,\mathcal{E})\in \mathscr{P}(\M_n,\C^Y)$ with $\rho$ pure is said to be \define{non-disturbing} if and only if $\im(\rho)$ is invariant with respect to the POVM elements $E_y$ associated with $\mathcal{E}$ for all $y\in Y\setminus N$, 
where $N=\{y\in Y\;|\;q_{y}=0\}$ and $q=\mathcal{E}(\rho)$. Otherwise, $(\rho,\mathcal{E})$ 
is said to be \define{disturbing}.
\ed

\br
The definition of a non-disturbing process in Definition~\ref{defn:NDPDP} is in agreement with the notion of a non-disturbing measurement on a state given in Section~9.4 of Ref.~\cite{Wilde2017} provided that we employ the square-root instrument associated with a POVM $\mathcal{E}$ with POVM elements $\{E_{y}\}$. This can be seen by the fact that if $\rho=|v\>\<v|$ for some unit vector $v\in\C^{n}$, then the assumption $\im(\rho)$ being invariant with respect to $E_{y}$ for $y\in Y\setminus N$ means that $E_{y}|v\>=c_{y}|v\>$ for some $c_{y}>0$, which implies $\sqrt{E_{y}}|v\>=\sqrt{c_{y}}|v\>$. Hence, $\sqrt{E_{y}}\rho\sqrt{E_{y}}=c_{y}\rho$ so that the updated state via the state-update rule for the square-root instrument (cf.\ Section IV.B.\ of Ref.~\cite{FuPa22}) equals $\rho$ (this corresponds to the case $\epsilon=0$ in equation (9.197) of Ref.~\cite{Wilde2017}). 
\er

\bt
\label{prop:PVMKN56}
Let $(\rho,\mathcal{E})\in \mathscr{P}(\M_n,\C^Y)$ with $\rho$ pure. Then the following statements hold.
\begin{enumerate}[i.]
\item \label{PVMKN561}
$I_{p}(\rho,\mathcal{E})=K_{p}(\rho,\mathcal{E})=0$ for all $p\in(0,1)$.
\item \label{PVMKN562}
If $(\rho,\mathcal{E})$ is non-disturbing, then $I_1(\rho,\mathcal{E})=K_{1}(\rho,\mathcal{E})=0$.
\item \label{PVMKN563}
If $(\rho,\mathcal{E})$ is disturbing, then $I_{1}(\rho,\mathcal{E})=-K_1(\rho,\mathcal{E})>0$.
\end{enumerate}
\et

\bprf
Let $|v\>\in\C^{n}$ be the unit vector such that $\rho=|v\>\<v|$ and let $\{E_y\}$ be the POVM elements associated with $\mathcal{E}$ as in~\eqref{eq:povmelts}. As in Theorem~\ref{prop:PVMKN}, we let $\rho_{p}^{y}=\frac{1}{2q_y}(\rho^pE_y\rho^{1-p}+\rho^{1-p}E_y\rho^{p})$ for all $p\in [0,1]$ and all $y\in Y\setminus N$. 

\noindent
\underline{Item \ref{PVMKN561}}: For all $p\in(0,1)$ we have $\rho^{y}_{p}=|v\>\<v|$. It then follows that $S(\rho^{y}_{p})=0$ for all $y\in Y\setminus N$. Hence, $I_{p}(\rho,\mathcal{E})=K_{p}(\rho,\mathcal{E})=0$ for all $p\in(0,1)$ by Theorem~\ref{prop:PVMKN}. 

Items \ref{PVMKN562} and \ref{PVMKN563} concern the case $p=1$ (or equivalently $p=0$), in which case we have
\[
\rho^{y}_{1}=\frac{1}{2q_{y}}\{\rho,E_{y}\}=
\frac{1}{2q_{y}}\Big(|v\>\<v|E_{y}+E_{y}|v\>\<v|\Big)
\]
for all $y\in Y\setminus N$.
Thus, 
\be \label{RHO1Y67x}
\rho^{y}_{1}|w\>=\frac{\<v|E_{y}|w\>}{2q_{y}}|v\>+\frac{\<v|w\>}{2q_{y}}E_{y}|v\>
\ee
for all $|w\>\in \C^n$. It then follows that $\mathrm{im}(\rho^{y}_{1})\subseteq\mathrm{span}\big\{|v\>, E_{y}|v\>\big\}$. Hence, $\rho^{y}_{1}$ is a self-adjoint matrix of rank at most 2 for all $y\in Y\setminus N$. 

\noindent
\underline{Item \ref{PVMKN562}}: Suppose $(\rho,\mathcal{E})$ is non-disturbing, so that $\im(\rho)$ is invariant with respect to $E_y$ for all $y\in Y\setminus N$. It then follows from \eqref{RHO1Y67x} that $\rho_1^{y}$ is a rank-1 projection since $\tr(\rho_1^y)=1$. Therefore, $S(\rho^{y}_{p})=0$ for all $y\in Y\setminus N$. It then follows that $I_1(\rho,\mathcal{E})=K_{1}(\rho,\mathcal{E})=0$ by Theorem~\ref{prop:PVMKN}.

\noindent
\underline{Item \ref{PVMKN563}}: Suppose $(\rho,\mathcal{E})$ is disturbing, so that there exists a $y\in Y\setminus N$ such that $\im(\rho)$ is \emph{not} invariant with respect to $E_y$. It then follows from \eqref{RHO1Y67x} that $\rho_1^{y}$ is of rank 2. We now show that $\rho_1^{y}$ is not positive by determining its non-zero eigenvalues. For this, let $|u\>=E_{y}|v\>$, so that $\rho^{y}_{1}=\frac{1}{2q_{y}}\big(|v\>\<u|+|u\>\<v|\big)$ and $\<v|u\>=\<u|v\>=\<v|E_{y}|v\>=q_{y}$. Writing an arbitrary eigenvector of $\rho^{y}_{1}$ as $|w\>=\alpha|v\>+\beta|u\>$, with $\alpha,\beta\in\C$, the eigenvalue equation $\rho^{y}_{p}|w\>=\lambda|w\>$ yields
\[
\lambda\alpha|v\>+\lambda\beta|u\>
=\frac{\alpha q_{y}+\beta\<u|u\>}{2q_{y}}|v\>+\frac{\alpha+\beta q_{y}}{2q_{y}}|u\>.
\]
By the linear independence of $\{|v\>,|u\>\}$, this guarantees 
\[
\lambda \alpha=\frac{\alpha q_{y}+\beta\<u|u\>}{2q_{y}}
\quad\text{ and }\quad
\lambda \beta = \frac{\alpha+\beta q_{y}}{2q_{y}}.
\]
These equations then yield the quadratic equation
\[
4\lambda^2-2\lambda+1-\frac{\<u|u\>}{q_{y}^2}=0
\]
in the variable $\lambda$. The solutions $\lambda_{\pm}$ are given by
\[
\lambda_{\pm}=\frac{1}{4}\left(1\pm\sqrt{4\left(\frac{\<u|u\>}{q_{y}^2}-1\right)+1}\right).
\]
Now let $\{|v\>,|v_1\>,\dots,|v_{n-1}\>\}$ be an orthonormal basis of $\C^n$ containing $|v\>$. By using the completeness relation 
\[
\mathds{1}_{n}=|v\>\<v|+\sum_{i=1}^{n-1}|v_{i}\>\<v_{i}|,
\]
we find that 
\[
\<u|u\>=\<v|E_{y}E_{y}|v\>
=\<v|E_{y}|v\>\<v|E_{y}|v\>+\sum_{i=1}^{n-1}\<v|E_{y}|v_{i}\>\<v_{i}|E_{y}|v\>
=q_{y}^2+\sum_{i=1}^{n-1}\big|\<v|E_{y}|v_{i}\>\big|^2> q_{y}^2,
\] 
which implies $\lambda_{-}<0$ and $\lambda_{+}>1$. Therefore, 
\[
S(\rho^{y}_{1})=-\lambda_{-}\log|\lambda_{-}|-\lambda_{+}\log(\lambda_{+})<0
\]
(cf.\ Figure~\ref{fig:newshannon}). This argument, together with the proof of item \ref{PVMKN562}, yields $S(\rho_1^y)=0$ whenever $\im(\rho)$ is invariant with respect to $E_y$  and $S(\rho_1^y)<0$ whenever $\im(\rho)$ is \emph{not} invariant with respect to $E_y$. It then follows from Theorem~\ref{prop:PVMKN} that $I_{1}(\rho,\mathcal{E})=-K_1(\rho,\mathcal{E})>0$, as desired.
\eprf

\br
In light of Theorem~\ref{prop:PVMKN56}, it follows that the symmetric bloom is the only symmetric $p$-bloom which distinguishes between disturbing and non-disturbing measurements. In particular, the Leifer--Spekkens state over time function does not distinguish between disturbing and non-disturbing measurements.
\er

\br
The proof of Theorem~\ref{prop:PVMKN56} also shows that $I_{p}(\rho,\mathcal{E})$ is not a continuous function of $p\in[0,1]$. This is because $I_{p}(\rho,\mathcal{E})=0$ for all $p\in(0,1)$, while $I_{p}(\rho,\mathcal{E})$  could be positive for $p\in\{0,1\}$. Incidentally, one could also view this result as an indication that the only symmetric $p$-bloom state over time function for which $I_{p}(\rho,\mathcal{E})$ does not identically vanish and satisfies $I_{p}(\rho,\mathcal{E})\ge0$ whenever $\rho$ is pure is the standard symmetric bloom, i.e., when $p=1$ (or equivalently $p=0$). 
\er

\section{A quantum entropic Bayes' rule} \label{S9}

The information-theoretic analogue of the classical Bayes rule associated with a pair of classical random variables $(X,Y)$ is the \emph{entropic Bayes' rule}~\cite{FuPa21}, namely,
\be \label{CENTPXBRX87}
H(Y|X)+H(X)=H(X|Y)+H(Y).
\ee
In this section, we generalize equation \eqref{CENTPXBRX87} to the dynamical conditional entropy $H_{\psi}$ associated with a hermitian state over time function $\psi$. For this, we first recall a quantum analogue of Bayesian inversion with respect to a hermitian state over time function, which was first introduced in Ref.~\cite{FuPa22a}. 

In this section, we relax the condition that a state over time function $\psi$ must be defined on a domain of the form $\mathscr{P}(\VA,\VB)$. In particular, we allow for the possibility that a state over time function $\psi$ may take as its input $(\rho,\mathcal{E})$, where $\mathcal{E}$ is not necessarily CPTP. For example, the symmetric bloom $\sbloom$ is well-defined for any input $(\rho,\mathcal{E})\in \VA\times \hom(\VA,\VB)$ (see Section V.\ in Ref~\cite{FuPa22a} for more details).   

\bd
Let $\psi$ be a hermitian state over time function and let $(\rho,\mathcal{E})\in \mathscr{P}(\VA,\VB)$. A $\dag$-preserving and trace-preserving element $\ov {\mathcal{E}}\in \hom(\VB,\VA)$ is said to be a \define{Bayes map} for the process $(\rho,\mathcal{E})$ 
with respect to the state over time function $\psi$ if and only if 
\be \label{UNIFIEDBAYES71}
\psi(\rho,\mathcal{E})=\gamma\left(\psi(\sigma,\ov {\mathcal{E}})\right),
\ee
where $\sigma=\mathcal{E}(\rho)$ and $\gamma:\VB\otimes \VA\to \VA\otimes \VB$ is the swap isomorphism. If such a Bayes map $\ov {\mathcal{E}}$ is in fact CPTP, then the process $(\sigma,\ov {\mathcal{E}})$ is said to be a \define{Bayesian inverse} of $(\rho,\mathcal{E})$ with respect to $\psi$.
\ed

\bx[The classical case]
Let $X$ and $Y$ be finite sets, let $\mathcal{E}:\C^X\to \C^Y$ be a classical channel, let $\rho\in \mathcal{S}(\C^X)$ and let $\mathcal{E}_{yx}$ be such that $\mathcal{E}(\delta_x)=\sum_{y\in Y}\mathcal{E}_{yx}\delta_y$. Then for all $p\in [0,1]$, the state over time $\spbloom{p}(\rho,\mathcal{E})\in \C^X\otimes \C^Y\cong \C^{X\times Y}$ is the classical state over time as in Example~\ref{CXST177}, i.e., the function given by
\[
\spbloom{p}(\rho,\mathcal{E})(x,y)=\mathcal{E}_{yx}\rho_x.
\]
Now let $\sigma=\mathcal{E}(\rho)$, and let $\ov {\mathcal{E}}:\C^Y\to \C^X$ be the classical channel given by $\ov {\mathcal{E}}(\delta_y)=\sum_{x\in X}\ov {\mathcal{E}}_{xy}\delta_x$, where $\ov {\mathcal{E}}_{xy}=\rho_x\mathcal{E}_{yx}/\sigma_y$, so that
\[
\spbloom{p}(\sigma,\ov {\mathcal{E}})(y,x)=\ov {\mathcal{E}}_{xy}\sigma_y.
\]
We then have
\be \label{CBYXRX17}
\spbloom{p}(\rho,\mathcal{E})(x,y)=\mathcal{E}_{yx}\rho_x=\ov {\mathcal{E}}_{xy}\sigma_y=\spbloom{p}(\sigma,\ov {\mathcal{E}})(y,x)=\gamma\left(\spbloom{p}(\sigma,\ov {\mathcal{E}})\right)(x,y).
\ee
It then follows that $\spbloom{p}(\rho,\mathcal{E})=\gamma\big(\spbloom{p}(\sigma,\ov {\mathcal{E}})\big)$, so that $(\sigma,\ov {\mathcal{E}})$ is a Bayesian inverse of $(\rho,\mathcal{E})$ with respect to the symmetric $p$-bloom $\spbloom{p}$ for all $p\in [0,1]$. Moreover, we see from \eqref{CBYXRX17} that the equation $\spbloom{p}(\rho,\mathcal{E})=\gamma\big(\spbloom{p}(\sigma,\ov {\mathcal{E}})\big)$ is a re-writing of the coordinate equation $\mathcal{E}_{yx}\rho_x=\ov {\mathcal{E}}_{xy}\sigma_y$, which is the classical Bayes' rule. 
\ex 

\bx
\label{ex:LSPetz}
A Bayesian inverse $(\sigma,\ov {\mathcal{E}})$ of $(\rho,\mathcal{E})$ exists with respect to $\psi_{\LS}$ for all processes $(\rho,\mathcal{E})\in \mathscr{P}(\VA,\VB)$. The associated Bayes map $\ov{\mathcal{E}}$ coincides with the \emph{Petz recovery map}~\cites{Pe84,FuPa22a}.
\ex

\bx
\label{ex:SBbayesmap}
A Bayes map $\ov {\mathcal{E}}$ of $(\rho,\mathcal{E})$ exists with respect to $\sbloom$ for all processes $(\rho,\mathcal{E})\in \mathscr{P}(\VA,\VB)$, and an explicit formula for the Bayes map is provided in Section V.A.\ of Ref.~\cite{FuPa22a}.
\ex

\bx[Disintegrations of deterministic evolution] \label{DINTXDT87}
Let $U\in \M_p\otimes \M_n$ be unitary, let $\mathcal{E}:\M_p\otimes \M_n\to \M_n$ be the map given by $\mathcal{E}=\tr_{\M_p}\circ \Ad_U$, and let $\rho=U^{\dag}(\tau \otimes \sigma)U$ for some density matrices $\tau\in \M_p$ and $\sigma\in \M_n$. Then the Bayesian inverse of $(\rho,\mathcal{E})$ with respect to the symmetric bloom state over time function is $(\sigma,\ov {\mathcal{E}})$, where $\ov {\mathcal{E}}=\Ad_{U^{\dag}}\circ \mathcal{G}$ and $\mathcal{G}:\M_n\to \M_p\otimes \M_n$ is the map given by $A\mapsto \tau\otimes A$. Such Bayesian inverses are quantum analogues of disintegrations, which were studied in detail in Refs.~\cites{PaRu19,PaBayes}. Their relationship to conditional expectations are provided in Ref.~\cite{GPRR21}. 
\ex

We now state and prove a quantum entropic Bayes' rule for the conditional entropy $H_{\psi}$ associated with a hermitian state over time function $\psi$. This result extends the classical entropic Bayes' rule of Ref.~\cite{FuPa21}. 

\bt[Quantum Entropic Bayes' Rule] \label{QEXTPXBR91}
Let $\psi$ be a hermitian state over time function, suppose $\ov {\mathcal{E}}$ is a Bayes map for the process $(\rho,\mathcal{E})$ with respect to $\psi$, and let $\sigma=\mathcal{E}(\rho)$. Then 
\be \label{QEXBRX17}
H_{\psi}(\rho,\mathcal{E})+S(\rho)=H_{\psi}(\sigma,\ov {\mathcal{E}})+S(\sigma).
\ee
In particular, if $\psi$ is the Leifer--Spekkens or symmetric bloom state over time function, then for every process $(\rho,\mathcal{E})$, there exists a Bayes map $\ov {\mathcal{E}}$ such that equation \eqref{QEXBRX17} holds.
\et

\bprf
Since $\ov {\mathcal{E}}$ is a Bayes map for $(\rho,\mathcal{E})$ with respect to $\psi$, we have
\[
\psi(\rho,\mathcal{E})=\gamma\left(\psi(\sigma,\ov {\mathcal{E}})\right).
\]
Thus,
\[
S_{\psi}(\rho,\mathcal{E})=S\left(\psi(\rho,\mathcal{E})\right)=S\left(\gamma(\psi(\sigma,\ov {\mathcal{E}}))\right)=S\left(\psi(\sigma,\ov {\mathcal{E}})\right)=S_{\psi}(\sigma,\ov {\mathcal{E}}),
\]
where the third equality follows from Proposition~\ref{prop:SEinv}, which applies since the swap map $\gamma$ is a $*$-isomorphism. Equation \eqref{QEXBRX17} then follows directly from the definition of conditional entropy.
The last statement then follows from this and Examples~\ref{ex:LSPetz} and~\ref{ex:SBbayesmap}.
\eprf

\section{Deterministic evolution} \label{S8}

Let $f:X\to Y$ be a surjective function between finite sets representing a deterministic classical channel, let $p:X\to [0,1]$ be a prior distribution on the inputs of $f$, and write $q$ as the pushforward probability distribution, the value of which is given by $q_{y}=\sum_{x\in X}f_{yx}p_{x}$ for each $y\in Y$. We saw in Example~\ref{CDXE771} that the dynamical versions of joint entropy, conditional entropy, mutual information, and information loss 
satisfy
\be\label{CINFX747}
S(p,f)=H(p) , \quad H(p,f)=0,\quad I(p,f)=H(q) \quad \text{and} \quad K(p,f)=H(p)-H(q) 
\ee
for such a classical deterministic process. 
We now generalize equations \eqref{CINFX747} to the quantum setting, where the analogue of a function is a deterministic quantum channel, i.e., unitary evolution followed by the partial trace with respect to a subsystem (justifications for calling such a composite ``deterministic'' are provided in Refs.~\cites{FuJa13,We17,Pa17}). What we find is that while the case of $f$ being a bijection generalizes with respect to the symmetric $p$-bloom $\spbloom{p}$ for all $p\in [0,1]$, the quantum analogues of equations \eqref{CINFX747} only hold in full generality with respect to the symmetric bloom $\spbloom{1}$. Moreover, while equations \eqref{CINFX747} follow immediately from the definitions of the classical dynamical information measures $S$, $H$, $I$, and $K$, in the quantum setting, the derivation of explicit formulas analogous to equations \eqref{CINFX747} for the information measures $S_{\psi}$, $H_{\psi}$, $I_{\psi}$, and $K_{\psi}$ requires a considerable amount of calculation. As such, the fact that equations \eqref{CINFX747} still end up holding in the quantum setting turns out to be a non-trivial result, thus yielding further justification for our use of the extended entropy function as a generalization of von~Neumann entropy.

We now prove the following result, which generalizes equations \eqref{CINFX747} in the case that $f$ is a bijection between finite sets. 

\bt \label{EnTxSTAR97}
Let $\mathcal{E}:\VA\to\VB$ be a $*$-isomorphism and let $\rho\in\mathcal{S}(\VA)$ be a state. 
Then, for all $p\in [0,1]$, the following statements hold.
\begin{enumerate}[i.]
\item \label{UNTEVX1}
$S_{p}(\rho,\mathcal{E})=I_{p}(\rho,\mathcal{E})=S(\rho)$. 
\item  \label{UNTEVX2}
$H_{p}(\rho,\mathcal{E})=K_{p}(\rho,\mathcal{E})=0$.
\end{enumerate}
\et
 
\bprf
These identities all follow from Theorem~\ref{BTXETXFX787}, Corollary~\ref{cor:TEMMA}, and Definition~\ref{DMXQIX981}. Namely, $S_{p}(\rho,\mathcal{E})
=S(\rho)$ is exactly what 
Corollary~\ref{cor:TEMMA}
claims. The rest follows from Definition~\ref{DMXQIX981}. Namely, 
\[
I_{p}(\rho,\mathcal{E})=S(\rho)+S\big(\mathcal{E}(\rho)\big)-S_{p}(\rho,\mathcal{E})
=S(\rho)+S(\rho)-S(\rho)=S(\rho), 
\]
\[
K_{p}(\rho,\mathcal{E})=S_{p}(\rho,\mathcal{E})-S\big(\mathcal{E}(\rho)\big)
=S(\rho)-S(\rho)=0,
\]
and
\[
H_{p}(\rho,\mathcal{E})=S_{p}(\rho,\mathcal{E})-S(\rho)
=S(\rho)-S(\rho)=0.
\]
This proves all the identities. 
\eprf

We now show that a quantum analogue of the identities in \eqref{CINFX747} hold in the setting of unitary evolution followed by partial trace for a certain class of input states. 
Unlike Theorem~\ref{EnTxSTAR97}, however, the following theorem only holds for the case of the symmetric bloom state over time function $\sbloom$.  

\bt \label{EXTSTZR109}
Let $U\in \M_p\otimes \M_n$ be unitary, let $\mathcal{E}:\M_p\otimes \M_n\to \M_n$ be the map given by $\mathcal{E}=\emph{tr}_{\M_p}\circ \Ad_U$, and let $\rho=U^{\dag}(\tau \otimes \sigma)U$ for some density matrices $\tau\in \M_p$ and $\sigma\in \M_n$. Then the following statements hold.
\begin{enumerate}[i.]
\item \label{EXTSTZR1091}
$S_{1}(\rho,\mathcal{E})=S(\rho)=S(\tau)+S(\sigma) \geq 0$ .
\item \label{EXTSTZR10912}
$H_{1}(\rho,\mathcal{E})=0$ .
\item \label{EXTSTZR10913}
$I_{1}(\rho,\mathcal{E})=S(\sigma)\geq 0$ .
\item \label{EXTSTZR10914}
$K_{1}(\rho,\mathcal{E})=S(\rho)-S(\sigma)=S(\tau)\geq 0$ .
\end{enumerate}
\et

\br
Let $(\rho,\mathcal{E})$ be as in the statement of Theorem~\ref{EXTSTZR109}, and let $\mathcal{G}:\M_n\to \M_p\otimes \M_n$ be the map given by $A\mapsto \tau\otimes A$. From Example~\ref{DINTXDT87} we know that $(\sigma,\Ad_{U^{\dag}}\circ \mathcal{G})$ is a Bayesian inverse with respect to the symmetric bloom $\sbloom$, and moreover, that $(\sigma,\Ad_{U^{\dag}}\circ \mathcal{G})$ is a non-commutative \emph{disintegration} of $(\rho,\mathcal{E})$ (for more on non-commutative disintegrations, see Refs.~\cites{PaRu19,PaBayes}). As such, Theorem~\ref{EXTSTZR109} is a statement about deterministic evolutions of quantum systems which are disintegrable, and Theorem~\ref{EXTSTZR109} shows that such disintegrable processes are precisely the processes for which a quantum analogue of the identities in \eqref{CINFX747} hold. In particular, if our initial state for $\mathcal{E}$ in Theorem~\ref{EXTSTZR109} is different from $\Ad_{U^{\dag}}$ applied to a product state, then Theorem~\ref{EXTSTZR109} does not necessarily hold. For example, in the case $n=p=2$ in Theorem~\ref{EXTSTZR109} with $U=\mathds{1}$, if $\rho_{\EPR}\in \M_4$ is taken as our initial state for $\mathcal{E}$ with $\rho_{\EPR}$ the EPR state given by \eqref{EPRXS}, then $(\rho_{\EPR},\mathcal{E})$ is \emph{not} distintegrable, Hence, $(\rho_{\EPR},\mathcal{E})$ does \emph{not} satisfy the conclusions of Theorem~\ref{EXTSTZR109}. In particular, it follows from Example~\ref{MXEXPXS771} that $H_1(\rho_{\EPR},\mathcal{E})<0$.
\er

\br
In the context of Theorem~\ref{EXTSTZR109}, item \ref{EXTSTZR10914} says that the information discrepancy $K_1(\rho,\mathcal{E})$ is the von~Neumann entropy difference of the initial and final states $\rho$ and $\sigma=\mathcal{E}(\rho)$ associated with the process $(\rho,\mathcal{E})$. And since the Hilbert--Schmidt adjoint of the partial trace is a $*$-homomorphism, the information discrepancy in this case coincides with the ``entropy change along a $*$-homomorphism'' defined in Ref.~\cite{AJP22}. 
\er

\bprf [Proof of Theorem~\ref{EXTSTZR109}]
By the isometric invariance and the additivity of the entropy function we have 
\[
S(\rho)=S(U^{\dag}(\tau\otimes \sigma)U)=S(\tau\otimes \sigma)=S(\tau)+S(\sigma),
\]
and since $\mathcal{E}(\rho)=\sigma$, it follows that $S(\mathcal{E}(\rho))=S(\sigma)$. Moreover, we have
\begin{eqnarray*}
S_1(\rho,\mathcal{E})=S(\sbloom(\rho,\mathcal{E}))\overset{\eqref{BNTX81}}=S(\tau)+S(\sigma),
\end{eqnarray*}
from which the theorem follows.
\eprf

\section{Concluding remarks} \label{S10}

In this work, we defined a dynamical entropy $S(\rho,\mathcal{E})$ associated with quantum processes $(\rho,\mathcal{E})$, where $\rho$ is a state and $\mathcal{E}$ is a CPTP map responsible for the dynamical evolution of $\rho$. We then used such an entropy $S(\rho,\mathcal{E})$ to define dynamical analogues of the quantum conditional entropy, the quantum mutual information, and an information measure we refer to as ``information discrepancy.'' Key to our formulation of dynamical entropy was an extension of von~Neumann entropy to arbitrary hermitian matrices using the real part of the analytic continuation of the logarithm to the negative real axis. Such an entropy function also yields a well-defined notion of entropy for quasi-probability distributions, which play a prevalent role in quantum theory~\cites{Wigner32,Ki33,Di45,Fe87}. 

We have shown that our information measures satisfy many properties of classical information measures associated with stochastic processes, such as the vanishing of conditional entropy under deterministic evolution. While classical information measures are always non-negative, the information measures we defined may be negative, which seems to be a characteristic aspect of quantum information versus classical information (see Table~\ref{table:infomeasures} for details). Our dynamical mutual information, however, seems to be an exception, since we find it to be non-negative in all known examples. This was in fact unexpected as our generalized entropy function does \emph{not} satisfy subadditivity on quasi-density matrices, contrary to the case of the usual von~Neumann entropy functional for density matrices. A proof that our dynamical mutual information is in fact a non-negative measure of quantum information still eludes us, and thus remains an open problem. If our dynamical mutual information is in fact non-negative and bounded from above (as suggested by Figure~\ref{fig:subaddSOT}), then one may define a form of channel capacity in analogy with the classical case by maximizing the mutual information of a channel over the set of input states. It would then be interesting to investigate the relation of such a notion of channel capacity with more standard notions of channel capacity from quantum information theory~\cites{AdCe97,Ll97,De05,Wilde2017}.

\begin{table}
\centering
\begin{tabular}{c|cccc}
&\begin{tabular}{c}classical\\ static\end{tabular}&\begin{tabular}{c}classical\\dynamic\end{tabular}&\begin{tabular}{c}quantum\\static\end{tabular}&\begin{tabular}{c}quantum\\dynamic\end{tabular}\\
\hline
entropy&\checkmark&\checkmark&\checkmark&\xmark\\
conditional entropy&\checkmark&\checkmark&\xmark&\xmark\\
mutual information&\checkmark&\checkmark&\checkmark&\checkmark?\\
information discrepancy&\checkmark&\checkmark&\xmark&\xmark
\end{tabular}
\caption{(On the non-negativity of information measures) This table depicts the four measures of information studied in this paper: entropy, conditional entropy, mutual information, and information discrepancy (called ``conditional information loss'' in the classical setting~\cite{FuPa21}). In the static setting, the given datum is a joint state. In the dynamic setting, the given datum is a process (an initial state together with a channel). In the present paper, we have introduced the information measures in the column ``quantum dynamic.'' 
A check mark (\checkmark) indicates that positivity holds for the information measure, while a cross mark (\xmark) indicates that positivity does not hold in general for that information measure. Note that the positivity of the dynamical mutual information is conjectural at present, which is why ``\checkmark?'' is written in that entry.}
\label{table:infomeasures}
\end{table}

While we have extensively studied such dynamical measures of quantum information from a mathematical viewpoint, their general operational interpretations are still lacking at this point (however, see Remark~\ref{rmk:CQI} and Theorem~\ref{prop:PVMKN56} for partial progress in this direction). It would also be desirable to find a more explicit and palatable connection between such information measures and fundamental aspects of quantum dynamics, such as causal correlations and entanglement in time~\cite{FJV15}. This is particularly relevant now, as a theory of quantum states over time has recently been developed in order to study these types of fundamental questions~\cites{FuPa22,FuPa22a}.

Furthermore, the theory of quantum states over time have recently been extended from one-step processes $(\rho,\mathcal{E})$ to $n$-step processes $(\rho,\mathcal{E}_1,\dots,\mathcal{E}_n)$~\cites{fullwood2023,LQDV,JSK23}, where
\[
\VA_{0}\xrightarrow{\mathcal{E}_1} \VA_{1}\xrightarrow{\mathcal{E}_2} \cdots\xrightarrow{\mathcal{E}_{n-1}}\VA_{n-1}\xrightarrow{\mathcal{E}_n}\VA_{n}
\]
is a sequence of CPTP maps responsible for the dynamical evolution of an initial state $\rho\in\mathcal{S}(\VA_{0})$. It would be interesting to investigate the behavior of the entropy of states over time associated with such $n$-step processes for increasing $n$, much like multipartite entanglement and information measures form an integral part of quantifying spatial correlations and their consequences in many-body quantum systems~\cites{ECP10,IMPTLRG15,LRSTKCKLG19,KaBr19}. Moreover, if our dynamical mutual information is in fact a non-negative measure of quantum information, then we suspect that it may provide a notion of proper time associated with the $n$-step process $(\rho,\mathcal{E}_1,\ldots,\mathcal{E}_n)$, which may be viewed as the flow of quantum information along a future-directed path in a causal set~\cites{BLMS87,HMS03}.

\bigskip
\noindent
{\bf Acknowledgements.} 
This work is supported by MEXT-JSPS Grant-in-Aid for Transformative Research Areas (A) ``Extreme Universe'', No.\ 21H05183.
We acknowledge support from the Blaumann Foundation. 
The figures and plots in this paper were created using Mathematica Version 13.0~\cite{Mathematica130}.
We thank Francesco Buscemi, Jonah Kudler-Flam, Takashi Matsuoka, Tadashi Takayanagi, Yusuke Taki, and Zixia Wei for fruitful discussions.
\appendix

\section{Some technical results} \label{BLXCXCS1971}
In this appendix, we prove some technical results which were used in the body of the text. We implement Dirac bra-ket notation to facilitate several proofs. In such a case, the matrix units $E_{ij}$ will be written as $|i\>\<j|$.

\blem[The swap lemma]\label{LSX17}
Let $B$ be an $m\times n$ matrix and let $C$ be an $n\times m$ matrix. Then
\be\label{ESX17}
\sum_{i,j}^{m}E_{ij}^{(m)}\otimes \left(CE_{ji}^{(m)}B\right)=\sum_{p,q}^{n}\left(BE_{pq}^{(n)}C\right)\otimes E_{qp}^{(n)},
\ee
\elem

\bprf
Indeed, 
\[
\begin{split}
\sum_{i,j}^{m}E_{ij}^{(m)}\otimes \left(CE_{ji}^{(m)}B\right)&=\sum_{i,j}^{m}\sum_{p,q}^{n}|i\>\<j|\otimes\Big(|q\>\<q|C|j\>\<i|B|p\>\<p|\Big)\\
&=\sum_{i,j}^{m}\sum_{p,q}^{n}\Big(|i\>\<i|B|p\>\<q|C|j\>\<j|\Big)\otimes|q\>\<p|\\
&=\sum_{p,q}^{n}\left(BE_{pq}^{(n)}C\right)\otimes E_{qp}^{(n)},
\end{split}
\]
as desired.
\eprf

\blem \label{MXMA1719}
Let $\VA$ be a matrix algebra. Then 
\be
\label{eq:musEij}
\mu^{*}_{\VA}\big(E_{ij}^{\VA}\big)=\sum_{k}E_{ik}^{\VA}\otimes E_{kj}^{\VA},
\ee
and for all $\rho\in \VA$ 
\be 
\label{MXMA17}
\mu_{\VA}^*(\rho)
=\sum_{i,j}\rho E_{ij}^{\VA}\otimes E_{ji}^{\VA}
=\sum_{i,j}E_{ij}^{\VA}\otimes E_{ji}^{\VA}\rho.
\ee
\elem

\bprf
The identity~\eqref{eq:musEij} is proved in~\cite[Lemma~3.39]{PaRuBayes}. Writing $\rho$ as $\rho=\sum_{i,j}\<i|\rho|j\>|i\>\<j|$, we then have
\[
\mu_{\VA}^*(\rho)=\sum_{i,j}\<i|\rho|j\>\mu_{\VA}^*\big(|i\>\<j|\big)
\overset{\eqref{eq:musEij}}=\sum_{i,j,k}|i\>\<i|\rho|j\>\<k|\otimes|k\>\<j|
=\sum_{j,k}\big(\rho|j\>\<k|\big)\otimes|k\>\<j|
=\sum_{i,j}\rho E_{ij}^{\VA}\otimes E_{ji}^{\VA},
\]
thus proving the first equality in~\eqref{MXMA17}. The second equality then follows from Lemma~\ref{LSX17}.
\eprf

\blem\label{LXSTAR77}
Suppose $A\in \M_{m}\otimes \M_m$ is of the form
\[
A=\sum_{i,j}E_{ij}\otimes a_{ij}E_{ji},
\]
where $E_{ij}$ are the standard matrix units in $\M_m$ and $a_{ij}\geq 0$ for all $i,j\in \{1,\dots,m\}$. Then, its spectrum is given by 
\[
\mathfrak{mspec}(A)=\big\{a_{ii} \hspace{1mm} | \hspace{1mm} i=1,\dots,m\big\}\cup \left\{\left.\pm \sqrt{a_{ij}a_{ji}} \hspace{1mm} \right| \hspace{1mm} 1\le i<j\leq m\right\}.
\]
In addition, if for any $i,j\in\{1,\dots,m\}$ such that $a_{ij}=0$ implies $a_{ji}=0$, then $A$ is diagonalizable.
\elem

\bprf
First suppose $a_{ij}>0$ for all $i,j\in \{1,\dots,m\}$, and let $e_i$ denote the standard basis vectors of $\C^m$. For each $i\in \{1,\dots,m\}$, let 
\[
v_{ii}=e_i\otimes e_i,
\]
and for each $1\le i<j\leq m$, let 
\[
v_{ij}^{\pm}=\sqrt{a_{ij}}e_i\otimes e_j\pm\sqrt{a_{ji}}e_j\otimes e_i.
\] 
Then
\[
A(v_{kk})=\left(\sum_{i,j}E_{ij}\otimes a_{ji}E_{ji}\right)(e_k\otimes e_k)=\sum_{i,j}E_{ij}(e_k)\otimes a_{ji}E_{ji}(e_k)=a_{kk}(e_{k}\otimes e_{k})=a_{kk}v_{kk}.
\]
Thus, $v_{kk}$ is an eigenvector of $A$ with eigenvalue $a_{kk}$ for all $k\in \{1,\dots,m\}$. We also have
\begin{eqnarray*}
A(v_{kl}^{\pm})&=&\left(\sum_{i,j}E_{ij}\otimes a_{ij}E_{ji}\right)\big(\sqrt{a_{kl}}e_k\otimes e_l\pm\sqrt{a_{lk}}e_l\otimes e_k\big) \\
&=&\sqrt{a_{kl}}\sum_{i,j}E_{ij}e_k\otimes a_{ij}E_{ji}e_{l}\pm\sqrt{a_{lk}}\sum_{i,j}E_{ij}e_l\otimes a_{ij}E_{ji}e_{k} \\
&=&\sqrt{a_{kl}}a_{lk}e_{l}\otimes e_{k}\pm \sqrt{a_{lk}}a_{kl}e_{k}\otimes e_{l} \\
&=&\sqrt{a_{kl}a_{lk}}\big(\sqrt{a_{lk}}e_{l}\otimes e_{k}\pm \sqrt{a_{kl}}e_{k}\otimes e_{l}\big) \\
&=&\pm \sqrt{a_{kl}a_{lk}} v_{kl}^{\pm}.
\end{eqnarray*}
Thus, $v_{kl}^{\pm}$ is an eigenvector of $A$ with eigenvalue $\pm\sqrt{a_{kl}a_{lk}}$ for all $k<l\leq m$. And since the set of vectors 
\be
\label{eq:eigvEA}
\big\{v_{ii} \hspace{1mm} | \hspace{1mm} i=1,\dots,n\big\}\cup \left\{\left. v_{ij}^{\pm} \hspace{1mm} \right| \hspace{1mm} 1\le i<j\leq n\right\}
\ee
is a basis of $\C^m\otimes \C^m$, it follows that 
\[
\mathfrak{mspec}(A)=\big\{a_{ii} \hspace{1mm} | \hspace{1mm} i=1,\dots,m\big\}\cup \left\{\left.\pm \sqrt{a_{ij}a_{ji}} \hspace{1mm} \right| \hspace{1mm} 1\le i<j\leq m\right\},
\] 
as desired. In particular, the characteristic polynomial of $A$ is given by
\be \label{CpX891}
p_A(z)=(z-a_{11})\cdots(z-a_{mm})\prod_{i<j\leq m}(z^2-a_{ij}a_{ji}).
\ee
Since $p_A(z)$ is a continuous function of the $a_{ij}$, it follows that \eqref{CpX891} defines the characteristic polynomial of $A$ even if $a_{ij}=0$ for some values of $i$ and $j$, thus proving the first result.

To prove the claim about diagonalizability, first note that if $a_{ij}>0$ for all $i,j$, the previous argument already provides a basis of eigenvectors given by~\eqref{eq:eigvEA}, thus proving that $A$ is diagonalizable. To complete the argument in the more general case, first note that if $a_{ii}=0$ for some $i\in\{1,\dots,m\}$, then $v_{ii}$ is still an eigenvector of $A$ (with eigenvalue $0$). Now, if there exist distinct $i,j\in\{1,\dots,m\}$ such that $a_{ij}=0$, then $a_{ji}=0$ by assumption. In this case, the collection of vectors $v_{ij}=e_{i}\otimes e_{j}$ are linearly independent eigenvectors of $A$ with eigenvalue $0$. This follows from a calculation similar to the one from above, namely
\[
A v_{ij}=\sum_{k,l}E_{kl}e_{i}\otimes a_{kl}E_{lk}e_{j}
=\sum_{k,l}\delta_{li}e_{k}\otimes a_{kl}\delta_{kj}e_{l}
=e_{j}\otimes a_{ji}e_{i}
=0
.
\]
It then follows that the vectors
\begin{equation}
\label{eq:Aijbasis}
\begin{aligned}
\big\{v_{ii}\;\big|\;1\le i\le m\big\}
&\cup
\big\{v_{ij}^{\pm}\;\big|\;1\le i<j\le m \hspace{2mm} \text{with} \hspace{2mm} a_{ij}\ne0\big\}\\
&\cup
\big\{v_{ij},v_{ji}\;\big|\;1\le i<j\le m \hspace{2mm} \text{with} \hspace{2mm}a_{ij}=0\big\}
\end{aligned}
\end{equation}
form a basis of eigenvectors of $A$, showing that $A$ is diagonalizable, thus completing the proof.
\eprf

\br
\label{rmk:DSOTRMK}
Although \eqref{eq:eigvEA} and \eqref{eq:Aijbasis} are indeed bases, they do not necessarily form orthogonal bases since $A$ in Lemma~\ref{LXSTAR77} need not be self-adjoint. Nevertheless, linear independence of \eqref{eq:eigvEA} and \eqref{eq:Aijbasis} guarantees that the matrix $P$ of eigenvectors is invertible so that $A$ can be diagonalized via $A=P \Lambda P^{-1}$, with $\Lambda$ the corresponding diagonal matrix of eigenvalues. 
\er

\blem\label{PXBLOOM17}
Let $\sigma=\sum_{i}^m\sigma_{i}E_{ii}$ be a diagonal density matrix in $\M_m$. Then 
\be \label{DIAGX77}
\pbloom{p}(\sigma,\id)=\sum_{i,j}^mE_{ij}\otimes \left(\sigma_{i}^{p}\sigma_{j}^{1-p}\right)E_{ji}
\ee
for all $p\in [0,1]$.
\elem

\bprf
Let $p\in [0,1]$. Since $\sigma$ is diagonal, it follows that $\sigma^{p}E_{ij}\sigma^{1-p}=\left(\sigma_{i}^{p}\sigma_{j}^{1-p}\right)E_{ij}$. We then have 
\begin{eqnarray*}
\pbloom{p}(\sigma,\id)&=&\left(\sigma^{p}\otimes \mathds{1}\right)\mathscr{J}[\id](\sigma^{1-p}\otimes \mathds{1})=\left(\sigma^{p}\otimes \mathds{1}\right)\left(\sum_{i,j}^mE_{ij}\otimes E_{ji}\right)(\sigma^{1-p}\otimes \mathds{1}) \\
&=&\sum_{i,j}^m\left(\sigma^{p}E_{ij}\sigma^{1-p}\right)\otimes E_{ji}=\sum_{i,j}^m\left(\sigma_{i}^{p}\sigma_{j}^{1-p}\right)E_{ij}\otimes E_{ji} 
=\sum_{i,j}^mE_{ij}\otimes \left(\sigma_{i}^{p}\sigma_{j}^{1-p}\right)E_{ji}, \\
\end{eqnarray*}
as desired.
\eprf

\blem \label{PHIADXU81}
Let $\rho\in \mathcal{S}(\M_m)$, and let $V\in \M_m$ be a unitary matrix such that $\rho=V\rho_dV^{\dag}$ is a unitary diagonalization, with $\rho_{d}$ diagonal. Then 
\be \label{DGXRSXP87}
\pbloom{p}(\rho,\Ad_U)=\Ad_{V\otimes UV}\big(\pbloom{p}(\rho_d,\id)\big)
\ee
for every unitary matrix $U\in \M_m$ and for all $p\in [0,1]$.
\elem

\bprf
Indeed,
\begingroup
\allowdisplaybreaks
\begin{eqnarray*}
\pbloom{p}(\rho,\Ad_U)&
=&(\rho^p\otimes \mathds{1})\left(\sum_{i,j}^mE_{ij}\otimes UE_{ji}U^{\dag}\right)(\rho^{1-p}\otimes \mathds{1}) \\
&=&\big(V\rho_{d}^{p}\otimes U\big)\left(\sum_{i,j}^{m}V^{\dag}E_{ij}V\otimes E_{ji}\right)\big(\rho_{d}^{1-p}V^{\dag}\otimes U^{\dag}\big)\\
&\overset{\eqref{ESX17}}=&\big(V\rho_{d}^{p}\otimes U\big)\left(\sum_{i,j}^{m}E_{ij}\otimes VE_{ji}V^{\dag}\right)\big(\rho_{d}^{1-p}V^{\dag}\otimes U^{\dag}\big)\\
&=&\big(V\otimes UV\big)\left(\sum_{i,j}^{m}\rho_{d}^{p}E_{ij}\rho_{d}^{1-p}\otimes E_{ji}\right)\big(V^{\dag}\otimes V^{\dag}U^{\dag}\big)\\
&=&\Ad_{V\otimes UV}\big(\pbloom{p}(\rho_d,\id)\big),
\end{eqnarray*}
\endgroup
as desired.
\eprf

\bn\label{LTZX81}
Let $\mathcal{E}:\M_p\otimes \M_n\to \M_n$ be the partial trace, let $\mu:\M_n\otimes \M_n\to \M_n$ be the multiplication map, and let $\rho=\sum_{k,l}^{p}E_{kl}^{(p)}\otimes \rho_{kl}\in \M_p\otimes \M_n$. Then 
\be \label{LTX87i7}
\rbloom(\rho,\mathcal{E})=\sum_{k,l}^pE_{kl}^{(p)}\otimes \mu^*(\rho_{kl})=\left(
\begin{array}{cccc}
\mu^*(\rho_{11}) & \cdots & \mu^*(\rho_{1p}) \\
\vdots & \ddots & \vdots \\
\mu^*(\rho_{p1}) & \cdots & \mu^*(\rho_{pp}) \\
\end{array}
\right).
\ee
In particular, if $\rho=\tau\otimes \sigma \in \M_p\otimes \M_n$, then 
\be \label{BLOOMSTRqX81}
\rbloom(\rho,\mathcal{E})=\tau\otimes \mu^*(\sigma),
\ee
and if we further assume that $\tau$ is self-adjoint, then
\be \label{BLOOMSTRqX83}
\sbloom(\rho,\mathcal{E})=\tau \otimes \sbloom(\sigma,\id).
\ee
\en

\bprf
Let $i=\mathcal{E}^*$, so that $i(B)=\mathds{1}\otimes B$ for all $B\in \M_n$. Then
\[
\rho \hspace{0.5mm} i(E_{ij}^{(n)})
=\left(\sum_{k,l}^{p}E_{kl}^{(p)}\otimes \rho_{kl}\right)(\mathds{1}\otimes E_{ij}^{(n)})
=\sum_{k,l}^pE_{kl}^{(p)}\otimes \rho_{kl}E_{ij}^{(n)}.
\]
Hence, 
\begin{eqnarray*}
\rbloom(\rho,\mathcal{E})&=&(\rho\otimes \mathds{1})\mathscr{J}[\mathcal{E}]\overset{\eqref{ANXNA111}}=\sum_{i,j}^n\rho \hspace{0.5mm} i(E_{ij}^{(n)})\otimes E_{ji}^{(n)} =\sum_{i,j}^n\left(\sum_{k,l}^pE_{kl}^{(p)}\otimes \rho_{kl}E_{ij}^{(n)}\right)\otimes E_{ji}^{(n)} \\
&=&=\sum_{k,l}^pE_{kl}^{(p)}\otimes \sum_{i,j}^{n}\rho_{kl}E_{ij}^{(n)}\otimes E_{ji}^{(n)}
\overset{\eqref{MXMA17}}=\sum_{k,l}^pE_{kl}^{(p)}\otimes \mu^*(\rho_{kl}),
\end{eqnarray*}
where the first equality follows from item \ref{PNXNP501} of Lemma~\ref{PNXNP1999}. This proves~\eqref{LTX87i7}.  

Now suppose $\rho=\tau\otimes \sigma$, let $E_{kl}$ denote the matrix units in $\M_p$, and let $\tau_{kl}\in\C$ be such that $\tau=\sum_{k,l}^{p}\tau_{kl}E_{kl}$. Then
\[
\rho=\tau\otimes \sigma=\left(\sum_{k,l}^{p}\tau_{kl}E_{kl}\right)\otimes \sigma=\sum_{k,l}^{p}E_{kl}\otimes \left(\tau_{kl}\sigma\right).
\]
By equation \eqref{LTX87i7}, we then have
\[
\rbloom(\rho,\mathcal{E})
=\sum_{k,l}^{p}E_{kl}\otimes \mu^*(\tau_{kl}\sigma)
=\sum_{k,l}^{p}\tau_{kl}E_{kl}\otimes \mu^*(\sigma)
=\tau\otimes \mu^*(\sigma).
\]
Thus, equation \eqref{BLOOMSTRqX81} holds. Now assume that $\tau$ is self-adjoint. Then 
\begin{eqnarray*}
\sbloom(\rho,\mathcal{E})&=&\frac{1}{2}\left(\rbloom(\rho,\mathcal{E})+\rbloom(\rho,\mathcal{E})^{\dag}\right)\overset{\eqref{LTX87i7}}=\frac{1}{2}\left(\tau\otimes \mu^*(\sigma)+(\tau\otimes \mu^*(\sigma))^{\dag}\right) \\
&=&\tau \otimes \left(\frac{1}{2}\left(\mu^*(\sigma)+\mu^*(\sigma)^{\dag}\right)\right) 
=\tau \otimes \sbloom(\sigma,\id),
\end{eqnarray*}
where the first equation follows from item \ref{PNXNP5} of Proposition~\ref{PNXNP1999} and the final equality follows from the fact that $\mu^*(\sigma)=\psi_1(\sigma,\id)$. Thus, equation \eqref{BLOOMSTRqX83} holds as well.
\eprf

For the next lemma, we make use of the following definition.

\bd
Let $(\VA,\VB)$ be a pair of matrix algebras, and let $\mathcal{E}\in \text{Hom}(\VA,\VB)$. The \define{right bloom} of $\mathcal{E}$ is the map $\bloom_{\mathcal{E}}\in \text{Hom}(\VA,\VA\otimes \VB)$ given by
\[
\bloom_{\mathcal{E}}=\left(\id\otimes \mathcal{E}\right)\circ \mu_{\VA}^*.
\]
\ed

\br
We note that it folows immediately from the definition of right bloom that $\rbloom(\rho,\mathcal{E})=\bloom_{\mathcal{E}}(\rho)$ for all processes $(\rho,\mathcal{E})$.
\er

\blem\label{PMZ87}
Let $\mathcal{E}:\VA\to\VB$ be a linear map and let $\mathcal{F}:\VA\to\VC$ be a $*$-isomorphism, where $\VA,\VB,\VC$ are multi-matrix algebras. Then $(\mathcal{F}\otimes\id_{\VB})\circ\bloom_{\mathcal{E}}=\bloom_{\mathcal{E}\circ\mathcal{F}^{-1}}\circ\mathcal{F}$ and $\bloom_{\mathcal{E}\circ\mathcal{F}^{-1}}=(\mathcal{F}\otimes\id_{\VB})\circ\bloom_{\mathcal{E}}\circ\mathcal{F}^{-1}$.
In particular, $\Ad_{U\otimes\mathds{1}_{\VB}}\circ\bloom_{\mathcal{E}}=\bloom_{\mathcal{E}\circ\Ad_{U^{\dag}}}\circ\Ad_{U}$ and $\bloom_{\mathcal{E}\circ\Ad_{U^{\dag}}}=\Ad_{U\otimes\mathds{1}_{\VB}}\circ\bloom_{\mathcal{E}}\circ\Ad_{U^{\dag}}$ whenever $U\in\VA$ is unitary.
\elem

\bprf
Set $E=\mathcal{E}^*$ and $F=\mathcal{F}^*$ to be the Hilbert--Schmidt adjoints of $\mathcal{E}$ and $\mathcal{F}$, respectively. Then the first claimed equation is equivalent to $\mu_{\VA}\circ(\id_{\VA}\otimes E)\circ(F\otimes\id_{\VB})=F\circ\mu_{\VC}\circ(\id_{\VC}\otimes(F^{-1}\circ E))$. This identity follows from 
\[
F\circ\mu_{\VC}\circ(\id_{\VC}\otimes(F^{-1}\circ E))
=\mu_{\VA}\circ(F\otimes F)\circ(\id_{\VC}\otimes(F^{-1}\circ E))
=\mu_{\VA}\circ(F\otimes E)
=\mu_{\VA}\circ(\id_{\VA}\otimes E)\circ(F\otimes\id_{\VB})
\]
by the interchange law for $\otimes$ and $\circ$ and because $F$ is a $*$-homomorphism. The second claimed equality follows by precomposing the first equality with $\mathcal{F}^{-1}$. 
\eprf

\blem\label{ENTXSTAR1971}
Let $U\in \M_p\otimes \M_n$ be unitary, let $\mathcal{E}:\M_p\otimes \M_n\to \M_n$ be the map given by $\mathcal{E}=\emph{tr}_{\M_p}\circ \Ad_U$, and let $\rho=U^{\dag}(\tau \otimes \sigma)U$ for some density matrices $\tau\in \M_p$ and $\sigma\in \M_n$. Then 
\be \label{BNTX81}
S\big(\rbloom(\rho,\mathcal{E})\big)=S\big(\sbloom(\rho,\mathcal{E})\big)=S(\tau)+S(\sigma).
\ee 
\elem

\bprf
Set $\mathcal{G}=\tr_{\M_p}$, let $\mu:\M_n\to \M_n$ denote the multiplication map, and denote the identity map on $\M_n$ by $\id$. Since $\mathcal{E}=\mathcal{G}\circ \Ad_U$, we have $\bloom_{\mathcal{E}}(\rho)=\Ad_{U^{\dag}\otimes \mathds{1}}(\bloom_{\mathcal{G}}(U\rho U^{\dag}))$ by 
Lemma~\ref{PMZ87}. Thus,
\begin{eqnarray*}
\rbloom(\rho,\mathcal{E})&=&\bloom_{\mathcal{E}}(\rho)=\Ad_{U^{\dag}\otimes \mathds{1}}(\bloom_{\mathcal{G}}(U\rho U^{\dag}))=\Ad_{U^{\dag}\otimes \mathds{1}}(\bloom_{\mathcal{G}}(\tau\otimes \sigma))\\
&=&\Ad_{U^{\dag}\otimes \mathds{1}}(\psi_1(\tau\otimes \sigma,\mathcal{G}))\overset{\eqref{BLOOMSTRqX81}}=\Ad_{U^{\dag}\otimes \mathds{1}}(\tau\otimes \mu^*(\sigma)).
\end{eqnarray*}
By the isometric invariance of the entropy function, we then have
\[
S\big(\rbloom(\rho,\mathcal{E})\big)=S\left(\tau\otimes \mu^*(\sigma)\right) 
=S(\tau)+S(\mu^*(\sigma))=S(\tau)+S(\sigma),
\]
where the final equality follows from the fact that $\mu^*(\sigma)=\rbloom(\sigma,\id)$, so that $S(\mu^*(\sigma))=S(\rbloom(\sigma,\id))=S(\sigma)$ by item \ref{UNTEVX1} of Theorem~\ref{EnTxSTAR97}. Similarly, we have
\begin{eqnarray*}
\sbloom(\rho,\mathcal{E})&=&\frac{1}{2}\left(\rbloom(\rho,\mathcal{E})+\rbloom(\rho,\mathcal{E})^{\dag}\right) \\
&=&\frac{1}{2}\left(\Ad_{U^{\dag}\otimes \mathds{1}}(\tau\otimes \mu^*(\sigma))+\Ad_{U^{\dag}\otimes \mathds{1}}(\tau\otimes \mu^*(\sigma))^{\dag}\right) \\
&=&\Ad_{U^{\dag}\otimes \mathds{1}}\left(\frac{1}{2}\left(\tau\otimes \mu^*(\sigma))+(\tau\otimes \mu^*(\sigma))^{\dag}\right)\right) \\
&=&\Ad_{U^{\dag}\otimes \mathds{1}}\left(\tau\otimes \frac{1}{2}\left(\mu^*(\sigma)+\mu^*(\sigma)^{\dag}\right)\right) \\
&=&\Ad_{U\otimes \mathds{1}}\left(\tau\otimes \sbloom(\sigma,\id)\right). 
\end{eqnarray*}
Thus, by the isometric invariance of the entropy function, we have
\[
S(\sbloom(\rho,\mathcal{E}))=S(\tau \otimes \sbloom(\sigma,\id))=S(\tau)+S(\sbloom(\sigma,\id))=S(\tau)+S(\sigma),
\]
where the final equality follows from item \ref{UNTEVX1} of Theorem~\ref{EnTxSTAR97}, thus concluding the proof.
\eprf

\addcontentsline{toc}{section}{\numberline{}Bibliography}
\bibliographystyle{plain}
\bibliography{arXiv-references}

\begin{bibdiv}
\begin{biblist}

\bib{AdCe97}{article}{
      author={Adami, Christoph},
      author={Cerf, Nicolas~J.},
       title={{von Neumann capacity of noisy quantum channels}},
        date={1997},
     journal={Phys. Rev. A},
      volume={56},
       pages={3470\ndash 3483},
      eprint={quant-ph/9609024},
         url={https://link.aps.org/doi/10.1103/PhysRevA.56.3470},
}

\bib{ABL64}{article}{
      author={Aharonov, Yakir},
      author={Bergmann, Peter~G.},
      author={Lebowitz, Joel~L.},
       title={Time symmetry in the quantum process of measurement},
        date={1964},
     journal={Phys. Rev.},
      volume={134},
       pages={B1410\ndash B1416},
         url={https://link.aps.org/doi/10.1103/PhysRev.134.B1410},
}

\bib{BaFr14}{article}{
      author={Baez, John~C.},
      author={Fritz, Tobias},
       title={A {B}ayesian characterization of relative entropy},
        date={2014},
     journal={Theory Appl. Categ.},
      volume={29},
      number={16},
       pages={422\ndash 457},
      eprint={1402.3067},
}

\bib{BFL11}{article}{
      author={Baez, John~C.},
      author={Fritz, Tobias},
      author={Leinster, Tom},
       title={A characterization of entropy in terms of information loss},
        date={2011},
     journal={Entropy},
      volume={13},
       pages={1945\ndash 1957},
      eprint={1106.1791},
}

\bib{Bo51}{book}{
      author={Bohm, David},
       title={{Quantum theory}},
   publisher={Prentice-Hall},
     address={Englewood Cliffs, {NJ}},
        date={1951},
        note={Also as reprint ed.: New York, NY, Dover Publications, 1989},
}

\bib{BLMS87}{article}{
      author={Bombelli, Luca},
      author={Lee, Joohan},
      author={Meyer, David},
      author={Sorkin, Rafael~D.},
       title={Space-time as a causal set},
        date={1987},
     journal={Phys. Rev. Lett.},
      volume={59},
       pages={521\ndash 524},
         url={https://link.aps.org/doi/10.1103/PhysRevLett.59.521},
}

\bib{Br21}{article}{
      author={Bradley, Tai-Danae},
       title={Entropy as a topological operad derivation},
        date={2021},
        ISSN={1099-4300},
     journal={Entropy},
      volume={23},
      number={9},
      eprint={2107.09581},
         url={https://doi.org/10.3390/e23091195},
}

\bib{BGWG21}{misc}{
      author={{Brandsen}, Sarah},
      author={{Geng}, Isabelle~J.},
      author={{Wilde}, Mark~M.},
      author={{Gour}, Gilad},
       title={{Quantum conditional entropy from information-theoretic
  principles}},
        date={2021},
        note={Preprint available at
  \href{https://arxiv.org/abs/2110.15330}{arXiv:2110.15330 [quant-ph]}},
}

\bib{CeAd97}{incollection}{
      author={Cerf, Nicolas~J.},
      author={Adami, Chris},
       title={Negative entropy in quantum information theory},
        date={1997},
   booktitle={New developments on fundamental problems in quantum physics
  ({O}viedo, 1996)},
      series={Fund. Theories Phys.},
      volume={81},
   publisher={Kluwer Acad. Publ., Dordrecht},
       pages={77\ndash 84},
         url={https://doi.org/10.1007/978-94-011-5886-2_11},
}

\bib{Ch75}{article}{
      author={Choi, Man~Duen},
       title={Completely positive linear maps on complex matrices},
        date={1975},
     journal={Linear Algebra Appl.},
      volume={10},
       pages={285\ndash 290},
         url={https://doi.org/10.1016/0024-3795(75)90075-0},
}

\bib{ChMa20}{article}{
      author={Chru{\'s}ci{\'n}ski, Dariusz},
      author={Matsuoka, Takashi},
       title={Quantum conditional probability and measurement induced
  disturbance of a quantum channel},
        date={2020},
        ISSN={0034-4877},
     journal={Rep. Math. Phys.},
      volume={86},
      number={1},
       pages={115\ndash 128},
         url={https://doi.org/10.1016/S0034-4877(20)30060-4},
}

\bib{CiKu23}{article}{
      author={Cipolloni, Giorgio},
      author={Kudler-Flam, Jonah},
       title={{Entanglement Entropy of Non-Hermitian Eigenstates and the
  Ginibre Ensemble}},
        date={2023Jan},
     journal={Phys. Rev. Lett.},
      volume={130},
       pages={010401},
      eprint={2206.12438},
         url={https://link.aps.org/doi/10.1103/PhysRevLett.130.010401},
}

\bib{CJS17}{article}{
      author={Couvreur, Romain},
      author={Jacobsen, Jesper~Lykke},
      author={Saleur, Hubert},
       title={Entanglement in nonunitary quantum critical spin chains},
        date={2017Jul},
     journal={Phys. Rev. Lett.},
      volume={119},
       pages={040601},
      eprint={1611.08506},
         url={https://link.aps.org/doi/10.1103/PhysRevLett.119.040601},
}

\bib{Cs08}{article}{
      author={Csisz{\'a}r, Imre},
       title={Axiomatic characterizations of information measures},
        date={2008},
     journal={Entropy},
      volume={10},
      number={3},
       pages={261\ndash 273},
         url={https://doi.org/10.3390/e10030261},
}

\bib{dePi67}{article}{
      author={de~Pillis, John},
       title={Linear transformations which preserve hermitian and positive
  semidefinite operators},
        date={1967},
     journal={Pac. J. Math.},
      volume={23},
      number={1},
       pages={129\ndash 137},
         url={http://dx.doi.org/10.2140/pjm.1967.23.129},
}

\bib{dRARDV11}{article}{
      author={del Rio, L{\'{\i}}dia},
      author={{\AA}berg, Johan},
      author={Renner, Renato},
      author={Dahlsten, Oscar},
      author={Vedral, Vlatko},
       title={The thermodynamic meaning of negative entropy},
        date={2011},
     journal={Nature},
      volume={474},
      number={7349},
       pages={61\ndash 63},
      eprint={1009.1630},
         url={https://doi.org/10.1038/nature10123},
}

\bib{De05}{article}{
      author={{Devetak}, Igor},
       title={The private classical capacity and quantum capacity of a quantum
  channel},
        date={2005},
     journal={IEEE Trans. Inf. Theory},
      volume={51},
      number={1},
       pages={44\ndash 55},
      eprint={quant-ph/0304127},
         url={https://doi.org/10.1109/TIT.2004.839515},
}

\bib{Di45}{article}{
      author={Dirac, P. A.~M.},
       title={On the analogy between classical and quantum mechanics},
        date={1945},
     journal={Rev. Mod. Phys.},
      volume={17},
       pages={195\ndash 199},
         url={https://link.aps.org/doi/10.1103/RevModPhys.17.195},
}

\bib{DHMTT22}{article}{
      author={Doi, Kazuki},
      author={Harper, Jonathan},
      author={Mollabashi, Ali},
      author={Takayanagi, Tadashi},
      author={Taki, Yusuke},
       title={Pseudoentropy in $\mathrm{dS}/\mathrm{CFT}$ and timelike
  entanglement entropy},
        date={2023},
     journal={Phys. Rev. Lett.},
      volume={130},
       pages={031601},
      eprint={2210.09457},
         url={https://link.aps.org/doi/10.1103/PhysRevLett.130.031601},
}

\bib{EPR}{article}{
      author={Einstein, Albert},
      author={Podolsky, Boris},
      author={Rosen, Nathan},
       title={Can quantum-mechanical description of physical reality be
  considered complete?},
        date={1935},
     journal={Phys. Rev.},
      volume={47},
       pages={777\ndash 780},
         url={https://link.aps.org/doi/10.1103/PhysRev.47.777},
}

\bib{ECP10}{article}{
      author={Eisert, Jens},
      author={Cramer, Marcus},
      author={Plenio, Martin~B.},
       title={Colloquium: Area laws for the entanglement entropy},
        date={2010},
     journal={Rev. Mod. Phys.},
      volume={82},
       pages={277\ndash 306},
      eprint={0808.3773},
         url={https://link.aps.org/doi/10.1103/RevModPhys.82.277},
}

\bib{Fa01}{book}{
      author={Farenick, Douglas~R.},
       title={Algebras of linear transformations},
      series={Universitext},
   publisher={Springer-Verlag, New York},
        date={2001},
        ISBN={0-387-95062-1},
         url={https://doi.org/10.1007/978-1-4613-0097-7},
}

\bib{Fe87}{incollection}{
      author={Feynman, Richard~P},
       title={Negative probability},
        date={1987},
   booktitle={Quantum implications: essays in honour of {D}avid {B}ohm},
   publisher={Routledge},
       pages={235\ndash 248},
         url={https://doi.org/10.4324/9780203392799},
}

\bib{FJV15}{article}{
      author={Fitzsimons, Joseph~F.},
      author={Jones, Jonathan~A.},
      author={Vedral, Vlatko},
       title={Quantum correlations which imply causation},
        date={2015},
     journal={Sci. Rep.},
      volume={5},
      number={1},
       pages={18281},
      eprint={1302.2731},
         url={https://doi.org/10.1038/srep18281},
}

\bib{Fo07}{book}{
      author={Folland, Gerald~B.},
       title={Real analysis: Modern techniques and their applications},
     edition={2},
   publisher={Wiley},
        date={2007},
}

\bib{FrCa22}{misc}{
      author={{Frembs}, Markus},
      author={{Cavalcanti}, Eric~G.},
       title={{Variations on the Choi--Jamiolkowski isomorphism}},
        date={2022},
        note={Preprint available at
  \href{https://arxiv.org/abs/2211.16533}{arXiv:2211.16533 [quant-ph]}},
}

\bib{Fu21AXM}{article}{
      author={Fullwood, James},
       title={An axiomatic characterization of mutual information},
        date={2023},
        ISSN={1099-4300},
     journal={Entropy},
      volume={25},
      number={4},
      eprint={2108.12647},
         url={https://www.mdpi.com/1099-4300/25/4/663},
}

\bib{fullwood2023}{misc}{
      author={Fullwood, James},
       title={Quantum dynamics as a pseudo-density matrix},
        date={2023},
        note={Preprint available at
  \href{https://arxiv.org/pdf/2304.03954.pdf}{arXiv:2304.03954 [quant-ph]}},
}

\bib{FuPa21}{article}{
      author={Fullwood, James},
      author={Parzygnat, Arthur~J.},
       title={The information loss of a stochastic map},
        date={2021},
        ISSN={1099-4300},
     journal={Entropy},
      volume={23},
      number={8},
      eprint={2107.01975},
         url={https://www.mdpi.com/1099-4300/23/8/1021},
}

\bib{FuPa22}{article}{
      author={Fullwood, James},
      author={Parzygnat, Arthur~J.},
       title={On quantum states over time},
        date={2022},
     journal={Proc. R. Soc. A},
      volume={478},
      eprint={2202.03607},
         url={https://doi.org/10.1098/rspa.2022.0104},
}

\bib{FuJa13}{article}{
      author={Furber, Robert},
      author={Jacobs, Bart},
       title={From {K}leisli categories to commutative {$C^*$}-algebras:
  probabilistic {G}elfand duality},
        date={2015},
     journal={Log. Methods Comput. Sci.},
      volume={11},
       pages={1\ndash 28},
      eprint={1303.1115},
}

\bib{GPRR21}{article}{
      author={Giorgetti, Luca},
      author={Parzygnat, Arthur~J.},
      author={Ranallo, Alessio},
      author={Russo, Benjamin~P.},
       title={Bayesian inversion and the {T}omita--{T}akesaki modular group},
        date={202303},
        ISSN={0033-5606},
     journal={Q. J. Math.},
      volume={00},
      eprint={2112.03129},
         url={https://doi.org/10.1093/qmath/haad014},
        note={haad014},
}

\bib{GdHJ89}{book}{
      author={Goodman, Frederick~M.},
      author={de~la Harpe, Pierre},
      author={Jones, Vaughan F.~R.},
       title={{Coxeter graphs and towers of algebras}},
      series={{Mathematical Sciences Research Institute Publications}},
   publisher={Springer-Verlag},
     address={New York},
        date={1989},
      volume={14},
        ISBN={0-387-96979-9},
         url={http://dx.doi.org/10.1007/978-1-4613-9641-3},
}

\bib{GoWi21}{article}{
      author={Gour, Gilad},
      author={Wilde, Mark~M.},
       title={Entropy of a quantum channel},
        date={2021},
     journal={Phys. Rev. Res.},
      volume={3},
       pages={023096},
      eprint={1808.06980},
         url={https://link.aps.org/doi/10.1103/PhysRevResearch.3.023096},
}

\bib{HMS03}{article}{
      author={Hawkins, Eli},
      author={Markopoulou, Fotini},
      author={Sahlmann, Hanno},
       title={Evolution in quantum causal histories},
        date={2003},
     journal={Class. Quantum Gravity},
      volume={20},
      number={16},
       pages={3839\ndash 3854},
      eprint={hep-th/0302111},
         url={https://dx.doi.org/10.1088/0264-9381/20/16/320},
}

\bib{Hi08}{book}{
      author={Higham, Nicholas~J.},
       title={Functions of matrices: theory and computation},
   publisher={Society for Industrial and Applied Mathematics},
     address={Philadelphia},
        date={2008},
         url={https://doi.org/10.1137/1.9780898717778},
}

\bib{HorOpp05}{article}{
      author={Horodecki, Micha\l},
      author={Oppenheim, Jonathan},
      author={Winter, Andreas},
       title={Quantum state merging and negative information},
        date={2007},
        ISSN={0010-3616},
     journal={Comm. Math. Phys.},
      volume={269},
      number={1},
       pages={107\ndash 136},
      eprint={quant-ph/0512247},
         url={https://doi.org/10.1007/s00220-006-0118-x},
}

\bib{HHPBS17}{article}{
      author={{Horsman}, Dominic},
      author={{Heunen}, Chris},
      author={{Pusey}, Matthew~F.},
      author={{Barrett}, Jonathan},
      author={{Spekkens}, Robert~W.},
       title={{Can a quantum state over time resemble a quantum state at a
  single time?}},
        date={2017},
     journal={Proc. R. Soc. A},
      volume={473},
      number={2205},
       pages={20170395},
      eprint={1607.03637},
}

\bib{Mathematica130}{misc}{
      author={Inc., Wolfram~Research{,}},
       title={Mathematica, {V}ersion 13.0},
         url={https://www.wolfram.com/mathematica},
        note={Champaign, IL, 2021},
}

\bib{IKMT22}{article}{
      author={Ishiyama, Yutaka},
      author={Kojima, Riku},
      author={Matsui, Sho},
      author={Tamaoka, Kotaro},
       title={{Notes on pseudo entropy amplification}},
        date={2022},
        ISSN={2050-3911},
     journal={Prog. Theor. Exp. Phys},
      volume={2022},
      number={9},
       pages={093B10},
      eprint={2206.14551},
         url={https://doi.org/10.1093/ptep/ptac112},
}

\bib{IMPTLRG15}{article}{
      author={Islam, Rajibul},
      author={Ma, Ruichao},
      author={Preiss, Philipp~M.},
      author={Tai, M.~Eric},
      author={Lukin, Alexander},
      author={Rispoli, Matthew},
      author={Greiner, Markus},
       title={Measuring entanglement entropy in a quantum many-body system},
        date={2015},
     journal={Nature},
      volume={528},
      number={7580},
       pages={77\ndash 83},
      eprint={1509.01160},
         url={https://doi.org/10.1038/nature15750},
}

\bib{Jam72}{article}{
      author={Jamio{\l}kowski, Andrzej},
       title={Linear transformations which preserve trace and positive
  semidefiniteness of operators},
        date={1972},
        ISSN={0034-4877},
     journal={Rep. Mathematical Phys.},
      volume={3},
      number={4},
       pages={275\ndash 278},
         url={https://doi.org/10.1016/0034-4877(72)90011-0},
}

\bib{JSK23}{misc}{
      author={Jia, Zhian},
      author={Song, Minjeong},
      author={Kaszlikowski, Dagomir},
       title={Quantum space-time marginal problem: global causal structure from
  local causal information},
        date={2023},
        note={Preprint available at
  \href{https://arxiv.org/pdf/2303.12819.pdf}{arXiv:2303.10544 [quant-ph]}},
}

\bib{KaBr19}{article}{
      author={Kato, Kohtaro},
      author={Brand{\~{a}}o, Fernando G. S.~L.},
       title={Quantum approximate {M}arkov chains are thermal},
        date={2019},
     journal={Comm. Math. Phys.},
      volume={370},
      number={1},
       pages={117\ndash 149},
      eprint={1609.06636},
         url={https://doi.org/10.1007/s00220-019-03485-6},
}

\bib{Kh57}{book}{
      author={Khinchin, Aleksandr~I.},
  translator={Silverman, R.A.},
  translator={Friedman, M.D.},
       title={Mathematical foundations of information theory},
   publisher={Dover Publications, Inc.},
        date={1957},
}

\bib{Ki33}{article}{
      author={Kirkwood, John~G.},
       title={Quantum statistics of almost classical assemblies},
        date={1933},
     journal={Phys. Rev.},
      volume={44},
       pages={31\ndash 37},
         url={https://link.aps.org/doi/10.1103/PhysRev.44.31},
}

\bib{Kr83}{book}{
      author={Kraus, Karl},
       title={States, effects, and operations: Fundamental notions of quantum
  theory},
      series={Lecture Notes in Physics Vol. 190},
   publisher={Springer Berlin Heidelberg},
        date={1983},
}

\bib{KNPPZ21}{article}{
      author={{Kukulski}, Ryszard},
      author={{Nechita}, Ion},
      author={{Pawela}, {\L}ukasz},
      author={{Pucha{\l}a}, Zbigniew},
      author={{{\.Z}yczkowski}, Karol},
       title={{Generating random quantum channels}},
        date={2021},
     journal={J. Math. Phys},
      volume={62},
      number={6},
       pages={062201},
      eprint={2011.02994},
         url={https://doi.org/10.1063/5.0038838},
}

\bib{KPP21}{article}{
      author={{Kurzyk}, Dariusz},
      author={{Pawela}, {\L}ukasz},
      author={{Pucha{\l}a}, Zbigniew},
       title={{Relating Entropies of Quantum Channels}},
        date={2021},
     journal={Entropy},
      volume={23},
      number={8},
       pages={1028},
      eprint={2106.09629},
         url={https://doi.org/10.3390/e23081028},
}

\bib{Le06}{article}{
      author={Leifer, Matthew~S.},
       title={Quantum dynamics as an analog of conditional probability},
        date={2006},
     journal={Phys. Rev. A},
      volume={74},
       pages={042310},
      eprint={0606022},
         url={https://link.aps.org/doi/10.1103/PhysRevA.74.042310},
}

\bib{Le07}{inproceedings}{
      author={{Leifer}, Matthew~S.},
       title={{Conditional Density Operators and the Subjectivity of Quantum
  Operations}},
        date={2007},
   booktitle={Foundations of probability and physics - 4},
      editor={{Adenier}, Guillaume},
      editor={{Fuchs}, Chrisopher},
      editor={{Khrennikov}, Andrei~Yu},
      series={American Institute of Physics Conference Series},
      volume={889},
       pages={172\ndash 186},
}

\bib{LeSp13}{article}{
      author={Leifer, Matthew~S.},
      author={Spekkens, Robert~W.},
       title={Towards a formulation of quantum theory as a causally neutral
  theory of {B}ayesian inference},
        date={2013},
     journal={Phys. Rev. A},
      volume={88},
       pages={052130},
      eprint={1107.5849},
         url={https://link.aps.org/doi/10.1103/PhysRevA.88.052130},
}

\bib{Leinster21}{book}{
      author={Leinster, Tom},
       title={Entropy and diversity: The axiomatic approach},
   publisher={Cambridge University Press},
        date={2021},
}

\bib{Li91}{inproceedings}{
      author={Lindblad, G{\"o}ran},
       title={Quantum entropy and quantum measurements},
        date={1991},
   booktitle={Quantum aspects of optical communications: Proceedings of a
  workshop held at the cnrs, paris, france 26--28 november 1990},
   publisher={Springer, Berlin, Heidelberg},
       pages={71\ndash 80},
         url={https://doi.org/10.1007/3-540-53862-3_168},
}

\bib{LQDV}{misc}{
      author={Liu, Xiangjing},
      author={Qiu, Yixian},
      author={Dahlsten, Oscar},
      author={Vedral, Vlatko},
       title={{Q}uantum {C}ausal {I}nference with {E}xtremely {L}ight {T}ouch},
        date={2023},
        note={Preprint available at
  \href{https://arxiv.org/pdf/2303.10544.pdf}{arXiv:2303.10544 [quant-ph]}},
}

\bib{Ll97}{article}{
      author={Lloyd, Seth},
       title={Capacity of the noisy quantum channel},
        date={1997},
     journal={Phys. Rev. A},
      volume={55},
       pages={1613\ndash 1622},
      eprint={quant-ph/9604015},
         url={https://link.aps.org/doi/10.1103/PhysRevA.55.1613},
}

\bib{LRSTKCKLG19}{article}{
      author={{Lukin}, Alexander},
      author={{Rispoli}, Matthew},
      author={{Schittko}, Robert},
      author={{Tai}, M.~Eric},
      author={{Kaufman}, Adam~M.},
      author={{Choi}, Soonwon},
      author={{Khemani}, Vedika},
      author={{L{\'e}onard}, Julian},
      author={{Greiner}, Markus},
       title={{Probing entanglement in a many-body-localized system}},
        date={2019},
     journal={Science},
      volume={364},
      number={6437},
       pages={256\ndash 260},
      eprint={1805.09819},
         url={https://www.science.org/doi/abs/10.1126/science.aau0818},
}

\bib{MaCh22}{inproceedings}{
      author={Matsuoka, Takashi},
      author={Chru{\'s}ci{\'n}ski, Dariusz},
       title={Compound state, its conditionality and quantum mutual
  information},
organization={Springer, Cham},
        date={2022},
   booktitle={International conference on quantum probability \& related
  topics},
       pages={135\ndash 150},
         url={https://doi.org/10.1007/978-3-031-06170-7_7},
}

\bib{Mo1920}{article}{
      author={Moore, Eliakim~H.},
       title={On the reciprocal of the general algebraic matrix},
        date={1920},
     journal={Bull. Am. Math. Soc.},
      volume={26},
      number={9},
       pages={394\ndash 395},
}

\bib{NTTTW21}{article}{
      author={Nakata, Yoshifumi},
      author={Takayanagi, Tadashi},
      author={Taki, Yusuke},
      author={Tamaoka, Kotaro},
      author={Wei, Zixia},
       title={New holographic generalization of entanglement entropy},
        date={2021},
     journal={Phys. Rev. D},
      volume={103},
       pages={026005},
      eprint={2005.13801},
         url={https://link.aps.org/doi/10.1103/PhysRevD.103.026005},
}

\bib{NiCh11}{book}{
      author={Nielsen, Michael~A.},
      author={Chuang, Isaac~L.},
       title={Quantum computation and quantum information},
     edition={10th Anniversary},
   publisher={Cambridge University Press, Cambridge},
        date={2011},
        ISBN={0-521-63235-8; 0-521-63503-9},
}

\bib{TTT21}{article}{
      author={Nishioka, Tatsuma},
      author={Takayanagi, Tadashi},
      author={Taki, Yusuke},
       title={Topological pseudo entropy},
        date={2021},
        ISSN={1126-6708},
     journal={J. High Energy Phys.},
      number={9},
       pages={Paper No. 015, 43},
      eprint={2107.01797},
         url={https://doi.org/10.1007/jhep09(2021)015},
}

\bib{Oc75}{article}{
      author={Ochs, Wilhelm},
       title={A new axiomatic characterization of the von~{N}eumann entropy},
        date={1975},
        ISSN={0034-4877},
     journal={Rep. Math. Phys.},
      volume={8},
      number={1},
       pages={109\ndash 120},
  url={http://www.sciencedirect.com/science/article/pii/0034487775900221},
}

\bib{Ohki15}{inproceedings}{
      author={Ohki, Kentaro},
       title={A smoothing theory for open quantum systems: The least mean
  square approach},
        date={2015},
   booktitle={{2015 54th IEEE Conference on Decision and Control (CDC)}},
       pages={4350\ndash 4355},
         url={https://doi.org/10.1109/CDC.2015.7402898},
}

\bib{Ohki17}{article}{
      author={Ohki, Kentaro},
       title={An invitation to quantum filtering and smoothing theory based on
  two inner products (building foundations for quantum statistical modeling)},
        date={2017},
     journal={RIMS K{\^o}ky{\^u}roku},
      volume={2018},
       pages={18\ndash 44},
         url={http://hdl.handle.net/2433/231710},
}

\bib{Ohya1983}{article}{
      author={Ohya, Masanori},
       title={Note on quantum probability},
        date={1983},
        ISSN={0024-1318},
     journal={Lett. Nuovo Cimento},
      volume={38},
      number={11},
       pages={402\ndash 404},
         url={https://doi.org/10.1007/BF02789599},
}

\bib{Oh83b}{article}{
      author={Ohya, Masanori},
       title={On compound state and mutual information in quantum information
  theory},
        date={1983},
     journal={IEEE Trans. Inf. Theory},
      volume={29},
      number={5},
       pages={770\ndash 774},
         url={https://doi.org/10.1109/TIT.1983.1056719},
}

\bib{OhPe93}{book}{
      author={Ohya, Masanori},
      author={Petz, D\'enes},
       title={Quantum entropy and its use},
      series={Texts and Monographs in Physics},
   publisher={Springer-Verlag, Berlin},
        date={1993},
        ISBN={3-540-54881-5},
         url={https://doi.org/10.1007/978-3-642-57997-4},
}

\bib{Pa17}{misc}{
      author={Parzygnat, Arthur~J.},
       title={Discrete probabilistic and algebraic dynamics: a stochastic
  commutative {G}elfand-{N}aimark theorem},
        date={2017},
        note={Preprint available at
  \href{https://arxiv.org/abs/1708.00091}{arXiv:1708.00091 [math.FA]}},
}

\bib{PaBayes}{misc}{
      author={Parzygnat, Arthur~J.},
       title={Inverses, disintegrations, and {B}ayesian inversion in quantum
  {M}arkov categories},
        date={2020},
        note={Preprint available at
  \href{https://arxiv.org/abs/2001.08375}{arXiv:2001.08375 [quant-ph]}},
}

\bib{PaQPL21}{inproceedings}{
      author={Parzygnat, Arthur~J.},
       title={Conditional distributions for quantum systems},
        date={2021},
   booktitle={Proceedings 18th {I}nternational {C}onference on {Q}uantum
  {P}hysics and {L}ogic},
      series={Electronic Proceedings in Theoretical Computer Science (EPTCS)},
}

\bib{AJP22}{article}{
      author={Parzygnat, Arthur~J.},
       title={A functorial characterization of von {N}eumann entropy},
        date={2022},
        ISSN={1245-530X},
     journal={Cah. Topol. G\'{e}om. Diff\'{e}r. Cat\'{e}g.},
      volume={63},
      number={1},
       pages={89\ndash 128},
      eprint={2009.07125},
}

\bib{PaBu22}{article}{
      author={{Parzygnat}, Arthur~J.},
      author={{Buscemi}, Francesco},
       title={Axioms for retrodiction: achieving time-reversal symmetry with a
  prior},
        date={2023},
        ISSN={2521-327X},
     journal={{Quantum}},
      volume={7},
       pages={1013},
      eprint={2210.13531},
         url={https://doi.org/10.22331/q-2023-05-23-1013},
}

\bib{FuPa22a}{article}{
      author={Parzygnat, Arthur~J.},
      author={Fullwood, James},
       title={{From time-reversal symmetry to quantum Bayes' rules}},
        date={2023Jun},
     journal={PRX Quantum},
      volume={4},
       pages={020334},
      eprint={2212.08088},
         url={https://link.aps.org/doi/10.1103/PRXQuantum.4.020334},
}

\bib{PaRuBayes}{article}{
      author={Parzygnat, Arthur~J.},
      author={Russo, Benjamin~P.},
       title={A non-commutative {B}ayes' theorem},
        date={2022},
        ISSN={0024-3795},
     journal={Linear Algebra Its Appl.},
      volume={644},
       pages={28\ndash 94},
      eprint={2005.03886},
  url={https://www.sciencedirect.com/science/article/pii/S0024379522000805},
}

\bib{PaRu19}{article}{
      author={Parzygnat, Arthur~J.},
      author={Russo, Benjamin~P.},
       title={Non-commutative disintegrations: existence and uniqueness in
  finite dimensions},
        date={2023},
     journal={J. Noncommut. Geom.},
      eprint={1907.09689},
         url={https://doi.org/10.4171/JNCG/493},
}

\bib{PTTW23}{misc}{
      author={Parzygnat, Arthur~J.},
      author={Takayanagi, Tadashi},
      author={Taki, Yusuke},
      author={Wei, Zixia},
       title={{SVD Entanglement Entropy}},
        date={2023},
}

\bib{Pa03}{book}{
      author={Paulsen, Vern},
       title={Completely bounded maps and operator algebras},
      series={Cambridge Studies in Advanced Mathematics},
   publisher={Cambridge University Press},
        date={2003},
         url={https://doi.org/10.1017/CBO9780511546631},
}

\bib{Pe55}{article}{
      author={Penrose, Roger},
       title={A generalized inverse for matrices},
        date={1955},
     journal={Math. Proc. Cambridge Philos. Soc.},
      volume={51},
      number={3},
       pages={406\ndash 413},
         url={https://doi.org/10.1017/S0305004100030401},
}

\bib{PGWPR06}{article}{
      author={P{\'{e}}rez-Garc{\'{\i}}a, David},
      author={Wolf, Michael~M.},
      author={Petz, Denes},
      author={Ruskai, Mary~Beth},
       title={Contractivity of positive and trace-preserving maps under lp
  norms},
        date={2006},
     journal={J. Math. Phys},
      volume={47},
      number={8},
       pages={083506},
      eprint={math-ph/0601063},
         url={https://doi.org/10.1063/1.2218675},
}

\bib{Pe71}{article}{
      author={Personick, Stewart~D.},
       title={Application of quantum estimation theory to analog communication
  over quantum channels},
        date={1971},
     journal={IEEE Trans. Inf. Theory},
      volume={17},
      number={3},
       pages={240\ndash 246},
         url={https://doi.org/10.1109/TIT.1971.1054643},
}

\bib{Pe84}{article}{
      author={Petz, D{\'e}nes},
       title={A dual in von~{N}eumann algebras with weights},
        date={1984},
     journal={Q. J. Math.},
      volume={35},
      number={4},
       pages={475\ndash 483},
}

\bib{ReWo14}{article}{
      author={Reeb, David},
      author={Wolf, Michael~M.},
       title={An improved {L}andauer principle with finite-size corrections},
        date={2014},
     journal={New J. Phys.},
      volume={16},
      number={10},
       pages={103011},
}

\bib{Re61}{incollection}{
      author={R{\'e}nyi, Alfr{\'e}d},
       title={On measures of entropy and information},
        date={1961},
   booktitle={Proc. 4th {B}erkeley {S}ympos. {M}ath. {S}tatist. and {P}rob.,
  {V}ol. {I}},
   publisher={Univ. California Press, Berkeley, Calif.},
       pages={547\ndash 561},
}

\bib{ReAh95}{article}{
      author={Reznik, Benni},
      author={Aharonov, Yakir},
       title={Time-symmetric formulation of quantum mechanics},
        date={1995},
     journal={Phys. Rev. A},
      volume={52},
       pages={2538\ndash 2550},
      eprint={quant-ph/9501011},
         url={https://link.aps.org/doi/10.1103/PhysRevA.52.2538},
}

\bib{RZF11}{article}{
      author={Roga, Wojciech},
      author={{\.Z}yczkowski, Karol},
      author={Fannes, Mark},
       title={Entropic characterization of quantum operations},
        date={2011},
     journal={Int. J. Quantum Inf.},
      volume={9},
      number={04},
       pages={1031\ndash 1045},
      eprint={1101.4105},
         url={https://doi.org/10.1142/S0219749911007794},
}

\bib{RyTa06}{article}{
      author={Ryu, Shinsei},
      author={Takayanagi, Tadashi},
       title={Holographic derivation of entanglement entropy from the anti--de
  sitter space/conformal field theory correspondence},
        date={2006},
     journal={Phys. Rev. Lett.},
      volume={96},
       pages={181602},
      eprint={hep-th/0603001},
         url={https://link.aps.org/doi/10.1103/PhysRevLett.96.181602},
}

\bib{SSW14}{article}{
      author={{Salek}, Sina},
      author={{Schubert}, Roman},
      author={{Wiesner}, Karoline},
       title={{Negative conditional entropy of postselected states}},
        date={2014},
     journal={Phys. Rev. A},
      volume={90},
      number={2},
       pages={022116},
      eprint={1305.0932},
}

\bib{Sc96}{article}{
      author={Schumacher, Benjamin},
       title={Sending entanglement through noisy quantum channels},
        date={1996},
     journal={Phys. Rev. A},
      volume={54},
       pages={2614\ndash 2628},
      eprint={quant-ph/9604023},
         url={https://link.aps.org/doi/10.1103/PhysRevA.54.2614},
}

\bib{NiSh96}{article}{
      author={Schumacher, Benjamin},
      author={Nielsen, Michael~A.},
       title={Quantum data processing and error correction},
        date={1996},
     journal={Phys. Rev. A},
      volume={54},
       pages={2629\ndash 2635},
         url={https://link.aps.org/doi/10.1103/PhysRevA.54.2629},
}

\bib{Sh48}{article}{
      author={Shannon, Claude~E.},
       title={A mathematical theory of communication},
        date={1948},
     journal={Bell Labs Tech. J.},
      volume={27},
      number={3},
       pages={379\ndash 423},
}

\bib{Ts22}{article}{
      author={Tsang, Mankei},
       title={Generalized conditional expectations for quantum retrodiction and
  smoothing},
        date={2022},
     journal={Phys. Rev. A},
      volume={105},
       pages={042213},
      eprint={1912.02711},
         url={https://link.aps.org/doi/10.1103/PhysRevA.105.042213},
}

\bib{Ts22b}{misc}{
      author={Tsang, Mankei},
       title={Operational meaning of a generalized conditional expectation in
  quantum metrology},
        date={2022},
         url={https://arxiv.org/abs/2212.13162},
        note={Preprint available at
  \href{https://arxiv.org/abs/2212.13162}{arXiv:2212.13162 [quant-ph]}},
}

\bib{TTC22}{article}{
      author={Tu, Yi-Ting},
      author={Tzeng, Yu-Chin},
      author={Chang, Po-Yao},
       title={{R{\'e}nyi entropies and negative central charges in
  non-Hermitian quantum systems}},
        date={2022},
     journal={SciPost Phys.},
      volume={12},
       pages={194},
      eprint={2107.13006},
         url={https://scipost.org/10.21468/SciPostPhys.12.6.194},
}

\bib{Wat55}{article}{
      author={Watanabe, Satosi},
       title={{Symmetry of Physical Laws. Part III. Prediction and
  Retrodiction}},
        date={1955},
     journal={Rev. Mod. Phys.},
      volume={27},
       pages={179\ndash 186},
         url={https://link.aps.org/doi/10.1103/RevModPhys.27.179},
}

\bib{Wa18}{book}{
      author={Watrous, John},
       title={The theory of quantum information},
   publisher={Cambridge University Press},
        date={2018},
         url={https://doi.org/10.1017/9781316848142},
}

\bib{We17}{inproceedings}{
      author={Westerbaan, Abraham},
       title={{Quantum Programs as Kleisli Maps}},
        date={2017},
   booktitle={{ Proceedings 13th International Conference on Quantum Physics
  and Logic, Glasgow, Scotland, 6-10 June 2016}},
      editor={Duncan, Ross},
      editor={Heunen, Chris},
      series={Electronic Proceedings in Theoretical Computer Science},
      volume={236},
   publisher={Open Publishing Association},
       pages={215\ndash 228},
         url={https://doi.org/10.4204/eptcs.236.14},
}

\bib{Wigner32}{article}{
      author={Wigner, Eugene},
       title={On the quantum correction for thermodynamic equilibrium},
        date={1932},
     journal={Phys. Rev.},
      volume={40},
       pages={749\ndash 759},
         url={https://link.aps.org/doi/10.1103/PhysRev.40.749},
}

\bib{Wilde2017}{book}{
      author={Wilde, Mark~M.},
       title={Quantum information theory},
     edition={2},
   publisher={Cambridge University Press},
        date={2017},
         url={https://doi.org/10.1017/9781316809976},
}

\bib{Woot87}{article}{
      author={Wootters, William~K.},
       title={A {W}igner-function formulation of finite-state quantum
  mechanics},
        date={1987},
        ISSN={0003-4916},
     journal={Ann. Physics},
      volume={176},
      number={1},
       pages={1\ndash 21},
         url={https://doi.org/10.1016/0003-4916(87)90176-X},
}

\end{biblist}
\end{bibdiv}

\Addresses

\end{document}